\documentclass[twocolumn]{aastex63}

\usepackage{comment}
\usepackage{amsmath}
\usepackage{rotating}
\usepackage{layouts}
\usepackage{tasks}
\usepackage{enumitem}

\newcommand\apogeesourcesfourarcmin{172}

\newcommand\kinematicsample{135}

\newcommand\nirspecsources{56}

\graphicspath{{./}{figures/}}

\submitjournal{ApJ}

\shorttitle{ONC Core Kinematics}
\shortauthors{Theissen et al.}
\begin{document}

\title{The 3-D Kinematics of the Orion Nebula Cluster: NIRSPEC-AO Radial Velocities of the Core Population}

\correspondingauthor{Christopher A. Theissen}
\email{ctheissen@ucsd.edu}

\author[0000-0002-9807-5435]{Christopher A. Theissen}
\altaffiliation{NASA Sagan Fellow}
\affil{Center for Astrophysics and Space Sciences, University of California, San Diego,
9500 Gilman Dr, La Jolla, CA 92093, USA}

\author[0000-0002-9936-6285]{Quinn M. Konopacky}
\affil{Center for Astrophysics and Space Sciences, University of California, San Diego,
9500 Gilman Dr, La Jolla, CA 92093, USA}

\author[0000-0001-9611-0009]{Jessica R. Lu}
\affil{Department of Astronomy, University of California, Berkeley, Berkeley, CA, 94720-3411, USA}

\author[0000-0002-6658-5908]{Dongwon Kim}
\affil{Department of Astronomy, University of California, Berkeley, Berkeley, CA, 94720-3411, USA}

\author[0000-0003-2122-6714]{Stella Y. Zhang}
\affil{Center for Astrophysics and Space Sciences, University of California, San Diego,
9500 Gilman Dr, La Jolla, CA 92093, USA}

\author[0000-0002-5370-7494]{Chih-Chun Hsu}
\affil{Center for Astrophysics and Space Sciences, University of California, San Diego,
9500 Gilman Dr, La Jolla, CA 92093, USA}

\author[0000-0002-1437-4463]{Laurie Chu}
\altaffiliation{NASA Postdoctoral Program Fellow}
\affil{NASA Ames Research Center, Space Science and Astrobiology Division, Mail Stop 245-3, Moffett Field, CA 94035 USA}

\author[0000-0002-2612-2933]{Lingfeng Wei}
\affil{Center for Astrophysics and Space Sciences, University of California, San Diego,
9500 Gilman Dr, La Jolla, CA 92093, USA}

\begin{abstract}

The kinematics and dynamics of stellar and substellar populations within young, still-forming clusters provides valuable information for constraining theories of formation mechanisms. Using Keck II NIRSPEC+AO data, we have measured radial velocities for \nirspecsources\ low-mass sources within 4\arcmin\ of the core of the Orion Nebula Cluster (ONC). We also re-measure radial velocities for \apogeesourcesfourarcmin\ sources observed with SDSS/APOGEE. These data are combined with proper motions measured using \textit{HST} ACS/WFPC2/WFC3IR and Keck II NIRC2, creating a sample of \kinematicsample\ sources with all three velocity components. The velocities measured are consistent with a normal distribution in all three components. 
We measure intrinsic velocity dispersions of ($\sigma_{v_\alpha}$, $\sigma_{v_\delta}$, $\sigma_{v_r}$) = ($1.64\pm0.12$, $2.03\pm0.13$, $2.56^{+0.16}_{-0.17}$) km s$^{-1}$. 
Our computed intrinsic velocity dispersion profiles are consistent with the dynamical equilibrium models from \citet{da-rio:2014:55} in the tangential direction, but not in the line of sight direction, possibly indicating that the core of the ONC is not yet virialized, and may require a non-spherical potential to explain the observed velocity dispersion profiles. 
We also observe a slight elongation along the north-south direction following the filament, which has been well studied in previous literature, and an elongation in the line of sight to tangential velocity direction. 
These 3-D kinematics will help in the development of realistic models of the formation and early evolution of massive clusters.
\end{abstract}


\section{Introduction} 
\label{sec:intro}

The Orion Nebula Cluster (ONC) represents one of the best laboratories for studies of cluster formation and dynamics. Since the vast majority of stars are expected to form in clusters \citep{allen:2007:492,carpenter:2000:3139,lada:2003:57}, understanding cluster formation is of paramount importance to constraining star formation theory. The ONC is one of the closest \citep[$d \approx 390$~pc;][]{kounkel:2017:142} examples of massive star formation, covering a large range of source masses \citep[0.1--50~$M_\Sun$][]{hillenbrand:1997:1733}. There is also evidence to suggest that the cluster is not yet dynamically relaxed \citep{furesz:2008:1109,tobin:2009:1103}, which is consistent with its youth \citep[$\sim$2.2 Myr;][]{reggiani:2011:a83}.

Early studies of the kinematics of the ONC identified mass segregation, which is expected for young clusters \citep{hillenbrand:1997:1733,hillenbrand:1998:540}. However, it remains an open question as to whether the observed mass segregation is primordial or dynamical. Given the estimated age of the ONC ($\sim$2.2~Myr), and the cluster crossing time ($\sim$2~Myr), it is thought that the mass segregation is primordial \citep{reggiani:2011:a83}. However, these studies relied on a limited sample of 3-D kinematic information, largely resulting from the high level of extinction in the region \citep[e.g.,][]{johnson:1965:964, walker:1983:642, jones:1988:1755, van-altena:1988:1744}.

The largest scale radial velocity (RV) studies of the ONC began with \citet{furesz:2008:1109} and \citet{tobin:2009:1103}, who observed 1215 and 1613 stars, respectively. These observations were taken using the Hectochelle multiobject echelle spectrograph \citep{szentgyorgyi:1998:242} on the 6.5-m MMT telescope, and the Magellan Inamori Kyocera Echelle \citep[MIKE;][]{bernstein:2003:1694,walker:2007:389} on the Magellan Clay telescope. These fiber-fed instruments obtain  high-resolution ($\lambda/\Delta\lambda \approx 35,000$), optical (5150--5300~\AA) spectra, but are limited in their ability to observe stars in highly embedded regions and/or crowded fields (minimum separations of 30\arcsec\ for Hectochelle and 14\arcsec\ for MIKE). In particular, the core of the ONC---within 1\arcmin\ of the Trapezium---had poor coverage (5 sources) due to the highly embedded and crowded nature of the core.

More recently, the Sloan Digital Sky Survey \citep[SDSS;][]{york:2000:1579}, SDSS-III \citep{eisenstein:2011:72} Apache Point Observatory Galactic Evolution Experiment \citep[APOGEE;][]{majewski:2017:94} conducted an infrared (IR) survey of the ONC \citep{da-rio:2016:59,da-rio:2017:105}. The APOGEE spectrograph \citep{wilson:2010:77351c} on the Sloan 2.5-m telescope \citep{gunn:1998:3040} is a fiber-fed instrument which obtains $H$-band spectra (1.51--1.68~$\mu$m) at a resolution of $\lambda/\Delta\lambda \approx 22,500$. This makes surveys using the APOGEE spectrograph less affected by extinction in embedded regions, but they are still limited in their ability to observe objects in crowded regions due to the 2\arcsec\ diameter fibers. The most recent compilation of APOGEE measurements (APOGEE-II) was given in \citet[][hereafter K18]{kounkel:2018:84}, which included measurements from SDSS Data Release 12 \citep[DR12;][]{alam:2015:12} and Data Release 14 \citep[DR14;][]{abolfathi:2018:42}. K18 presented high-precision ($\sim$0.2~km s$^{-1}$) RVs for 7774 sources in the ONC, but only included 12 sources within 1\arcmin\ of the Trapezium.

Here, we present a study of the 3-D kinematics of the ONC sources within 4\arcmin\ of the Trapezium. Our sample consists of \nirspecsources\ sources observed with the Near InfarRed echelle SPECtrograph \citep[NIRSPEC;][]{mclean:1998:566,mclean:2000:1048} on the Keck II 10-meter telescope coupled with adaptive optics (AO), and a reanalysis of \apogeesourcesfourarcmin\ sources observed with SDSS/APOGEE. This combined ONC sample represents the largest sample to date of RVs within the core of the ONC ($\lesssim 1\arcmin$; 41 sources). In Section~\ref{sec:data}, we discuss the literature data and new observations used in this study. Section~\ref{sec:nirspec} describes the methods used in reducing and forward-modeling the NIRSPEC+AO (NIRSPAO) data. We discuss a reanalysis of the SDSS/APOGEE data using our forward-modeling pipeline in Section~\ref{sec:apogee}. A detailed study of the 3-D kinematics of the ONC core is undertaken in Section~\ref{sec:kinematics}. Lastly, a discussion of our results is given in Section~\ref{sec:discussion}.

\section{Data}
\label{sec:data}

\begin{figure}[!htbp]
\centering
\includegraphics[width=\linewidth]{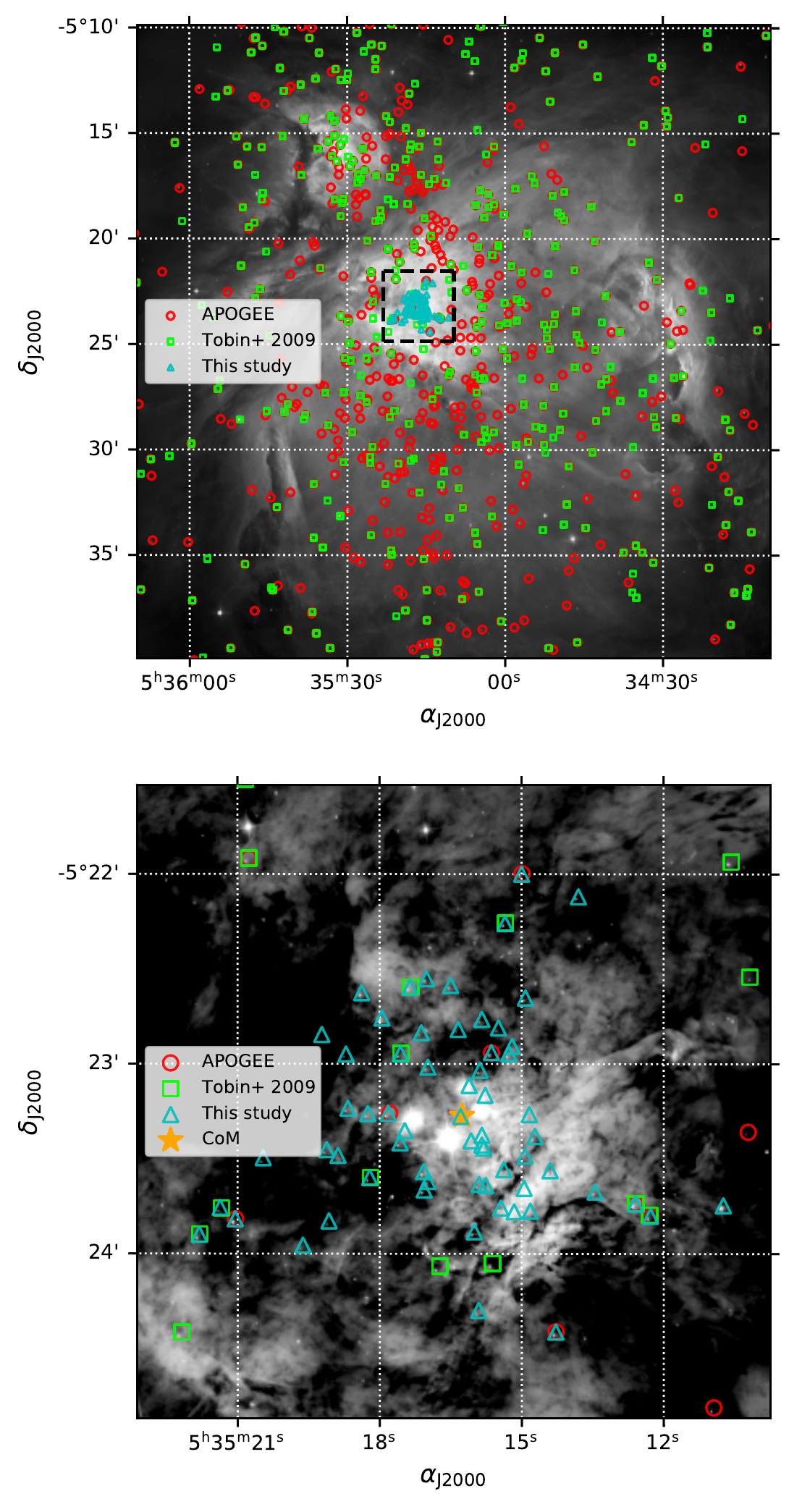}
\caption{
\emph{Top}: \emph{HST} ACS $R$-band image of the ONC \citep{robberto:2013:10}. Plotted are K18 APOGEE sources (red circles), T09 sources (green squares), and NIRSPAO sources from this study (cyan triangles). The black dashed box indicates the area shown in the bottom figure.
\emph{Bottom}: Close up image of the Trapezium. Sources are plotted with markers indicated in the legend. Our sources primarily represent the reddest sources closest to the Trapezium. The center-of-mass coordinates from \citet{da-rio:2014:55} are indicated with the orange star.
\label{fig:onc}
}
\end{figure}

We obtained high-resolution near-infrared (NIR) spectra of sources closest to the core of the ONC---surrounding the Trapezium---using Keck/NIRSPAO between 2015--2020. Targets were initially chosen from a preliminary catalog of proper motions (PMs) computed using Keck Near Infrared Camera 2 (NIRC2; PI: K. Matthews) data. These sources were then cross-referenced with the \citet{hillenbrand:2000:236} study of the low-mass members of the ONC. Figure~\ref{fig:onc} shows the targets from this study, as well as sources with RVs from the optical survey of \citet[][hereafter T09]{tobin:2009:1103} and the NIR survey using SDSS/APOGEE presented in \citet[][hereafter K18]{kounkel:2018:84}. %From these data, we obtained  to study the 3-D kinematics of the core of the ONC
For the remainder of this study, we use the center-of-mass (CoM) coordinates determined by \citet[$\alpha_\mathrm{J2000} = 05$:35:16.26; $\delta_\mathrm{J2000} = -05$:23:16.4;][]{da-rio:2014:55} to represent the ``center" of the ONC.

%\clearpage

\subsection{NIRSPAO Observations}
\label{sec:observations}

All objects in our curated sample were observed using NIRSPEC on Keck II, in conjunction with the laser guide star (LGS) adaptive optics (AO) system \citep{van-dam:2006:310,wizinowich:2006:297}. 
We used the instrument in its high-spectral resolution AO mode with a slit width $\times$ slit length of $0.041\arcsec \times 2.26\arcsec$. Observations were carried out in the $K$- or $N7$-band to capitalize on the strong CO bands within that regime ($\sim$2.3~$\mu$m, orders 32 \& 33). We additionally used a cross-disperser angle of 35.65$^\circ$ and an echelle angle of 63$^\circ$. The resolution in this setup is $R \approx$ 25,000, as determined by the width of unresolved OH sky lines, and the wavelengths covered are approximately 2.044--2.075~$\mu$m (order 37), 2.100--2.133~$\mu$m (order 36), 2.160--2.193~$\mu$m (order 35), 2.224--2.256~$\mu$m (order 34), 2.291--2.325~$\mu$m (order 33), and 2.362--2.382~$\mu$m (order 32), with some portions of the bands beyond the edges of the detector. For this work, all analysis was done using orders 32 and 33, the orders containing the CO bandheads in addition to numerous telluric features, and all NIRSPAO data presented come from these orders. We note that NIRSPEC underwent an upgrade during 2018, and our setup changed slightly post-upgrade. The most significant change is a larger wavelength coverage for each order due to a larger Hawaii 2RG detector ($2048\times2048$ post-upgrade versus $1024\times1024$ pre-upgrade; \citealt{martin:2018:107020a}), and a higher resolution \citep[$R \approx$ 35,000;][]{hsu:2021:arxiv:2107.01222}.

Typical observations consisted of four spectra taken in an ABBA dither pattern along the length of the slit. In a few cases, more or less than four spectra were taken. Either before or after each target was observed, an A0V calibrator star at similar airmass was observed for a telluric reference. Table~\ref{tbl:observations} gives the log of our spectroscopic observations, listing the targets observed, the date of observation, the number of spectra, and the integration time for each spectrum. Etalon lamp and flat field frames were also taken each night for use in data reduction (Section~\ref{sec:nirspec}).

\startlongtable
\begin{deluxetable}{cccccc}
\tablecaption{Log of NIRSPAO-LGS Observations \label{tbl:observations}}
\tablecolumns{6}
\tabletypesize{\scriptsize}
%\tablenum{5}
%\tablewidth{0pt}
\tablehead{
\colhead{Target} & \colhead{Date of} & \colhead{A0V Star} & \colhead{Exposure Time} & \colhead{No. of} & \colhead{Filter} \\
\colhead{Name\tablenotemark{a}} & \colhead{Obs. (UT)} &
\colhead{Standard} & \colhead{(s $\times$ coadds)} & \colhead{Frames} & 
}
\startdata
HC322 & 2015 Dec 23 & HD 37887 & $300 \times 1$ & 4 & $K$  \\
HC296 & 2015 Dec 23 & HD 37887 & $1200\times 1$ & 4 & $K$  \\
HC259 & 2015 Dec 23 & HD 37887 & $90  \times 1$ & 4 & $K$  \\
HC213 & 2015 Dec 23 & HD 37887 & $60  \times 1$ & 4 & $K$  \\
HC306 & 2015 Dec 24 & HD 37887 & $180 \times 1$ & 4 & $K$  \\
HC287 & 2015 Dec 24 & HD 37887 & $600 \times 1$ & 4 & $K$  \\
HC291 & 2015 Dec 24 & HD 37887 & $600 \times 1$ & 4 & $K$  \\
HC252 & 2015 Dec 24 & HD 37887 & $300 \times 1$ & 4 & $K$  \\
HC250 & 2015 Dec 24 & HD 37887 & $1200\times 1$ & 4 & $K$  \\
HC244 & 2015 Dec 24 & HD 37887 & $600 \times 1$ & 4 & $K$  \\
HC261 & 2015 Dec 24 & HD 37887 & $900 \times 1$ & 4 & $K$  \\
HC248 & 2016 Dec 14 & HD 37887 & $600 \times 1$ & 4 & $K$  \\
HC223 & 2016 Dec 14 & HD 37887 & $300 \times 1$ & 4 & $K$  \\
HC219 & 2016 Dec 14 & HD 37887 & $600 \times 1$ & 4 & $K$  \\
HC324 & 2016 Dec 14 & HD 37887 & $1200\times 1$ & 3 & $K$  \\
HC295 & 2018 Feb 11 & HD 37887 & $450 \times 1$ & 5 & $K$  \\
HC313 & 2018 Feb 11 & HD 37887 & $180 \times 1$ & 4 & $K$  \\
HC332 & 2018 Feb 11 & HD 37887 & $300 \times 1$ & 4 & $K$  \\
HC331 & 2018 Feb 11 & HD 37887 & $450 \times 1$ & 4 & $K$  \\
HC337 & 2018 Feb 11 & HD 37887 & $60  \times 1$ & 4 & $K$  \\
HC375 & 2018 Feb 11 & HD 37887 & $180 \times 1$ & 4 & $K$  \\
HC338 & 2018 Feb 11 & HD 37887 & $120 \times 1$ & 4 & $K$  \\
HC425 & 2018 Feb 12 & HD 37887 & $60  \times 1$ & 4 & $K$  \\
HC713 & 2018 Feb 12 & HD 37887 & $90  \times 1$ & 4 & $K$  \\
HC408 & 2018 Feb 12 & HD 37887 & $450 \times 1$ & 4 & $K$  \\
HC410 & 2018 Feb 12 & HD 37887 & $600 \times 1$ & 4 & $K$  \\
HC436 & 2018 Feb 12 & HD 37887 & $90  \times 1$ & 4 & $K$  \\
HC442 & 2018 Feb 13 & HD 37887 & $450 \times 1$ & 4 & $K$  \\
HC344 & 2018 Feb 13 & HD 37887 & $60  \times 1$ & 4 & $K$  \\
HC522 & 2019 Jan 12 & HD 37887 & $450 \times 1$ & 4 & $K$  \\
HC145 & 2019 Jan 12 & HD 37887 & $600 \times 1$ & 4 & $K$  \\
HC202 & 2019 Jan 12 & HD 37887 & $120 \times 1$ & 4 & $K$  \\
HC188 & 2019 Jan 12 & HD 37887 & $600 \times 1$ & 4 & $K$  \\
HC302 & 2019 Jan 13 & HD 37887 & $450 \times 1$ & 4 & $N7$  \\
HC275 & 2019 Jan 13 & HD 37887 & $450 \times 1$ & 2 & $N7$  \\
HC245 & 2019 Jan 13 & HD 37887 & $180 \times 1$ & 4 & $N7$  \\
HC258 & 2019 Jan 13 & HD 37887 & $180 \times 1$ & 4 & $N7$  \\
HC344 & 2019 Jan 13 & HD 37887 & $120 \times 1$ & 4 & $N7$  \\
HC370 & 2019 Jan 16 & HD 37887 & $180 \times 1$ & 4 & $N7$  \\
HC389 & 2019 Jan 16 & HD 37887 & $120 \times 1$ & 4 & $N7$  \\
HC386 & 2019 Jan 16 & HD 37887 & $120 \times 1$ & 4 & $N7$  \\
HC398 & 2019 Jan 16 & HD 37887 & $120 \times 1$ & 4 & $N7$  \\
HC413 & 2019 Jan 16 & HD 37887 & $180 \times 1$ & 4 & $N7$  \\
HC253 & 2019 Jan 16 & HD 37887 & $120 \times 1$ & 4 & $N7$  \\
HC288 & 2019 Jan 17 & HD 37887 & $450 \times 1$ & 4 & $N7$  \\
HC420 & 2019 Jan 17 & HD 37887 & $450 \times 1$ & 4 & $N7$  \\
HC412 & 2019 Jan 17 & HD 37887 & $450 \times 1$ & 4 & $N7$  \\
HC288 & 2019 Jan 17 & HD 37887 & $450 \times 1$ & 4 & $N7$  \\
HC282 & 2019 Jan 17 & HD 37887 & $450 \times 1$ & 3 & $N7$  \\
HC277 & 2019 Jan 17 & HD 37887 & $180 \times 1$ & 4 & $N7$  \\
HC217 & 2020 Jan 18 & HD 37887 & $120 \times 1$ & 4 & $N7$  \\ 
HC229 & 2020 Jan 18 & HD 37887 & $450 \times 1$ & 4 & $N7$  \\
HC229 & 2020 Jan 18 & HD 37887 & $450 \times 1$ & 4 & $N7$  \\
HC228 & 2020 Jan 19 & HD 37887 & $120 \times 1$ & 4 & $N7$  \\
HC224 & 2020 Jan 19 & HD 37887 & $90  \times 1$ & 4 & $N7$  \\
HC135 & 2020 Jan 19 & HD 37887 & $180 \times 1$ & 4 & $N7$  \\
HC440 & 2020 Jan 20 & HD 37887 & $120 \times 1$ & 4 & $N7$  \\
HC450 & 2020 Jan 20 & HD 37887 & $300 \times 1$ & 4 & $N7$  \\
HC277 & 2020 Jan 20 & HD 37887 & $300 \times 1$ & 4 & $N7$  \\
HC204 & 2020 Jan 20 & HD 37887 & $300 \times 1$ & 4 & $N7$  \\
HC229 & 2020 Jan 20 & HD 37887 & $450 \times 1$ & 4 & $N7$  \\
HC214 & 2020 Jan 20 & HD 37887 & $450 \times 1$ & 4 & $N7$  \\
HC215 & 2020 Jan 21 & HD 37887 & $300 \times 1$ & 4 & $N7$  \\
HC240 & 2020 Jan 21 & HD 37887 & $300 \times 1$ & 4 & $N7$  \\
HC546 & 2020 Jan 21 & HD 37887 & $300 \times 1$ & 4 & $N7$  \\
HC504 & 2020 Jan 21 & HD 37887 & $120 \times 1$ & 4 & $N7$  \\
HC703 & 2020 Jan 21 & HD 37887 & $300 \times 1$ & 4 & $N7$  \\
HC431 & 2020 Jan 21 & HD 37887 & $120 \times 1$ & 3 & $N7$  \\
HC229 & 2020 Jan 21 & HD 37887 & $450 \times 1$ & 2 & $N7$  \\
\enddata
\tablenotetext{a}{Identifier from \citet{hillenbrand:2000:236}.}
\end{deluxetable}

\section{NIRSPAO Reduction}
\label{sec:nirspec}

Reduction of NIRSPAO data was done using a modified version of the NIRSPEC Data Reduction Pipeline (NSDRP\footnote{\url{https://www2.keck.hawaii.edu/koa/nsdrp/nsdrp.html}}\footnote{\url{https://github.com/Keck-DataReductionPipelines/NIRSPEC-Data-Reduction-Pipeline}}). The NSDRP was specifically designed for point source extraction, and has been used extensively to obtain ``quick looks" while observing at the Keck facility and for the Keck Observatory Archive \citep[KOA;][]{berriman:2005:627, berriman:2010:119, tran:2012:845129}. Our updated version\footnote{\url{https://github.com/ctheissen/NIRSPEC-Data-Reduction-Pipeline}} includes the following modifications:
\begin{enumerate}
 \item Use with the $K$-AO observing mode.
 \item Spatial rectification using the object trace rather than the order edge traces.
 \item Spectral rectification and wavelength calibration using etalon lamps.
 \item Cosmic ray cleaning of flats.
 \item Bad pixel cleaning using methods ported from \texttt{fixpix\_rs.pro} which is a utility from REDSPEC \citep{kim:2015:,prato:2015:12}.
\end{enumerate}
In addition to the above modifications, we also removed fringes from the flat field images prior to median combining using the wavelet method of \citet{rojo:2006:553}. This helps to mitigate beat patterns that can appear between fringing in the flats and science data.

Data for each night was reduced using standard procedures following the NSDRP documentation\footnote{\url{https://www2.keck.hawaii.edu/koa/nsdrp/nsdrp.html}}. 
We provide a summary of the steps involved in the reduction process here.
\begin{enumerate}
    \item Flat frames are median combined into a master flat frame.
    \item The master flat is used to find order edges using pre-determined dispersions based on the grating equation for NIRSPEC.
    \item Each frame (i.e., object, etalon/arc lamp) is cleaned for cosmic rays using Laplacian edge detection \citep{van-dokkum:2001:1420}.
    \item Each frame is flat normalized and orders are extracted individually using the edges traces found from the master flat frame.
    \item Each order is cleaned for bad pixels using the \texttt{fixpix} routine.
    \item Each order is spatially rectified using the object trace, determined by a Gaussian profile fit along each column.
    \item Order edges are trimmed to remove bad pixels.
    \item Each order is spectrally rectified using either the etalon or the arc lamp frame. The spectral trace is done by fitting Gaussians to the emission line traces, and then finding the optimal spectral tilt (y-direction).
    \item The object is extracted using box extraction, using optimal object and background regions found from Gaussian fits to the profile. The average sky (calculated using background regions adjacent to the 2-D object spectrum) is subtracted from the object spectrum.
    \item Flux and noise are calculated using standard methods in the NSDRP.
    \item The wavelength solution is calibrated for each order using a synthesized etalon or sky spectrum and fitting to lines found in each order.
\end{enumerate}

Initial wavelength solutions were obtained by mapping pixels to the etalon lamp wavelengths; however, etalon lamps only provide uniform spacing in the frequency domain, with the initial absolute or starting position unknown. For example, it is unknown whether the first etalon fringe starts at 20000~\AA\ or 20010~\AA\, but the spacing of $c\lambda^{-1}$ is absolute. Therefore, the wavelength solution is re-calibrated using telluric features within each frame, which are anchored to an absolute rest-frame, as part of our forward-modeling framework (Section~\ref{sec:mcmc}).

\subsection{Forward-Modeling NIRSPEC Data}
\label{sec:mcmc}

Our data were forward-modeled using the Spectral Modeling Analysis and RV Tool \citep[SMART\footnote{\url{https://github.com/chihchunhsu/smart}};][]{hsu:2021:}. Details for the fitting routine using SMART are given in \citet{hsu:2021:arxiv:2107.01222}. Here, we briefly outline our methods.

Reduced NIRSPAO data were modeled using an iterative approach. The first step was obtaining an absolute wavelength solution to orders 32 and 33. For our initial wavelength solution, we use the global wavelength solution provided by the NSDRP---which is a quadratic polynomial\footnote{\url{https://www2.keck.hawaii.edu/koa/nsdrp/documents/NSDRP_Software_Design.pdf}} of the form
\begin{equation}
\lambda(p, M) = r_0 + r_1 p + r_2 p^2 + \frac{r_3}{M} + \frac{r_4 p}{M} + \frac{r_5 p^2}{M},
\end{equation} 
\noindent where $\lambda(p, M)$ is the wavelength falling on column pixel $p$ of order $M$ and coefficients $r_n$. This initial wavelength solution is obtained from fitting to the etalon lamps, which provides uniform offsets in frequency space. However, as mentioned previously, the absolute wavelength solution, or starting position, is unknown for the etalon spectrum.

To obtain a more precise a wavelength calibration, we performed a cross-correlation between our telluric spectrum from the A-star calibrator to the high-resolution telluric spectrum from \citet{moehler:2014:16}. First, we modeled and removed the continuum of our A-star calibrator using a quadratric polynomial, essentially leaving just the flat imprinted telluric absorption spectrum. Then, we scaled the flux of the high-resolution telluric model to the A-star flux. Next, the telluric spectrum from the A0V star was cross-correlated with the telluric model using a window of 100 pixels, and a step size of 20 pixels, calculating the best fit cross-correlation shift for each window along the entire spectrum. Then, a 4th order polynomial---i.e., $\lambda(p) = a_i + b_i p + c_i p^2 + d_i p^3 + e_i p^4$, where $i$ is the iteration number---was fit to the best wavelength shifts for all windows used in that iteration. The initial wavelength solution was given using the 2nd order polynomial provided by the NSDRP, with the additional coefficients set to zero (i.e., $d_0$, $e_0$). The best-fit coefficients for each pass (e.g., $\delta_{a_0}$) were added to the previous solution, e.g., $a_1 = a_0 + \delta_{a_1}$, $b_1 = b_0 + \delta_{b_1}$. This loop was repeated until the wavelength solution converged to the smallest residuals between the telluric spectrum and telluric model. One thing to note is that different iterations used different pixel window and step sizes as finer granularity is required as the wavelength solution gets closer to the optimal fit. For instance, the second pass used a step size of 10 pixels and a window size of 150 pixels.

The aforementioned fitting procedure was done for each frame independently, resulting in an absolute wavelength solution for each frame. We cross-correlated a single A star frame to A star calibration data taken over 14 nights (both A and B nods), and found a RMS between frames of 0.004 \AA\ (0.058 km s$^{-1}$). Although this systematic uncertainty is typically much smaller than our measurement uncertainty, we add this systematic in quadrature with our measurement uncertainties per frame.

Next, we modeled the spectrum using a forward-modeling approach based on the method provided by \citet{blake:2010:684} and also \citet{butler:1996:500, blake:2007:1198, blake:2008:l125, blake:2010:684}, and \citet{burgasser:2016:25}, utilizing a Markov Chain Monte Carlo (MCMC) method built on the \textit{emcee} \citep{foreman-mackey:2013:306} to sample the parameter space. 

The flux from the source can be modeled using the following equation,
\begin{equation}\label{nirspecmodel}
\begin{aligned}
F_M[p] ={} & C[p(\lambda)]\times \\
& \left[\left(\left(M\left[p^{\ast}\left(\lambda \left[1 + \frac{V_r}{c}\right]\right),\, T_\mathrm{eff}, \log g, \mathrm{[M/H]} \right]\right) \right.\right. \\
& \ast \kappa_R(v\sin{i}) + C_\mathrm{veil} \bigg) \times T[p^{\ast}(\lambda), AM, PWV]\bigg] \\
& \ast \kappa_G(\Delta \nu_\mathrm{inst}) + C_\mathrm{flux},
\end{aligned}
\end{equation}
where $\ast$ indicates convolution, $C[p(\lambda)]$ is a quadratic polynomial which is used for continuum correction---this is meant to correct and scale for variations induced by the instrumental profile on the observed flux and the absolute flux of the model, $M$ is the photospheric model of the source. We fixed $\log g = 4$ as this is approximately the expected gravity for low-mass stars at the age of the ONC which are still contracting onto the main sequence, and consistent with other studies \citep[e.g.,][]{kounkel:2018:84}. Additionally, we show later that RV variations with $\log g$ tend to be small.
We also fixed [M/H] $= 0$ based on the average metallicity of the ONC \citep[e.g.,][]{dorazi:2009:973}. 
$\kappa_R(v\sin{i})$ is the rotational broadening kernel (using the methods of \citealt{gray:1992:} and a limb-darkening coefficient of 0.6), $V_r$ and $c$ are the heliocentric RV and speed of light, respectively, $T$ is the telluric spectrum from \citet{moehler:2014:16}, and $AM$ and $PWV$ are the airmass and precipitable water vapor of the telluric spectrum, respectively. Veiling is parameterized by $C_\mathrm{veil}$, which is an additive gray body flux to the stellar model flux (continuum), $r_\lambda = F_{\lambda, \mathrm{cont}}/F_{\lambda, \ast}$ (or $C_\mathrm{veil}/F_{\lambda, \ast}$), to represent potential veiling along the line of sight, which will weaken the depths of the photospheric absorption lines \citep[e.g.,][]{fischer:2011:73, muzerolle:2003:l149, muzerolle:2003:266}. Extinction effects, which reduce the intensity of the emission lines, are multiplicative in nature, and therefore get folded into the fit for $C[p(\lambda)]$. We note that due to the small range of wavelengths used, effects due to extinction should be minimal, and would mostly impact stellar parameters rather than radial velocities.

Here, $p^{\ast}=p(\lambda) + C_\lambda$, where $p(\lambda)$ is a 4$^\mathrm{th}$ order polynomial mapping of pixel to wavelength based on the telluric spectrum from the absolute wavelength calibration to the A star, and $C_\lambda$ is a small constant offset/correction to the zeroth-order term. We keep all coefficients in the 4$^\mathrm{th}$ order polynomial constant, which were derived from the A star calibrator, save for the zeroth order term, which we fit for a small constant offset (nuisance parameter) to account for small differences in the absolute wavelength calibration versus the observed data. This is necessary as observations not taken along the exact same pixels will shift by a small amount due to the spatial curvature of the flux along the detector. $\kappa_G(\Delta \nu_\mathrm{inst})$ is the spectrograph line spread function (LSF), modeled as a normalized Gaussian. We include an additive flux offset, $C_\mathrm{flux}$, as an additional nuisance parameter to account for small differences in the absolute flux calibration. We fit for separate $C_\lambda$ and $C_\mathrm{flux}$ parameters for each order. For stellar models, we chose the PHOENIX-ACES-AGSS-COND-2011 stellar models \citep{husser:2013:a6}, which have been used previously for modeling ONC young stellar objects (YSOs) observed in the NIR with SDSS/APOGEE \citep{kounkel:2018:84}

The log-likelihood function, assuming normally distributed parameters and noise, is
\begin{equation}\label{eqn:likelihood}
\ln \mathcal{L} = -\frac{1}{2} \sum_p \left[\left(\frac{D[p] - F_M[p]}{\sigma [p] \times C_\mathrm{noise}}\right)^2 + \ln[2 \pi (\sigma[p] \times C_\mathrm{noise})^2]\right],
\end{equation}
\noindent where $D[p]$ is the data, $\sigma[p]$ is the noise, or flux uncertainty, and $C_\mathrm{noise}$ is a scaling parameter for the noise to account for systematic errors between the observations and the models.

We simultaneously fit for all of the above parameters, using the affine-invariant ensemble sampler \textit{emcee}, with the kernel-density estimator (KDE) described in \citet{farr:2014:024014}. Uniform priors were used across the parameter ranges shown in Table~\ref{tbl:forwardmodel}. We did an initial fit using 100 walkers and 400 steps, discarding the first 300 steps. Typical convergence occurred after the initial 200 steps. These fits were then masked for bad pixels outside of three standard deviations of the median difference between the model and the data, effectively removing bad pixels and cosmic rays that were not removed by the \texttt{fixpix} utility. Next, another fit was performed on the masked data with 100 walkers, 300 steps, and a burn-in of 200 steps. Walkers were initialized within 10\% of the best-fit parameters from the initial fit. Convergence typically occurred within 100 steps. Heliocentric RVs were corrected for barycentric motion using the \textit{astropy} function \texttt{radial\_velocity\_correction}.

\begin{figure*}[!htbp]
\centering
\includegraphics[width=\textwidth]{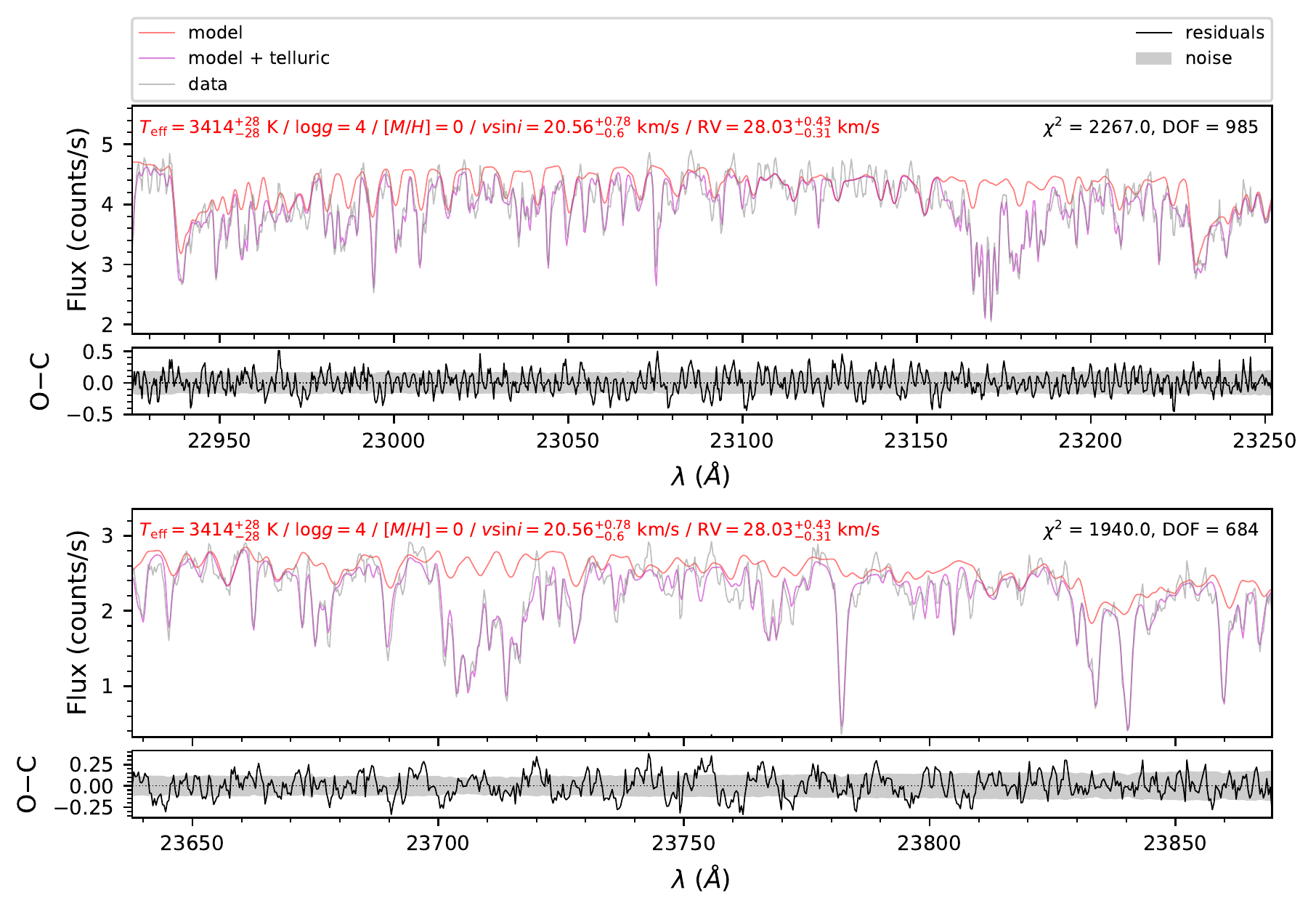}
\caption{Best-fit forward model of a single frame for [HC2000] 244, orders 33 (top) and 32 (bottom). Plotted are the data (light gray lines), stellar model (red lines), and stellar model multiplied by the telluric spectrum (magenta lines). The bottom plot under each order shows the residuals (black lines) and the uncertainty in the flux (gray shaded regions). Best-fit parameters are listed in the top right corners, with $\log g$ and metallicity ([M/H]) fixed at 4 and 0, respectively. The residuals are dominated by fringing.
}
\label{fig:fit}
\end{figure*}

\begin{figure*}[!htbp]
\centering
\includegraphics[width=\textwidth]{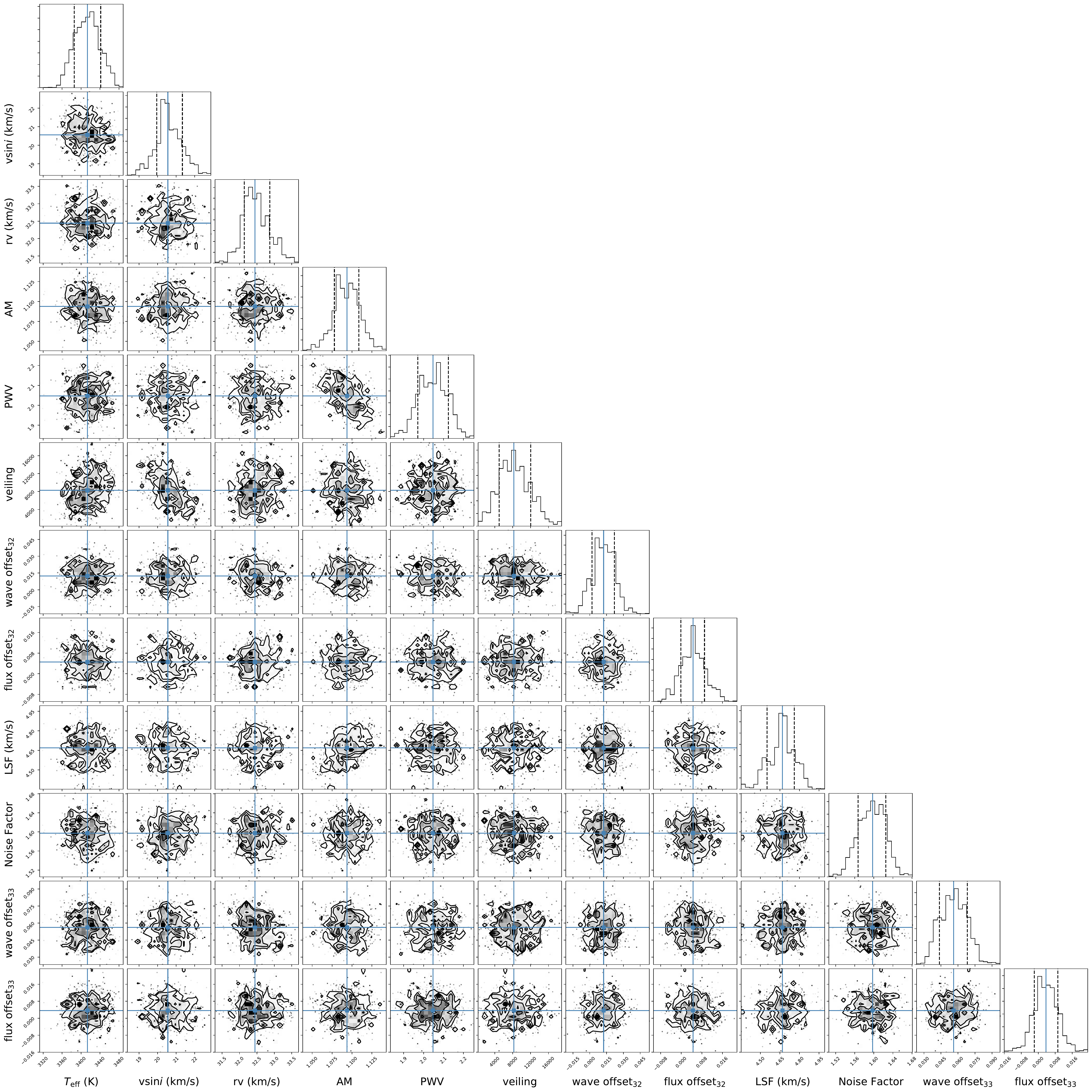}
\caption{Corner plot of a single frame for [HC2000] 244. This corner plot corresponds to the fit in Figure~\ref{fig:fit}. The subscripts in the parameter labels refer to the orders. The solid line shows the median of each distribution, and the dotted lines show the 16th and 84th percentiles.
}
\label{fig:corner}
\end{figure*}

An example fit is shown in Figure~\ref{fig:fit}, with the observed---telluric-corrected---spectrum shown with the gray line, the best-fit stellar model shown with the red line (model parameters indicated in red text), and the best-fit stellar model convolved with the best-fit telluric model shown with the purple line. The gray spectrum and purple model are compared in our MCMC routine. The bottom plots show the residuals between the observed spectrum and best-fit model (black line), and the total uncertainty (noise computed from the NSDRP with the scaling factor in equation~(\ref{eqn:likelihood})). The largest contributor to the residuals is fringing, which is an ongoing project to model, and will be addressed in a future study. However, previous studies of modeling NIRSPEC fringing have shown it does not significantly effect RV determinations \citep{blake:2010:684}. Figure~\ref{fig:corner} shows the corner plot for the MCMC run for Figure~\ref{fig:fit} (last 100 steps $\times$ 100 walkers, 10000 data points). The parameters listed on the x- and y-axis are the same as those from equations~\ref{nirspecmodel} and \ref{eqn:likelihood}. This plot indicates that there is little to no correlations between the majority of parameters, with the exception of $v\sin{i}$ and veiling, which shows a slight correlation. As can be seen, some of these distributions are non-Gaussian, which is why we report median value with 16th and 84th percentiles as uncertainties.

\begin{deluxetable}{lcc}
\tablecaption{Forward-Modeled Parameter Ranges \label{tbl:forwardmodel}}
\tablecolumns{3}
\tablehead{
\colhead{Description} & \colhead{Symbol} & \colhead{Bounds} %\\
}
\startdata
Stellar Effective Temp. & $T_\mathrm{eff}$ & (2300, 7000) K \\
Rotational Velocity & $v\sin{i}$ & (1, 100) km s$^{-1}$ \\
Radial Velocity & $V_r$ & ($-100$, 100) km s$^{-1}$ \\
Airmass & $AM$ & (1, 3) \\
Precip. Water Vapor & $PWV$ & (0.5, 30) mm \\
Line Spread Function & $\Delta \nu_\mathrm{inst}$ & (1, 100) km s$^{-1}$ \\
Flux Offset Param. & $C_\mathrm{flux}$ & ($10^{-15}$, $10^{15}$) cnts s$^{-1}$\\
Noise Factor & $C_\mathrm{noise}$ & (1, 50) \\
Wave. Offset Param. & $C_\lambda$ & ($-10$, 10) \AA \\
\enddata
\end{deluxetable}

Many of our sources converged to relatively high veiling parameters (i.e., $r_\lambda > 0.2$), and we note that temperature should be highly degenerate with veiling ratio due to the strong dependence of the fits on the CO bands, which could be weakened either by higher effective temperatures or higher veiling ratios. 
However, we do not have sufficient data to put constraints on the veiling parameter for each source, as that would required a 3-D extinction map of the ONC \citep[e.g.,][]{schlafly:2015:116}, and precise distances to individual sources. However, even with the \textit{Gaia} satellite \citep{gaia-collaboration:2016:a1} the embedded nature of the ONC makes parallax measurements extremely difficult \citep{kim:2019:109}.
Therefore, this study is focused on the kinematics of the ONC, however, a future study will investigate the $T_\mathrm{eff}$ (and mass) dependence of kinematics in the ONC core. 

\begin{figure}[!htbp]
\centering
\includegraphics[width=\linewidth]{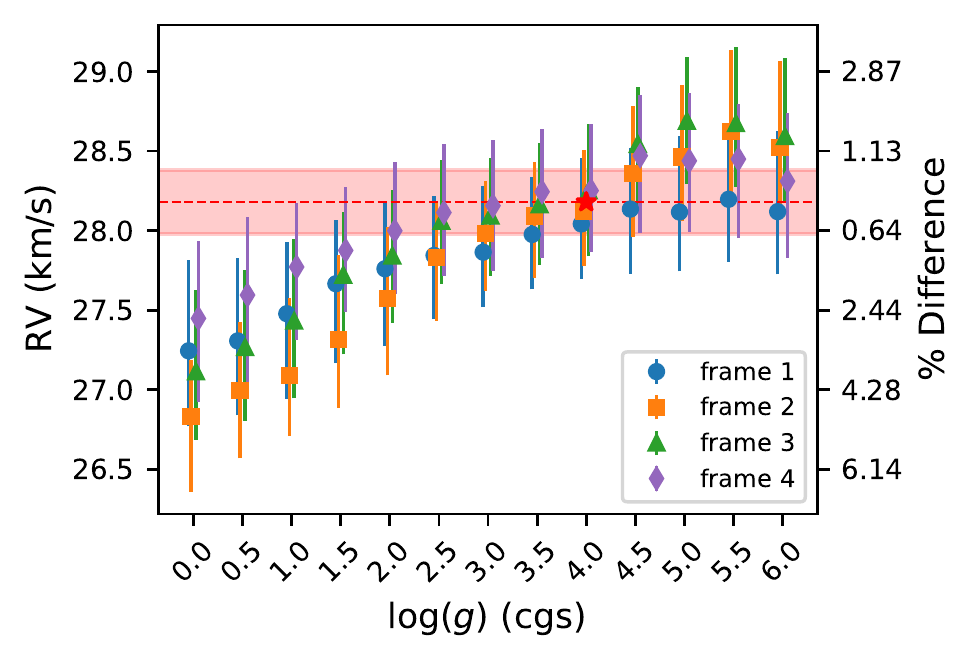}
\includegraphics[width=\linewidth]{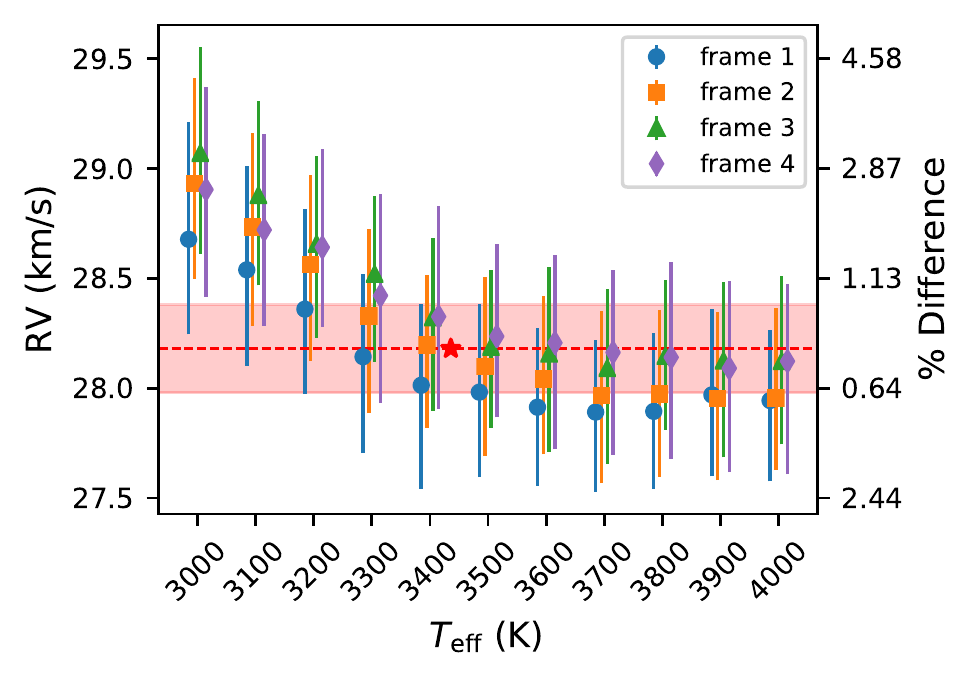}
\caption{Best-fit RVs for [HC2000] 244 while keeping $\log g$ (top) and temperature (bottom) constant. Each marker shows one of the four frames. 
\textit{Top}: The RV variation within $\log g \pm 0.5$ dex of our adopted values (28.18~km s$^{-1}$ and $\log g = 4$; red star and dashed line, uncertainties shown with the filled area) is less than a 1\%. Even at much lower/higher gravities the RV variation is not more than a few percent.  At the age of the ONC, the expected surface gravity values are $\log g \sim 3$--4.
\textit{Bottom}: The RV variation within a few hundred kelvin of our adopted values (28.18~km s$^{-1}$ and 3436~K; red star and dashed line, uncertainties shown with the filled area) is less than a few percent.
}
\label{fig:RVvariation}
\end{figure}

Our RV measurements are relatively robust to changes in stellar parameters, since they are strongly anchored to the CO bandheads and the absolute calibration of the telluric spectrum. To illustrate this, we fit [HC2000] 244 (the same source shown in Figure~\ref{fig:fit}) using the same procedure outlined above and holding the $\log g$ constant from 0 to 6 in steps of 0.5 (the resolution of the model grid). Figure~\ref{fig:RVvariation} (top) shows the RV variation due to different $\log g$ values. For $\log g$ values that differ by more than $\pm$1 dex from the nominal value RV variations are more than 1\%, however, those are extremely large variations in $\log g$ that are inconsistent with the youth of these targets. For RV values within $\pm$0.5 dex, the variation is less than 0.5\%. To be conservative, we chose to add this systematic uncertainty to our combined measurement uncertainty across all frames for each single object.

We performed a similar test to the one above, this time holding temperature constant. Figure~\ref{fig:RVvariation} (bottom) shows the RV variation due to different temperatures. Within a few hundred kelvin, the RV variation is less than 3\%. However, within $\pm$100 K this value is less than 0.5\%, which is within the systematic uncertainty found in the $\log g$ variation. Therefore, no additional systematic is required since the $\pm$0.5\% systematic accounts for both the $\log g$ and $T_\mathrm{eff}$ variations. In summary, our reported uncertainties are the summed quadrature of our measurement uncertainty from the MCMC fits, the 0.058 km s$^{-1}$ systematic uncertainty between calibration frames, and the 0.5\% variation found from differing $\log g$ and $T_\mathrm{eff}$.

\begin{figure}
\centering
\includegraphics[width=\linewidth]{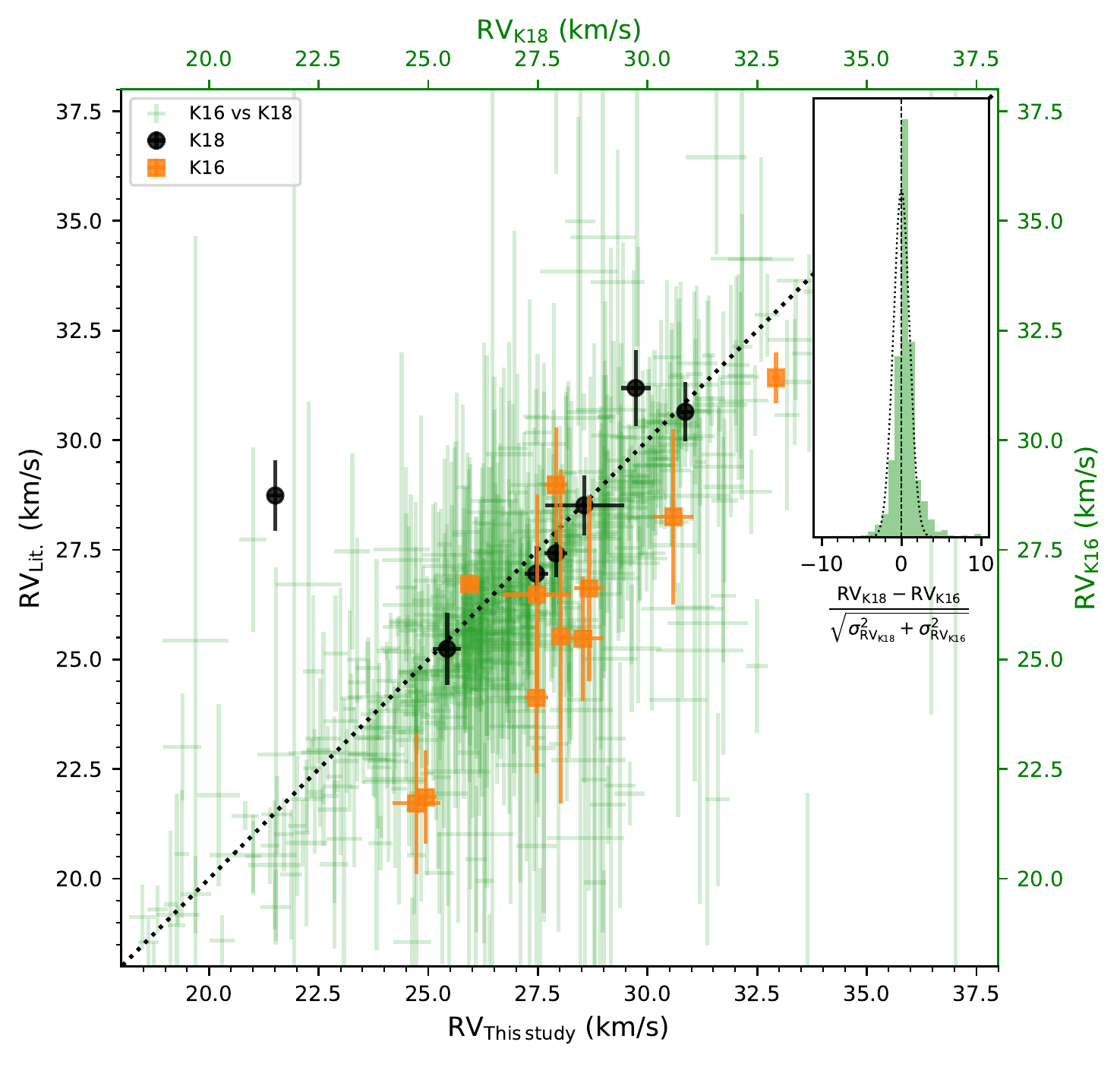}
\caption{RV comparison between NIRSPEC RVs from this study and APOGEE RVs from K18 (black circles) and optical RVs from \citet[][orange squares]{kounkel:2016:8}. K16 versus K18 are shown with translucent green markers (662 sources). The black dotted line indicates where RV measurements are equal. The large outlier, [HC2000] 546, is a potential RV variable binary. The inset plot shows the distribution of $(\mathrm{RV}_\mathrm{K18} - \mathrm{RV}_\mathrm{K16}) / \sqrt{\sigma_{\mathrm{RV}_\mathrm{K18}}^2 + \sigma_{\mathrm{RV}_\mathrm{K16}}^2}$ (green markers), which has $\mu = 0.58$ km s$^{-1}$ and $\sigma = 1.83$ km s$^{-1}$. This indicates that the K16 RVs are, on average, smaller than the K18 APOGEE RVs, and that the uncertainties in at least one of the surveys are underestimated. The inset plot shows a normal distribution with $\mu=0$ and $\sigma=1$ for comparison (dotted line).} 
\label{fig:rvcompare}
\end{figure}

Lastly, we compare our derived RVs to those from APOGEE using the results of K18, and \citet{tobin:2009:1103}, using the reanalyzed RVs from \citet[][hereafter K16]{kounkel:2016:8}. Figure~\ref{fig:rvcompare} shows the results of our RV comparison. In total, there were 11 matches between our NIRSPEC targets and the T09/K16 sample, and 7 matches to APOGEE (using a 0.5\arcsec crossmatch radius). Our results are consistent with the K18 APOGEE results with only 2 exceptions differentiating by more than 1-$\sigma$ of their combined uncertainty, possibly the result of spectroscopic binaries, 2MASS J05352104$-$0523490 and 2MASS J05351498$-$0521598. In comparison to optical RVs from T09/K16, our measured RVs are generally smaller than those measured from optical data, by $\sim$1.8 km s$^{-1}$ on average. We also compared K16 to K18 RVs, again using a 0.5\arcsec crossmatch radius, finding 586 sources in common (Figure~\ref{fig:rvcompare}, green pluses). The distribution of the uncertainty weighted difference between these two measurements, $(\mathrm{RV}_\mathrm{K18} - \mathrm{RV}_\mathrm{K16}) / \sqrt{\sigma_{\mathrm{RV}_\mathrm{K18}}^2 + \sigma_{\mathrm{RV}_\mathrm{K16}}^2}$, has a mean value of $\mu = 0.58$~km s$^{1}$ and a standard deviation of $\sigma = 1.83$~km s$^{1}$. This is consistent with the K16 values being, on average, smaller than the K18 APOGEE RVs. The width of the distribution also indicates that one, or both, of the uncertainties are underestimated. In general, RVs derived from NIR data versus optical are likely to suffer from fewer systematics in this highly embedded region.

\begin{deluxetable*}{ccccccccccc}
\tablecaption{NIRSPEC Forward-Modeling Results \label{tbl:kinematics}}
\tabletypesize{\scriptsize}
\tablecolumns{11}
\tablehead{
\colhead{[HC2000]} & \colhead{ID\tablenotemark{a}} & \colhead{$\alpha_{J2000}$} & \colhead{$\delta_{J2000}$} & \colhead{$\overline{RV}$\tablenotemark{b}} & \colhead{$\mu_{\alpha\cos\delta}$} & \colhead{$\mu_\delta$} &
\colhead{$T_\mathrm{eff}$\tablenotemark{c}} & \colhead{$v\sin{i}$} & \colhead{Veiling\tablenotemark{d}} & \colhead{Note\tablenotemark{e}} \\
\colhead{} & \colhead{} & \colhead{(deg)} &
\colhead{(deg)} & \colhead{(km s$^{-1}$)} & \colhead{(mas yr$^{-1}$)} & 
\colhead{(mas yr$^{-1}$)} & \colhead{(K)} & \colhead{(km s$^{-1}$)} & \colhead{$\left(\dfrac{F_{\mathrm{O33}, \mathrm{cont}}}{F_{\mathrm{O33}, \ast}}\right)$} 
}
\startdata
135 & \nodata & $83.80950000$ & $-5.40686111$ & $28.56 \pm 0.91$ & \nodata & \nodata & $3610 \pm 300$ & $49.33 \pm 2.26$ & 0.24 \\
188 & 559 & $83.83175000$ & $-5.39925000$ & $29.74 \pm 0.27$ & $0.41 \pm 0.52$ & $-0.83 \pm 0.18$ & $3475 \pm 73$ & $6.90 \pm 1.72$ & 0.01 \\
202 & 615 & $83.81666667$ & $-5.39805556$ & $25.69 \pm 0.75$ & $2.80 \pm 0.06$ & $-1.56 \pm 0.08$ & $3393 \pm 204$ & $42.27 \pm 2.95$ & 0.75 & E1\\
204 & \nodata & $83.84087500$ & $-5.39830556$ & $24.73 \pm 0.55$ & \nodata & \nodata & $3712 \pm 50$ & $48.16 \pm 2.56$ & 0.28 \\
215 & \nodata & $83.80116667$ & $-5.39669444$ & $27.91 \pm 0.27$ & \nodata & \nodata & $3641 \pm 33$ & $9.03 \pm 1.41$ & 0.20 \\
217 & \nodata & $83.83770833$ & $-5.39694444$ & $29.73 \pm 0.35$ & \nodata & \nodata & $3614 \pm 33$ & $47.35 \pm 0.89$ & 0.23 & V \\
219 & 87 & $83.81316667$ & $-5.39630556$ & $24.08 \pm 0.48$ & $1.87 \pm 0.06$ & $-0.65 \pm 0.02$ & $3486 \pm 57$ & $20.78 \pm 1.40$ & 0.43 \\
220 & 23 & $83.81175000$ & $-5.39625000$ & $33.47 \pm 0.23$ & $-1.66 \pm 0.40$ & $0.08 \pm 0.13$ & $3484 \pm 19$ & $9.04 \pm 0.30$ & 0.02 \\
223 & 164 & $83.81433333$ & $-5.39597222$ & $27.45 \pm 0.23$ & $1.10 \pm 0.08$ & $-0.68 \pm 0.23$ & $3359 \pm 16$ & $9.74 \pm 0.69$ & 0.32 \\
224 & \nodata & $83.79475000$ & $-5.39575000$ & $25.94 \pm 0.15$ & \nodata & \nodata & $3859 \pm 21$ & $24.94 \pm 0.58$ & 0.00 \\
228 & \nodata & $83.80250000$ & $-5.39561111$ & $27.46 \pm 0.27$ & \nodata & \nodata & $3782 \pm 30$ & $40.37 \pm 0.78$ & 0.46 \\
229 & 536 & $83.83904167$ & $-5.39594444$ & $28.30 \pm 0.21$ & $0.12 \pm 0.14$ & $0.67 \pm 0.27$ & $3613 \pm 79$ & $12.19 \pm 1.21$ & 0.15 \\
240 & \nodata & $83.80604167$ & $-5.39455556$ & $27.11 \pm 0.44$ & \nodata & \nodata & $3640 \pm 65$ & $14.15 \pm 1.28$ & 3.89 \\
244 & 180 & $83.82108333$ & $-5.39438889$ & $28.18 \pm 0.11$ & $-0.02 \pm 0.15$ & $0.83 \pm 0.40$ & $3436 \pm 17$ & $21.29 \pm 0.83$ & 0.04 \\
245 & \nodata & $83.81229167$ & $-5.39425000$ & $31.05 \pm 0.21$ & \nodata & \nodata & $3869 \pm 18$ & $17.03 \pm 0.35$ & 0.00 \\
248 & 200 & $83.81566667$ & $-5.39400000$ & $26.56 \pm 0.78$ & $1.81 \pm 0.43$ & $-2.83 \pm 0.20$ & $3438 \pm 48$ & $39.39 \pm 3.82$ & 0.35 & E2 \\
250 & 197 & $83.81625000$ & $-5.39388889$ & $29.93 \pm 0.46$ & $1.70 \pm 0.40$ & $-3.85 \pm 1.18$ & $2975 \pm 91$ & $15.90 \pm 1.86$ & 0.39 & E2 \\
253 & \nodata & $83.82587500$ & $-5.39330556$ & $28.67 \pm 0.36$ & \nodata & \nodata & $4000 \pm 117$ & $6.21 \pm 1.48$ & 2.66 \\
258 & 521 & $83.81000000$ & $-5.39269444$ & $28.02 \pm 0.19$ & $-0.11 \pm 0.23$ & $-1.65 \pm 0.03$ & $3568 \pm 41$ & $7.71 \pm 1.03$ & 0.82 \\
259 & \nodata & $83.82112500$ & $-5.39277778$ & $27.53 \pm 0.08$ & \nodata & \nodata & $3501 \pm 44$ & $32.49 \pm 1.95$ & 0.54 \\
261 & 206 & $83.81408333$ & $-5.39261111$ & $26.39 \pm 0.20$ & $-0.72 \pm 0.24$ & $0.87 \pm 0.12$ & $3374 \pm 14$ & $1.49 \pm 0.31$ & 0.00 \\
275 & 65 & $83.81220833$ & $-5.39141667$ & $30.88 \pm 0.01$ & $-1.18 \pm 0.14$ & $1.19 \pm 0.06$ & $3860 \pm 17$ & $17.59 \pm 0.30$ & 0.00 \\
277A & 530 & $83.83525000$ & $-5.39158333$ & $25.88 \pm 0.36$ & $-0.81 \pm 0.12$ & $0.14 \pm 0.06$ & $3444 \pm 51$ & $12.90 \pm 2.65$ & 0.36 & B \\
282 & 44 & $83.82866667$ & $-5.39136111$ & $26.96 \pm 0.28$ & $-0.85 \pm 0.24$ & $1.35 \pm 0.03$ & $3394 \pm 22$ & $5.59 \pm 1.05$ & 1.06 \\
288 & 71 & $83.82966667$ & $-5.39086111$ & $26.61 \pm 0.17$ & $-0.42 \pm 0.04$ & $-0.52 \pm 0.41$ & $3666 \pm 54$ & $18.20 \pm 0.53$ & 0.25 \\
291 & 211 & $83.81600000$ & $-5.39044444$ & $29.06 \pm 0.22$ & $1.10 \pm 0.45$ & $-1.63 \pm 0.07$ & $3181 \pm 33$ & $15.83 \pm 0.72$ & 0.42 & B \\
295 & \nodata & $83.82320833$ & $-5.39025000$ & $21.85 \pm 0.93$ & \nodata & \nodata & $3662 \pm 305$ & $22.69 \pm 7.18$ & 0.76 \\
302 & 221 & $83.81133333$ & $-5.38969444$ & $29.24 \pm 0.11$ & $-0.24 \pm 0.40$ & $0.71 \pm 0.13$ & $3541 \pm 21$ & $13.73 \pm 1.12$ & 0.48 \\
306A & \nodata & $83.81600000$ & $-5.38958333$ & $29.12 \pm 0.50$ & \nodata & \nodata & $4201 \pm 229$ & $25.92 \pm 1.36$ & 1.12  & B\\
306B & \nodata & $83.81600000$ & $-5.38958333$ & $21.15 \pm 0.54$ & \nodata & \nodata & $3473 \pm 35$ & $40.15 \pm 2.53$ & 0.64  & B\\
313 & 198 & $83.82279167$ & $-5.38919444$ & $26.35 \pm 0.19$ & $1.07 \pm 0.86$ & $3.21 \pm 0.20$ & $4206 \pm 235$ & $10.04 \pm 0.93$ & 0.64 & E2 \\
322 & \nodata & $83.81787500$ & $-5.38794444$ & $24.90 \pm 0.27$ & \nodata & \nodata & $3414 \pm 18$ & $37.47 \pm 0.92$ & 0.07 \\
324 & 226 & $83.81183333$ & $-5.38777778$ & $30.77 \pm 0.34$ & $0.70 \pm 0.55$ & $-1.87 \pm 0.04$ & $3449 \pm 36$ & $5.12 \pm 1.21$ & 0.01\\
331 & 121 & $83.82604167$ & $-5.38769444$ & $31.83 \pm 0.69$ & $1.26 \pm 0.24$ & $1.03 \pm 0.04$ & $4034 \pm 452$ & $18.92 \pm 1.17$ & 0.64 \\
332A & 183 & $83.82425000$ & $-5.38766667$ & $25.43 \pm 0.32$ & $-1.61 \pm 0.17$ & $0.60 \pm 0.38$ & $4012 \pm 37$ & $16.48 \pm 0.43$ & 0.01 & BC \\
370 & 227 & $83.81616667$ & $-5.38388889$ & $31.24 \pm 0.17$ & $0.01 \pm 0.74$ & $-0.46 \pm 0.09$ & $3802 \pm 45$ & $14.43 \pm 0.34$ & 0.66 \\
375 & 560 & $83.82075000$ & $-5.38361111$ & $33.19 \pm 0.45$ & $0.78 \pm 0.13$ & $1.15 \pm 0.04$ & $4225 \pm 244$ & $26.95 \pm 0.70$ & 0.41 \\
386 & \nodata & $83.81362500$ & $-5.38241667$ & $27.44 \pm 0.18$ & \nodata & \nodata & $3673 \pm 13$ & $11.20 \pm 0.10$ & 0.11 \\
388 & 532 & $83.82316667$ & $-5.38244444$ & $30.59 \pm 0.46$ & $0.65 \pm 0.23$ & $0.35 \pm 0.04$ & $4279 \pm 142$ & $17.83 \pm 1.46$ & 0.52 \\
389 & \nodata & $83.81516667$ & $-5.38233333$ & $30.86 \pm 0.15$ & \nodata & \nodata & $3566 \pm 34$ & $26.89 \pm 0.46$ & 0.68 \\
408 & 143 & $83.83008333$ & $-5.38075000$ & $23.52 \pm 0.94$ & $-1.51 \pm 0.71$ & $-1.07 \pm 0.10$ & $3660 \pm 105$ & $50.79 \pm 1.85$ & 0.02 \\
410 & 62 & $83.82133333$ & $-5.38058333$ & $21.42 \pm 1.91$ & $-0.85 \pm 0.29$ & $0.32 \pm 0.12$ & $3910 \pm 94$ & $57.77 \pm 2.25$ & 0.03 \\
412 & 151 & $83.81808333$ & $-5.38030556$ & $32.93 \pm 0.32$ & $1.21 \pm 0.20$ & $0.51 \pm 0.27$ & $3563 \pm 28$ & $31.96 \pm 1.61$ & 0.80 \\
413 & 35 & $83.81454167$ & $-5.38016667$ & $25.51 \pm 0.41$ & $-0.43 \pm 0.04$ & $-0.27 \pm 0.74$ & $3643 \pm 44$ & $28.79 \pm 0.55$ & 0.05 \\
420 & 154 & $83.81600000$ & $-5.37941667$ & $25.91 \pm 0.24$ & $-0.38 \pm 0.31$ & $-1.82 \pm 0.25$ & $3510 \pm 35$ & $24.16 \pm 0.67$ & 0.00 \\
425 & \nodata & $83.82479167$ & $-5.37930556$ & $24.94 \pm 0.25$ & \nodata & \nodata & $3576 \pm 22$ & $21.83 \pm 0.32$ & 0.00 \\
431 & 700 & $83.81216667$ & $-5.37752778$ & $33.49 \pm 0.31$ & $1.68 \pm 0.60$ & $0.05 \pm 0.94$ & $3603 \pm 22$ & $18.80 \pm 0.97$ & 0.65 \\
436 & \nodata & $83.82658333$ & $-5.37708333$ & $32.93 \pm 0.08$ & \nodata & \nodata & $3658 \pm 347$ & $16.30 \pm 0.90$ & 0.11 \\
440 & \nodata & $83.82233333$ & $-5.37661111$ & $28.52 \pm 0.48$ & \nodata & \nodata & $3715 \pm 88$ & $23.57 \pm 1.07$ & 1.37 \\
450 & \nodata & $83.82087500$ & $-5.37586111$ & $26.23 \pm 0.24$ & \nodata & \nodata & $3674 \pm 54$ & $12.14 \pm 0.74$ & 0.02 \\
504 & 64 & $83.81395833$ & $-5.37100000$ & $27.48 \pm 0.78$ & $-0.29 \pm 0.15$ & $-0.32 \pm 0.02$ & $3969 \pm 295$ & $30.65 \pm 1.86$ & 3.79 \\
522A & 110 & $83.81787500$ & $-5.36955556$ & $24.75 \pm 0.47$ & $0.03 \pm 0.10$ & $-1.97 \pm 0.14$ & $3441 \pm 24$ & $33.48 \pm 1.98$ & 0.31 & B \\
522B & 642 & $83.81787500$ & $-5.36955556$ & $24.54 \pm 0.28$ & $-0.69 \pm 0.11$ & $0.48 \pm 0.33$ & $3170 \pm 39$ & $21.94 \pm 1.20$ & 0.80 & B \\
546 & 567 & $83.81250000$ & $-5.36666667$ & $21.51 \pm 0.18$ & $2.25 \pm 0.09$ & $0.96 \pm 0.11$ & $3793 \pm 56$ & $15.32 \pm 1.18$ & 0.27 & V\\
703 & 137 & $83.80750000$ & $-5.36863889$ & $33.39 \pm 0.06$ & $-1.51 \pm 0.90$ & $-1.03 \pm 0.52$ & $3584 \pm 27$ & $9.14 \pm 0.41$ & 1.24 \\
713 & \nodata & $83.82795833$ & $-5.38247222$ & $22.95 \pm 0.31$ & \nodata & \nodata & $4988 \pm 192$ & $6.75 \pm 1.54$ & 1.60 
\enddata
\tablenotetext{a}{ID number from \citet{kim:2019:109}.}
\tablenotetext{b}{Reported uncertainties also include the 0.058 km s$^{-1}$ systematic uncertainty between calibration frames, and the 0.5\% variation found from differing $\log g$ and $T_\mathrm{eff}$.}
\tablenotetext{c}{Continuum veiling causes extreme degeneracies with $T_\mathrm{eff}$. Caution should be taken when using derived temperatures with high veiling.}
\tablenotetext{d}{Order 33 veiling parameter.}
\tablenotetext{e}{In this column, 
B = double star, previously known in the literature~\citep{hillenbrand:1997:1733, robberto:2013:10}; 
BC = new binary candidates, previously reported as single in the literature; 
E1 = escape group 1;
E2 = escape group 2;
V = RV variable source.}
\end{deluxetable*}

\section{APOGEE Reanalysis}
\label{sec:apogee}

It is useful to assess the fidelity of the parameters derived in our pipeline by applying it to an independent dataset. We chose to apply our pipeline to APOGEE $H$-band data. These data have independent measurements of $T_\mathrm{eff}$, $\log g$, and RVs of ONC sources from K18. We chose to do a subset of the entire K18 catalog, selecting objects within 4\arcmin\ of the ONC CoM (\apogeesourcesfourarcmin\ sources). We used a similar version of our aforementioned pipeline (see Section~\ref{sec:mcmc}), with the flux for each chip modeled using equation~(\ref{nirspecmodel}).
The LSF was modeled as a sum of Gauss-Hermite functions \citep{nidever:2015:173}, obtained using the \textit{apogee}\footnote{\url{https://github.com/jobovy/apogee}} code \citep{bovy:2016:49}. It should be noted that our model choice of PHOENIX-ACES-AGSS-COND-2011 \citep{husser:2013:a6} is the same as that used in K18.

We fit all three chips simultaneously (A: 1.647--1.696~$\mu$m, B: 1.585--1.644~$\mu$m, and C: 1.514--1.581~$\mu$m), allowing each chip to have separate nuisance parameters (e.g., $C0_\mathrm{flux,A}$ and $C1_\mathrm{flux,A}$ for chip A, $C0_\mathrm{flux,B}$ and $C1_\mathrm{flux,B}$ for chip B), similar to our simultaneous modeling of separate NIRSPEC orders. In general, the APOGEE sources tend to be much brighter than our NIRSPEC sources; however, the median $H$-band veiling ratio for APOGEE sources is 0.58, and these sources are more susceptible to confusion due to the size of the fiber \citep[2\arcsec\ diameter;][]{majewski:2017:94}.
The results of our fits are listed in Table~\ref{tbl:apogeeresults}, including measured veiling ratios and noting sources where a nearby companion could confuse results. We show comparisons of our derived $T_\mathrm{eff}$ and RVs to those from K18 in Figure~\ref{fig:apogeecomparisons}. Our derived temperatures are overall consistent with K18, although there is a small systematic shift from low to high temperatures. Our RVs are consistent with those from K18 ($\overline{\Delta \mathrm{RV}} = 0.23 \pm 0.43$~km s$^{-1}$), with a number of sources having measured RVs where K18 only provided upper limits. In total, we provide RV measurements for 87 sources that previously had no measurement in K18. We note that although these sources have no definitive measurement in K18, many of these sources have RV measurements from the SDSS/APOGEE processing pipeline \citep{nidever:2015:173}. However, K18 mention that these estimates tend to be unreliable for sources with $T_\mathrm{eff} < 3000$~K, and potentially also for YSOs, where veiling must be accounted for.

\begin{figure*}[!htbp]
\centering
\includegraphics[angle=90]{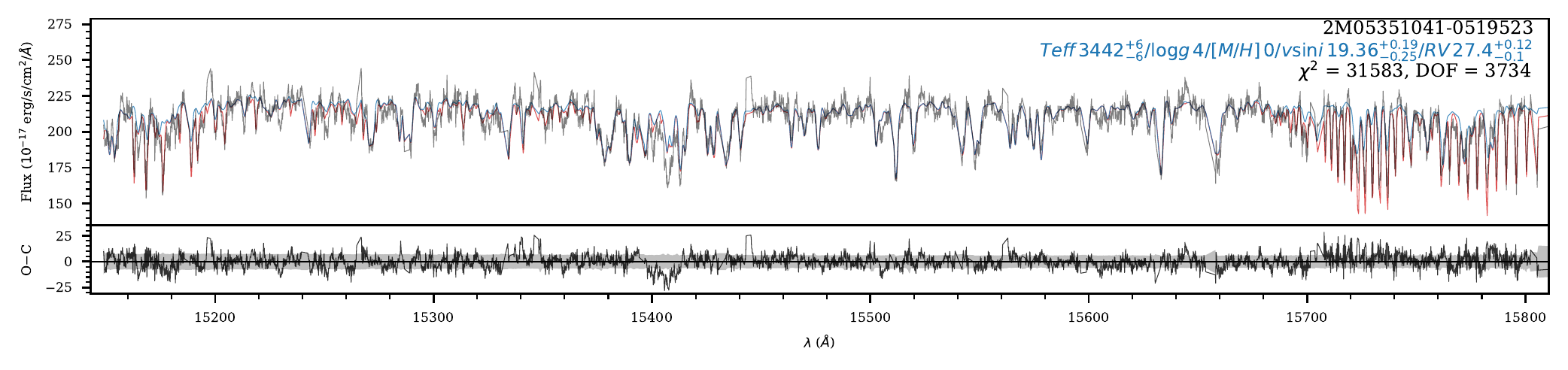}
\includegraphics[angle=90]{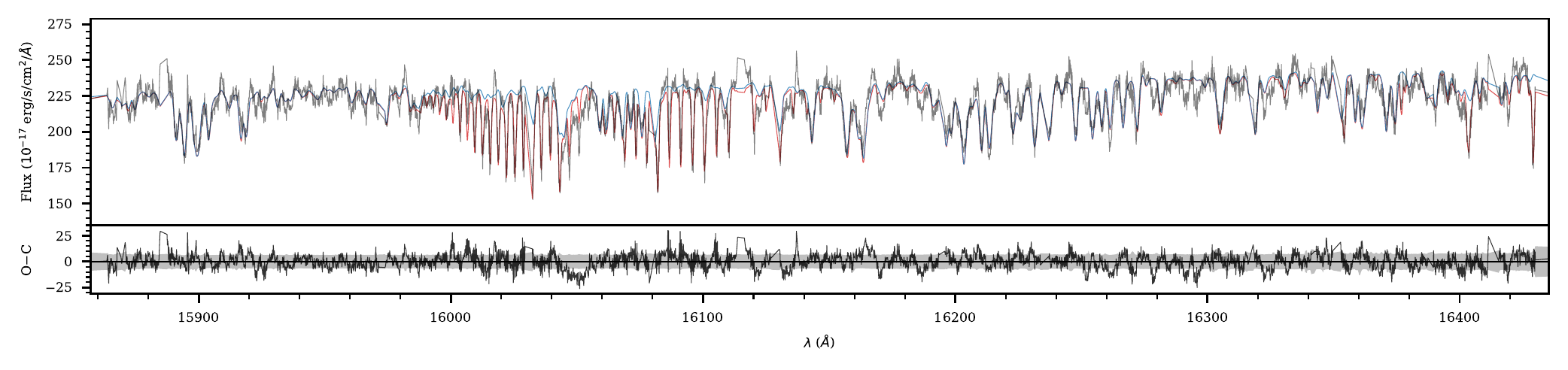}
\includegraphics[angle=90]{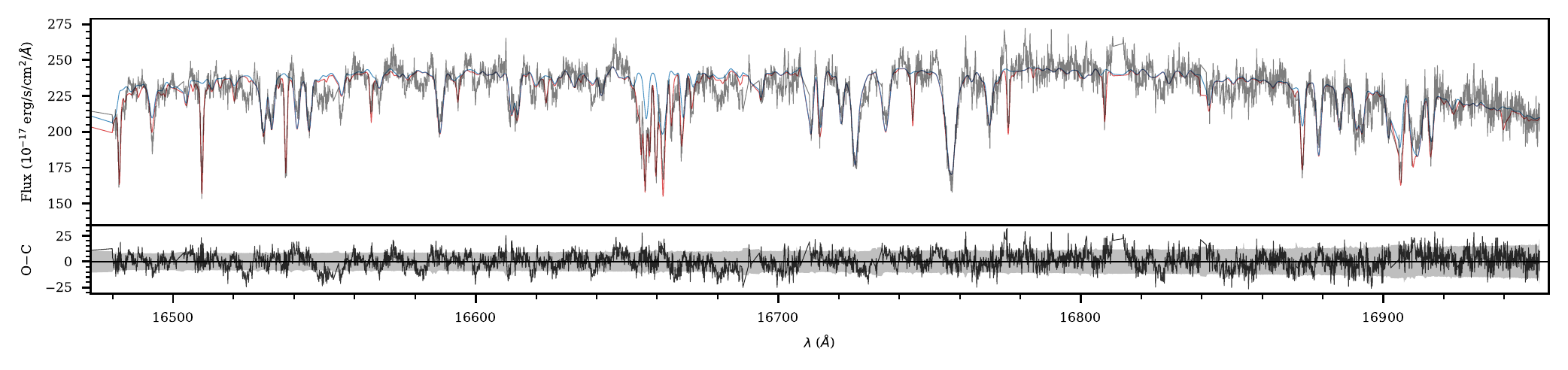}
\caption{Example fit to APOGEE data for all three chips. Raw APOGEE data are shown in gray, the best-fit stellar model is shown in blue (parameters given in the upper right hand corner), and the best-fit stellar model convolved with the best-fit telluric model is shown in red. Residuals are shown in the lower plot (black line), along with the flux uncertainty (gray region).
}
\label{fig:apogeespectra}
\end{figure*}

\begin{figure*}
\centering
\includegraphics[width=\linewidth]{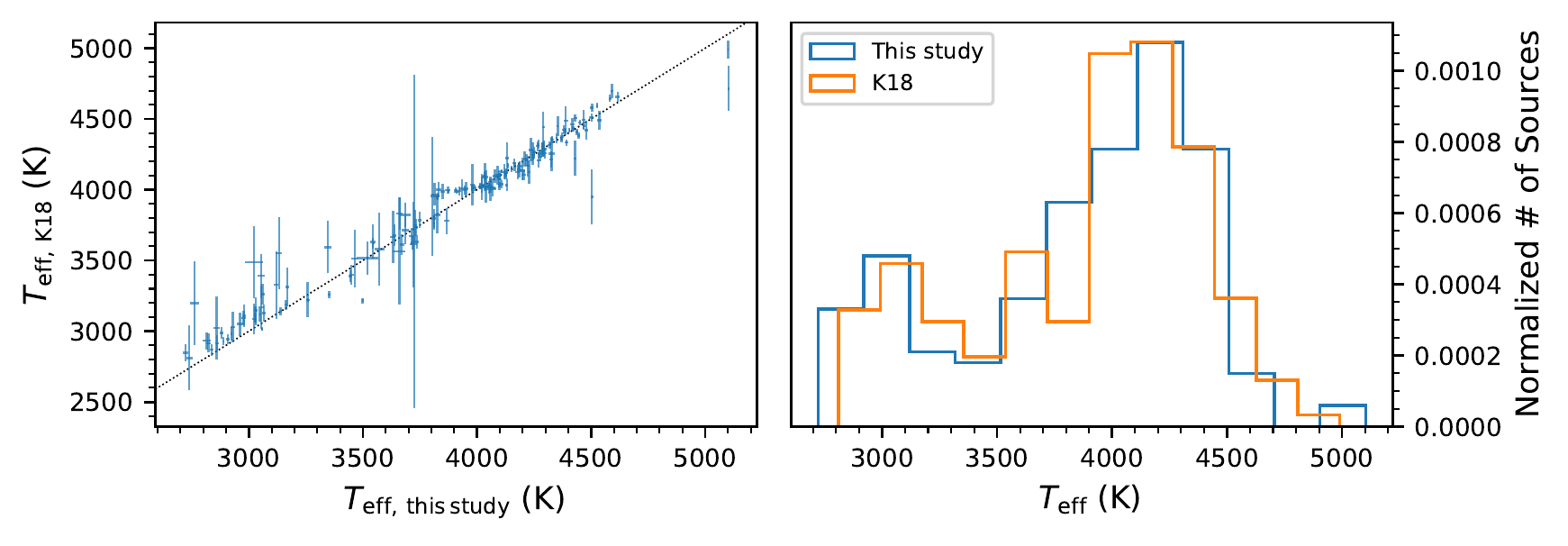}
\includegraphics[width=\linewidth]{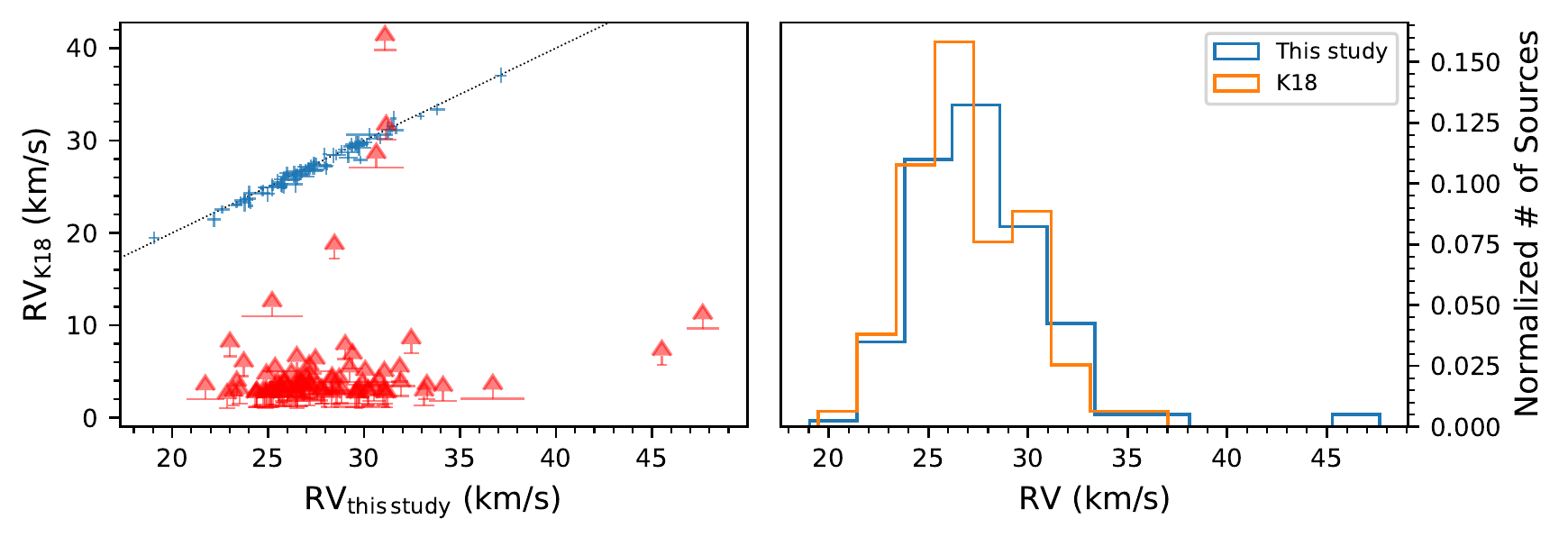}
\caption{Comparison between our derived parameters using APOGEE spectra and those derived in K18 (168 sources). Black dotted lines indicate where agreement is one-to-one.
\textit{Top}: $T_\mathrm{eff}$ comparison. There is a slight offset, with our derived $T_\mathrm{eff}$ being slightly lower than K18 for $T_\mathrm{eff} < 3600$, and slightly higher than K18 for $T_\mathrm{eff} > 3600$.
\textit{Bottom}: RV comparison. Red arrows represent lower limits as reported by K18, while we provide measurements with errorbars. 
}
\label{fig:apogeecomparisons}
\end{figure*}

\section{The Kinematic Structure of the ONC Core}
\label{sec:kinematics}

Three-dimensional kinematic studies of the ONC core have been primarily focused on the Trapezium Stars, as they are the brightest objects in the highly embedded region \citep[e.g.,][]{olivares:2013:106}. From Figure~\ref{fig:onc} it can be seen that very few previous studies have obtained RVs for sources within the direct vicinity of the Trapezium stars. Here, we analyze the 3-D kinematics of sources that make up the ``core" of the ONC.

\subsection{Tangential Velocities}

Our measurements provide velocities along the line of sight; however, to measure 3-D velocities we require tangential motions. A number of studies have measured the PMs of sources within the ONC \citep[e.g.,][]{parenago:1954:3, jones:1988:1755, van-altena:1988:1744, gomez:2005:1166, dzib:2017:139, kuhn:2019:32, kim:2019:109, platais:2020:272}. The two most recent catalogs produced by \citet{kim:2019:109} and \citet{platais:2020:272} both use imaging data from the \textit{Hubble Space Telescope} (\textit{HST}). We chose to adopt the values from \citet{kim:2019:109} as their combination of \textit{HST} imaging with data from the Keck Near Infrared Camera 2 (NIRC2; PI: K. Matthews) provide a baseline of $\sim$20 years. Additionally, the uncertainties published by \citet{platais:2020:272} tend to be very small, and are possibly underestimated due to the fact that they were unable to determine systematic uncertainties. This likely explains their much smaller errors versus \citet{kim:2019:109}, even though a shorter time baseline of $\sim$11 years was used.

The \citet{kim:2019:109} catalog includes PM measurements for 701 sources with typical uncertainties $\lesssim$1~mas yr$^{-1}$. However, to convert PMs into tangential velocities we require a distance, or distance distribution, to the ONC. a number of studies have investigated the distance to the ONC using Very-long-baseline interferometry \citep[VLBI;][]{menten:2007:515, sandstrom:2007:1161, kounkel:2017:142}, and, more recently, \textit{Gaia} Data Release 2 \citep{gaia-collaboration:2018:a1, kounkel:2018:84, grossschedl:2018:a106, kuhn:2019:32}. The majority of these measurements are consistent with an average distance to the ONC of $\sim$390~pc. For our study, we adopted the VLBI trigonometric distance of $388\pm5$~pc from \citet{kounkel:2017:142}, which is consistent with \citet[$389\pm3$~pc, \textit{Gaia} DR2;]{kounkel:2018:84}, \citet[$403^{+7}_{-6}$~pc, \textit{Gaia} DR2;]{kuhn:2019:32}, \citet[$397\pm16$~pc, \textit{Gaia} DR2;]{grossschedl:2018:a106}, and \citep[$389^{+24}_{-21}$~pc, VLBI;][]{sandstrom:2007:1161}

\begin{figure*}
\centering
\includegraphics[width=\linewidth]{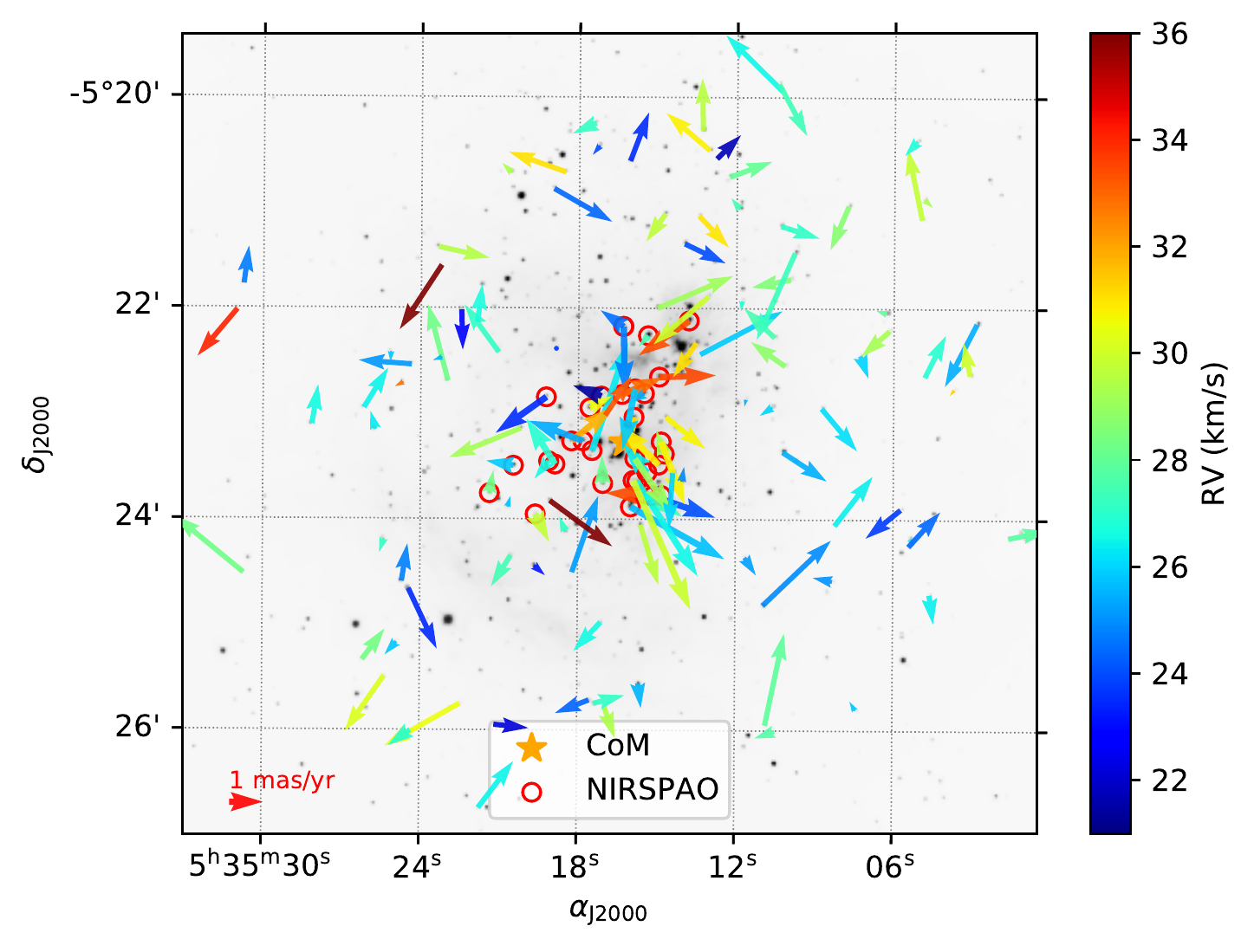}
\caption{Diagram of PMs and line of sight velocities for sources within a $4\arcmin \times 4\arcmin$ window centered on the core of the ONC. The background image is a CTIO/Blanco ISPI $K_s$-band image from \citet{robberto:2010:950}. Sources with NIRSPAO RVs are denoted with red circles, while the remaining sources have RVs measured from APOGEE data. The CoM for the ONC is indicated by the orange star.}
\label{fig:vel_quiver}
\end{figure*}

\begin{figure*}
\centering
\includegraphics[width=\textwidth]{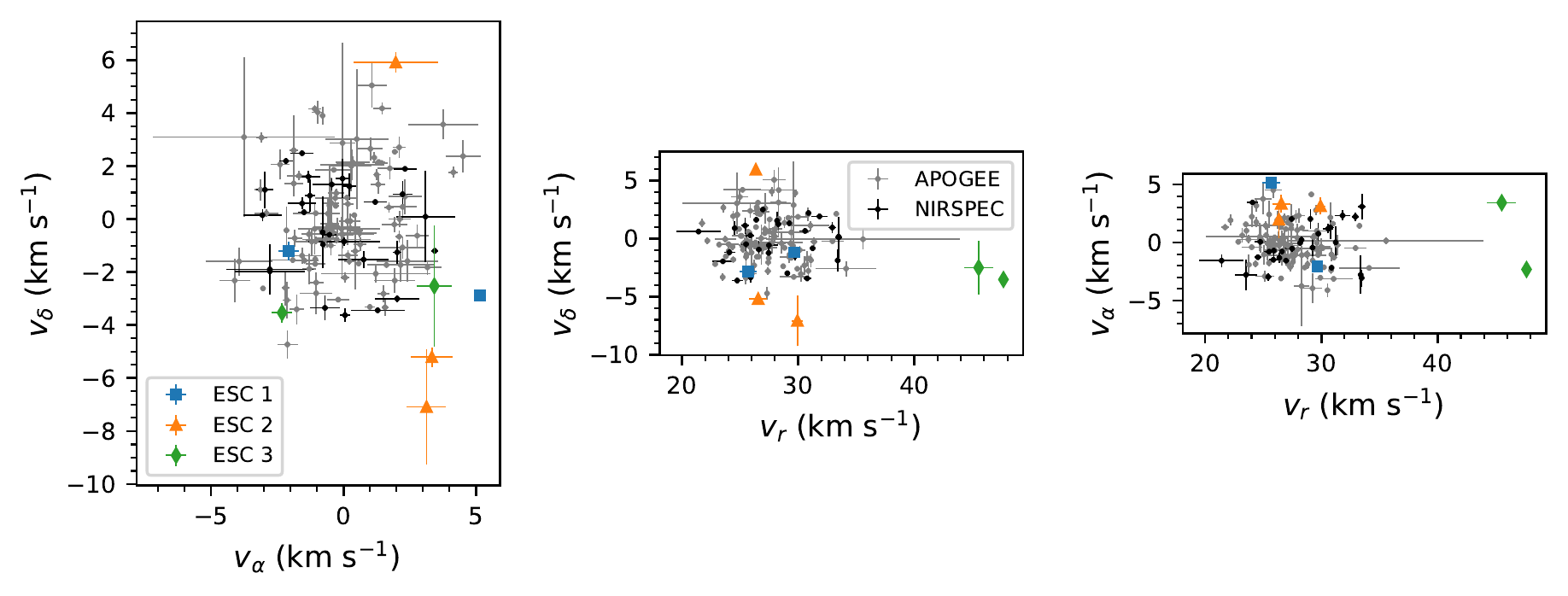}
\caption{Plots of each velocity component with respect to one another. All axes have equal scaling. Shown are APOGEE measurements (gray points), NIRSPEC measurements (black points), and kinematic outliers (see Section~\ref{outliers}) from escape groups 1 (blue squares), 2 (orange triangles), and 3 (green diamonds).}
\label{fig:3dvelocities}
\end{figure*}

Using the above distance estimate, we converted PMs to tangential velocities, combining errors in PMs and the distance to the ONC using standard error propagation. The combined sample with all three components of motion totaled \kinematicsample\ sources. Figure~\ref{fig:vel_quiver} shows a map of the ONC with vectors displaying the PM of sources in our sample and colors representing the measured RVs.
All three components of motion for our subsample are shown in Figure~\ref{fig:3dvelocities}.  These velocities were used to investigate the 3-D kinematics of the ONC core region.  There are a number of kinematic outliers, both in tangential velocity space (as discussed in \citealt{kim:2019:109}), and in radial velocity space, which will be discussed in Section~\ref{outliers}. 

\subsection{Intrinsic Velocity Dispersion Calculation}
\label{model}

To determine velocity dispersion, we utilized a similar Bayesian framework to \citet{kim:2019:109}, where each $i$-th kinematic measurement is parameterized as $v_{(\alpha, \delta, r)_i} \pm \epsilon_{{(\alpha, \delta, r)}_i}$. We assume each kinematic measurement is drawn from a multivariate Gaussian distribution with mean values of $\bar v_\alpha$, $\bar v_\delta$, and $\bar v_r$, and standard deviations of $\sqrt{\sigma_{v_{(\alpha,\delta,r)}}^2 + \epsilon_{(\alpha, \delta, r)_i}^2}$, where $\sigma_{v_\alpha}$, $\sigma_{v_\delta}$, and $\sigma_{v_r}$ are the intrinsic velocity dispersions (IVDs). The log-likelihood for the three dimensional kinematics of the $i$-th object is given by
\begin{equation}
\ln L_i = -\frac{1}{2}\left[\ln(|\boldsymbol\Sigma_i|) + (\boldsymbol v_i - \boldsymbol\mu)^{{{\!\mathsf{T}}}}\boldsymbol\Sigma^{-1}_i(\boldsymbol v_i  - \boldsymbol\mu) + 3\ln(2\pi)\right],
\end{equation}
\noindent where the input kinematic measurement vector for the $i$-th object $\boldsymbol v_i$ is defined as
\begin{equation}
\boldsymbol v_i = \begin{bmatrix}
v_{\alpha_i}\\
v_{\delta_i}\\
v_{r_i}
\end{bmatrix},
\end{equation}
\noindent while the vectors for the mean velocities and dispersions are defined as
\begin{align}
\boldsymbol\mu = \begin{bmatrix}
\bar v_\alpha\\
\bar v_\delta\\
\bar v_r
\end{bmatrix}, \hspace{35pt} &
\boldsymbol\sigma = \begin{bmatrix}
\sigma_{v_\alpha}\\
\sigma_{v_\delta}\\
\sigma_{v_r}
\end{bmatrix},
\end{align}
\noindent and the covariance matrix $\boldsymbol\Sigma_i$ for the $i$-th object is defined as
\begin{equation}
\boldsymbol\Sigma_i = \begin{bmatrix}
\sigma_{v_\alpha}^2 + \epsilon_{\alpha_i}^2 & \begin{split}\rho_1 (\sigma_{v_\alpha} & \sigma_{v_\delta} +\\ &\epsilon_{\alpha_i}\epsilon_{\delta_i})\end{split} & \begin{split}\rho_2 (\sigma_{v_\alpha} & \sigma_{v_r} +\\ & \epsilon_{\alpha_i}\epsilon_{r_i})\end{split}\\
\begin{split}\rho_1 (\sigma_{v_\alpha} & \sigma_{v_\delta} + \\ & \epsilon_{\alpha_i}\epsilon_{\delta_i})\end{split} & \sigma_{v_\delta}^2 + \epsilon_{\delta_i}^2 & \begin{split}\rho_3 (\sigma_{v_\delta} & \sigma_{v_r} + \\ & \epsilon_{\delta_i}\epsilon_{r_i})\end{split}\\
\begin{split}\rho_2 (\sigma_{v_\alpha} & \sigma_{v_r} + \\ & \epsilon_{\alpha_i}\epsilon_{r_i})\end{split} & \begin{split}\rho_3 (\sigma_{v_\delta} & \sigma_{v_r} +\\ & \epsilon_{
\delta_i}\epsilon_{r_i})\end{split} & \sigma_{v_r}^2 + \epsilon_{r_i}^2
\end{bmatrix}.
\end{equation}
\noindent The correlation coefficients are given by $\rho_{1,2,3}$, and should be 0 if the covariance is completely uncorrelated. Using Bayes's theorem, for a set of $N$ measurements $\boldsymbol D \equiv \{\boldsymbol v_i, \boldsymbol \epsilon_i\}^N_{i=1}$, the log of the posterior probability is given as 
\begin{equation}
\ln P(\boldsymbol \mu, \boldsymbol \sigma|\boldsymbol D) \varpropto \sum_i^N \left[ \ln L_i(\boldsymbol v_i, \boldsymbol \epsilon_i|\boldsymbol \mu, \boldsymbol \sigma) + \ln p(\boldsymbol \mu, \boldsymbol \sigma)\right],
\end{equation}
\noindent where $p(\boldsymbol \mu, \boldsymbol \sigma) \equiv p(\boldsymbol \mu)p(\boldsymbol \sigma)$ is the prior on the means and dispersions. Similar to \citet{kim:2019:109}, we adopted a flat prior on the mean vector, $\boldsymbol \mu$, and a Jeffreys prior on the dispersion vector, $p(\boldsymbol \sigma) \varpropto \boldsymbol \sigma^{-1}$. To determine best-fit parameters for $\bar v_\alpha$, $\bar v_\delta$, $\bar v_r$, $\sigma_{v_\alpha}$, $\sigma_{v_\delta}$, $\sigma_{v_r}$, $\rho_1$, $\rho_2$, and $\rho_3$, we utilized the Monte Carlo Markov Chain (MCMC) affine-invariant ensemble sampler \textit{emcee} \citep{foreman-mackey:2013:306}.

\subsection{Kinematic Outliers}
\label{outliers}

Previous studies have identified subpopulations within the ONC with kinematics consistent with high-velocity escaping stars \citep[e.g.,][]{kim:2019:109,kuhn:2019:32,platais:2020:272}. \citet{kim:2019:109} and \citet{kuhn:2019:32} used outliers from an expected normal distribution to determine kinematic outliers. \citet{kim:2019:109} also used the ONCs 2-D mean-square escape velocity ($\approx$3.1~mas yr$^{-1}$, or 6.1~km s$^{-1}$ at a distance of 414~pc) to identify potential escaping sources. It should be noted that \citet{kim:2019:109} used a distance of $414\pm7$~pc from \citet{menten:2007:515}, whereas our adopted value of $388\pm5$~pc would give a mean-square escape velocity of 5.7~km s$^{-1}$. In practice this should not alter results since outliers are computed from the bulk properties of the cluster, and our smaller distance would equally impact all tangential velocities the same.

To determine high-probability escaping or evaporating sources, we identified sources whose velocities deviate from the Gaussian velocity distribution model (see Section~\ref{model}). We used a methodology similar to \citet{kim:2019:109} and \citet{kuhn:2019:32} plotting sources on $Q$-$Q$ plots, where data quantiles ($Q_\mathrm{data}$) are plotted against theoretical quantiles ($Q_\mathrm{theo}$) corresponding to a Gaussian distribution. These two quantiles are defined as 
\begin{equation}
Q_{\mathrm{data},i} = \frac{v_{{(\alpha,\delta,r)}_i} - \bar v_{(\alpha,\delta,r)}}{\sqrt{\sigma_{(\alpha,\delta,r)}^2 + \epsilon_{{(\alpha,\delta,r)}_i}^2}},
\end{equation}
\begin{equation}
Q_{\mathrm{theo},i} = \sqrt{2}\ \mathrm{erf}^{-1}\left(\frac{2(r_i-0.5)}{n}-1\right),
\end{equation}
\noindent where erf$^{-1}$ is the inverse of the error function, $r_i$ is the rank of the $i$-th measurement, and $n$ is the number of measurements. 

\begin{figure*}
\centering
\includegraphics[width=\textwidth]{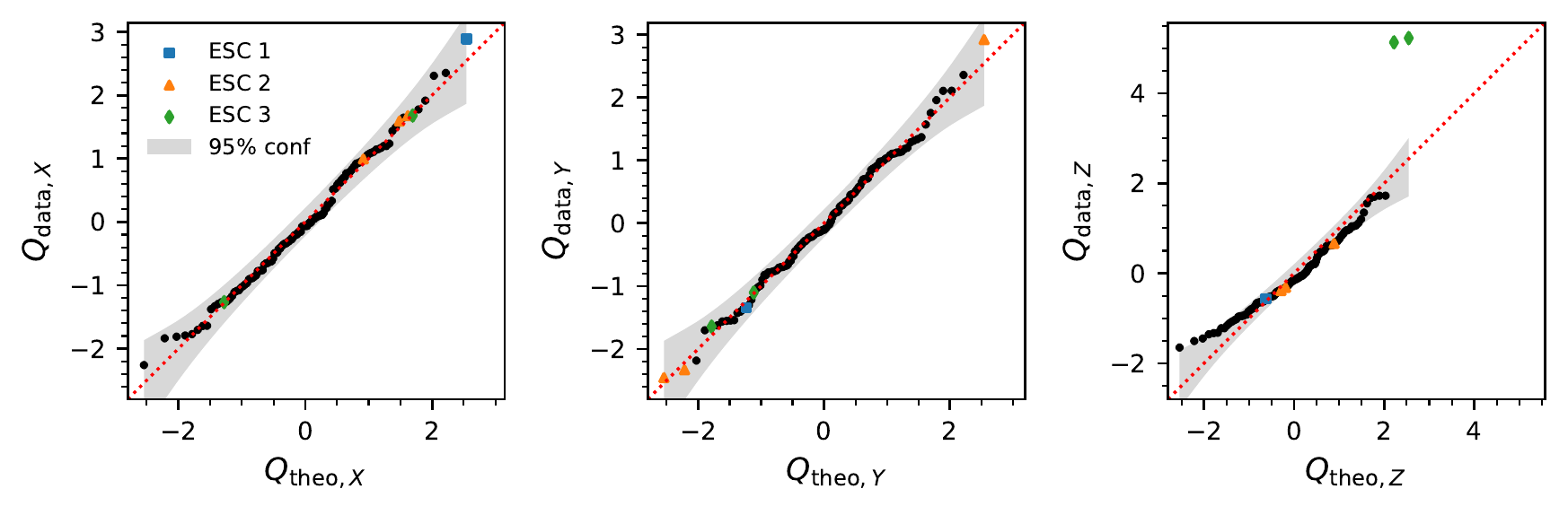}
\includegraphics[width=\textwidth]{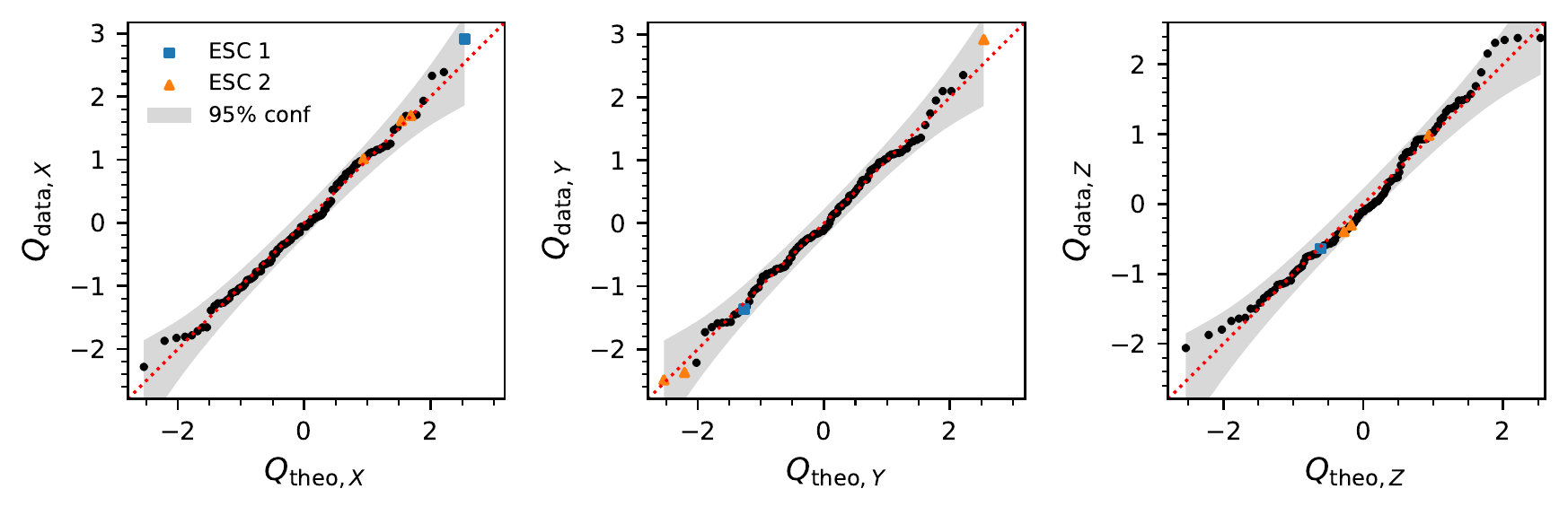}
\includegraphics[width=\textwidth]{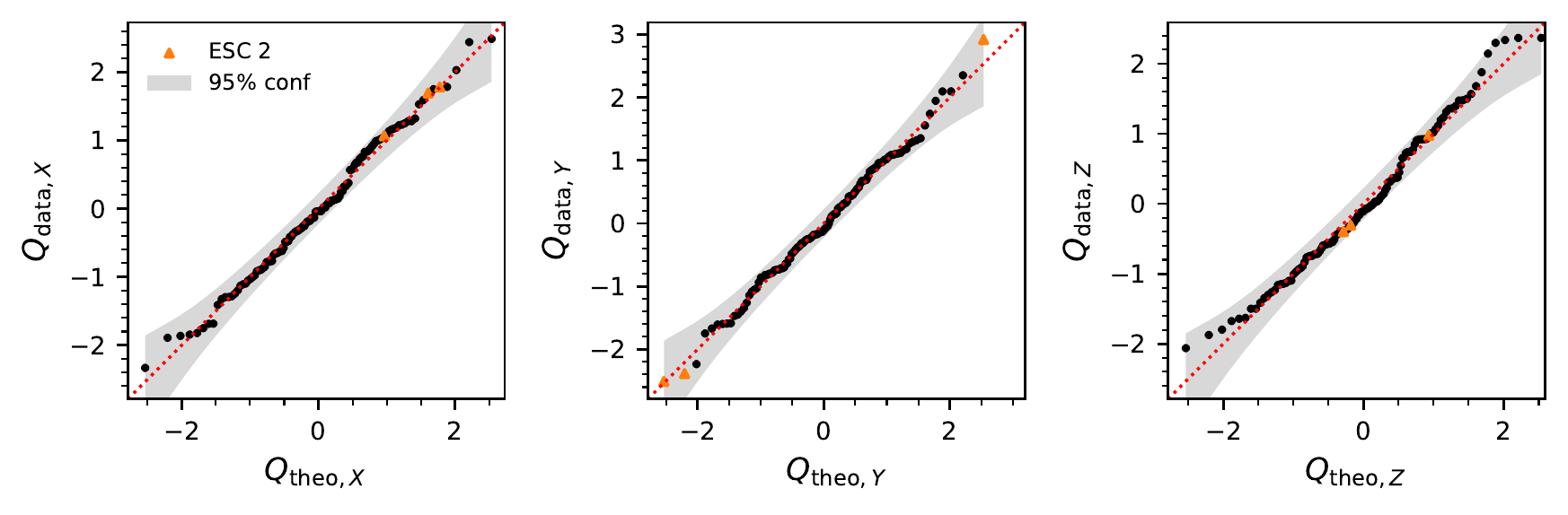}
\caption{$Q$-$Q$ plots showing the normality of the three velocity components (from left to right: $v_\alpha$, $v_\delta$, and $v_r$). High velocity (blue squares) and low velocity (orange triangles) escaping sources identified by \citet{kim:2019:109} are indicated. Additionally, velocity outliers from this study are indicated with green diamonds.
\textit{Top}: Analysis including all sources.
\textit{Middle}: Analysis after removing the escape group 3 sources.
\textit{Bottom}: Analysis after removing all sources in escape groups 1 and 3.
}
\label{fig:qq}
\end{figure*}

The means and IVDs are computed using the method outlined in Section~\ref{model}. Figure~\ref{fig:qq} shows the quantiles for all sources (top plot), quantiles after removing outliers in RV space (middle plot), and quantiles after removing large outliers in RV and tangential kinematic space (bottom plot). In each plot, the gray band indicates the 95\% confidence interval from bootstrapping. 
\citet{kim:2019:109} identified two separate groups of kinematic outliers, sources with highly significant deviations within the $Q$-$Q$ plot ($Q_{\mathrm{data},X} \geq \pm$3; escape group 1), and sources with low-significance escape velocities defined from the angular escape speed (sources with $\mu_{\mathrm{tot}} > 3.1$~mas yr$^{-1}$; escape group 2). We additionally identify sources with RVs that significantly deviate from the sample ($Q_{\mathrm{data},Z} \geq \pm$3), and label this escape group 3 (see Figures~\ref{fig:3dvelocities} and \ref{fig:qq}).
\citet{kim:2019:109} and \citet{kuhn:2019:32} estimated the escape speed for the ONC to be $\approx 6.1$ km s$^{-1}$ using the virial theorem, and all of the outliers in escape groups 1 and 3 have velocities in excess of this speed. However, it should be noted that velocity outliers in RV space could be potential binaries rather than escaping/evaporating sources.

Of the escaping stars identified in \citet{kim:2019:109}, one of the sources in escape group 1 and three of the sources in escape group 2 were observed with NIRSPAO, and none were in our reanalyzed APOGEE subsample (identified in Table~\ref{tbl:kinematics}). This is not surprising as all of the sources in escape groups 1 and 2 are within 2.5\arcmin\ of the ONC CoM (16 of 18 sources within 1\arcmin). The majority of sources in escape groups 1 and 2 tend to be fainter (14 of 18 sources with F139 mag $\gtrsim 12$), or have a nearby source within 2\arcsec.
All of the sources in escape groups 1 and 2 have RVs that are consistent with the bulk RV distribution of the ONC. Conversely, none of the sources in escape group 3 (line of sight velocity outliers) are within escape groups 1 and 2, and both escape group 3 sources are APOGEE sources. This indicates that the high velocity components for these sources are primarily in either the tangential or line of sight direction, but not both.

\subsection{Intrinsic Velocity Dispersions (IVDs)}
\label{ivds}

Using the framework outlined in Section~\ref{model}, we computed intrinsic velocity dispersions after removing all sources in escape groups 1 and 3. The values from our model were:
{\begin{align*}
\bar v_\alpha &= 0.01 \pm 0.16~\mathrm{km~s}^{-1},\\
\bar v_\delta &= -0.09 \pm 0.18~\mathrm{km~s}^{-1},\\
\bar v_r &= 27.45_{-0.22}^{+0.21}~\mathrm{km~s}^{-1},\\
\sigma_{v_\alpha} &= 1.64 \pm 0.12~\mathrm{km~s}^{-1},\\
\sigma_{v_\delta} &= 2.03 \pm 0.13~\mathrm{km~s}^{-1},\\
\sigma_{v_r} &= 2.56_{-0.17}^{+0.16}~\mathrm{km~s}^{-1},\\
\rho_1 &= 0.06 \pm 0.09,\\
\rho_2 &= -0.11 \pm 0.09,\\
\rho_3 &= -0.05 \pm 0.09.
\end{align*}}
\noindent The correlation coefficients are consistent with 0, indicating little to no correlation between velocity components.

Our computed velocity dispersions are roughly consistent with other studies in both the tangential (plane of the sky) direction: ($\sigma_{v_\alpha}$, $\sigma_{v_\delta}$) = ($1.63 \pm 0.04$, $2.20 \pm 0.06$)~km s$^{-1}$ (K19); ($\sigma_{v_\alpha}$, $\sigma_{v_\delta}$) = ($1.85 \pm 0.04$, $2.45 \pm 0.06$)~km s$^{-1}$ \citep{platais:2020:272}; and ($\sigma_{v_\alpha}$, $\sigma_{v_\delta}$) = ($1.79 \pm 0.12$, $2.32 \pm 0.10$)~km s$^{-1}$ \citep{jones:1988:1755},
and in the line of sight direction: $\sigma_{v_r} \approx 3.1$~km s$^{-1}$ \citep{furesz:2008:1109}; $\sigma_{v_r} \approx 2.1$--2.4~km s$^{-1}$ \citep{tobin:2009:1103}; and $\sigma_{v_r} \approx 2.2$~km s$^{-1}$ \citep{da-rio:2017:105}. However, our derived value of $\sigma_{v_r} = 2.56_{-0.17}^{+0.16}$~km s$^{-1}$ is slightly higher than the value measured in \citet{tobin:2009:1103} and \citet{da-rio:2017:105}. This is possibly due to our closer proximity to the ONC core than previous studies, or potentially the target selection of fainter, redder (lower-mass) sources. This may also indicate kinematic structure along the line of sight direction, similar to the velocity elongation seen in the north-south direction along the filament \citep[e.g.,][]{jones:1988:1755}.
It is also possible that this component potentially suffers from the impact of unresolved binaries, which we evaluate in the next section. 

We identified one new binary candidate within our sample ([HC2000] 332A), along with six other targets being components of apparent doubles (candidate binaries). At least one source ([HC2000] 546) shows significant RV variability between APOGEE and NIRSPAO measurements, although no secondary component was detected in the AO imaging. There is only one epoch of APOGEE data and one epoch of NIRSPAO data for [HC2000] 546, therefore, it is not possible to determine the orbital parameters of this potential binary without additional observations. Two of our sources have multi-epoch NIRSPAO observations ([HC2000] 229 and [HC2000] 217), however, RVs computed at each epoch are consistent with one another within errors.

\subsubsection{Unresolved Binarity}

Radial velocities are instantaneous velocity measurements, in contrast to PM measurements which are taken over baselines of years. The effects of binary orbital motion therefore influence RV measurements, but are typically averaged out over the long time baselines of PM measurements. Consequently, it is important for us to quantify the potential effects that unresolved binaries may have on the line of sight velocity dispersion(s).

\citet{raghavan:2010:1} did an extensive multiplicity study of solar-type stars, both in the field and young sources determined from chromospheric activity. Their findings were that the overall field multiplicity of solar-type stars is $44\% \pm 4\%$, with the multiplicity fraction of younger sources being statistically equivalent $40\% \pm 3\%$. Additionally, \citet{raghavan:2010:1} also noted a declining trend in the multiplicity fraction with redder color (lower primary mass). Down to the M dwarf regime, the multiplicity fraction for the field is estimated to be $\sim$20\%--25\%, declining with lower primary mass \citep{fischer:1992:178,clark:2012:119,ward-duong:2015:2618,bardalez-gagliuffi:2019:205}.

In the ONC, estimates of visual binaries range from $\sim$3\%--30\% \citep{kohler:2006:461, reipurth:2007:2272, duchene:2018:1825, de-furio:2019:95, jerabkova:2019:a57}. Over the specific range investigated by \citet{duchene:2018:1825}---0.3--2~$M_\Sun$ with companions within 10--60 au---they found the ONC binary fraction approximately 10\% higher than the field population estimates from \citet{raghavan:2010:1} and \citet{ward-duong:2015:2618}. 
Combined, this would place the binary fraction between 30\%--50\% for systems with primary masses between 0.1--1~$M_\Sun$. 
These results are roughly consistent with modeling estimates \citep[e.g.,][]{kroupa:1999:495,kroupa:2000:615,kroupa:2001:699}.

With such a large potential binary fraction in the ONC, it is important quantify how orbital motion can affect our measured line of sight velocity dispersion. This is primarily important to determine whether the observed anisotropy between the line of sight component and the tangential components is explained with orbital motion, or if there is a true elongation along the line of sight velocities.

\subsubsection{Simulating the Effect of Binaries}

To determine the effects of unresolved binarity on the IVD, we performed a test similar to \citet{da-rio:2017:105} and \citet{karnath:2019:46}. First, we generate a random intrinsic dispersion drawn from a uniform sample between 1--4 km s$^{-1}$. Next, we convolve this dispersion with the measurement error distribution. We use our observed measurement error distribution from the APOGEE+NIRSPEC dataset, and create an inverse cumulative distribution function which we randomly sample to build the error distribution. Lastly, we convolve this distribution with a velocity distribution from a set of synthetic binaries, and compare the final distribution to our observed distribution.

To generate our synthetic systems, we first generate a distribution of stars with masses between 0.1--1~$M_\Sun$. We apply a binary fraction (discussed below) to our sample, and use the mass ratio distribution from \citet{reggiani:2013:a124} to determine component masses within each system. Next, for binary systems, we apply the eccentricity distribution from \citet{duchene:2013:269}, and the period distribution from \citet{raghavan:2010:1}. 

For our simulations we generate 135 systems (the same number of sources in our 3-D kinematics sample). These synthetic systems are created using the \texttt{velbin} package \citep{cottaar:2014:a20,foster:2015:136}. This sampling assumes random orbits with random orientations and provides a one-dimensional radial velocity distribution. 

\begin{figure}
\centering
\includegraphics[width=\linewidth]{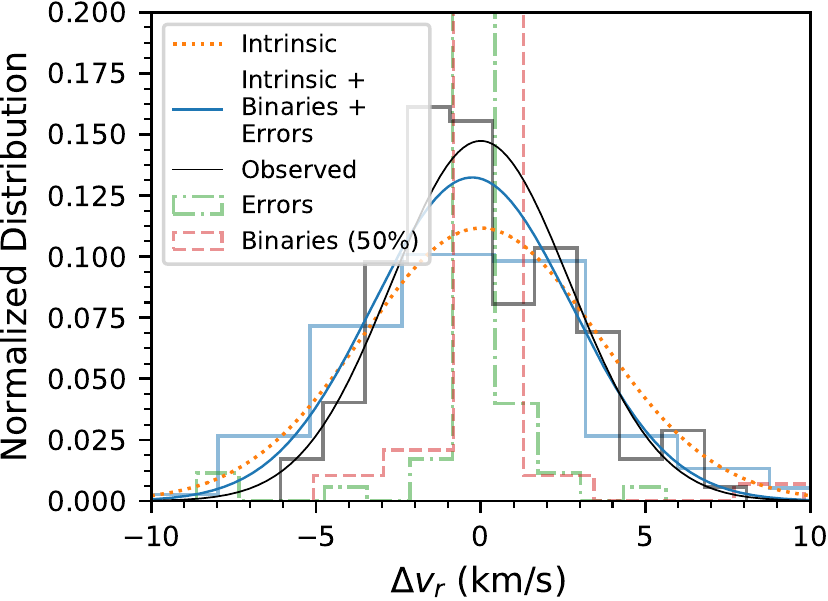}
\caption{Example of a single model generated in our binary simulation. The gray distribution and black solid line indicate our observations. The orange dotted line shows the intrinsic IVD distribution, and the green dash-dotted and red dashed lines indicate the distributions of uncertainties and binaries, respectively. The blue line and distribution show the final distribution accounting for binaries and uncertainties, which is compared to observations. Binaries and measurement uncertainties work to widen the dispersion, making the observed velocity dispersion larger than the intrinsic velocity dispersion.}
\label{fig:syntheticvelocities}
\end{figure}

This distribution is then compared to our observed APOGEE+NIRSPEC RV distribution, requiring that the standard deviation of the synthetic distribution be within 2-$\sigma$ of the observed distribution's standard deviation.
An example of one randomly drawn distribution compared to our observed distribution is shown in Figure~\ref{fig:syntheticvelocities}. The binary contribution to the resulting distribution is typically far less than the intrinsic dispersion and measurement error distribution, a result also noted by \citet{da-rio:2017:105}.

\begin{figure}
\centering
\includegraphics[width=\linewidth]{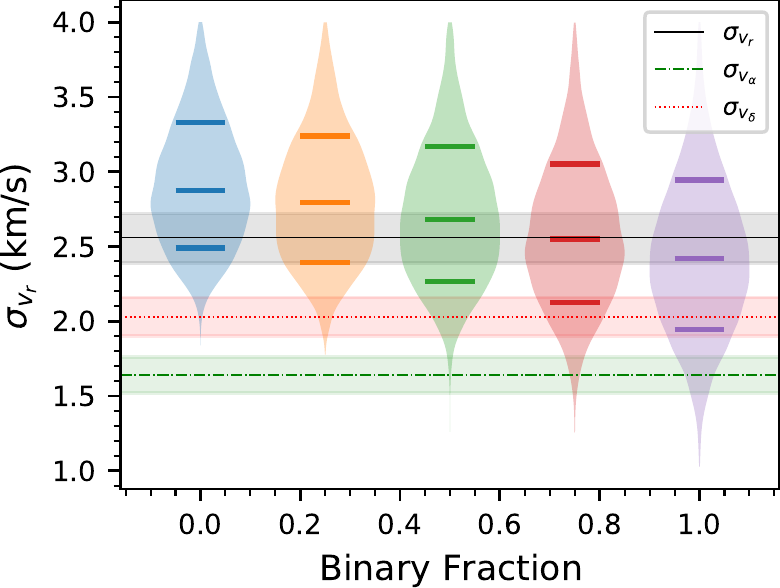}
\caption{Violin plots showing the distributions of our binary simulations. Distributions are generated at binary fractions of 0\%, 25\%, 50\%, 75\%, and 100\%, and lines within each distribution indicate 16th, 50th, and 84th percentiles. Our measured IVD (black line and gray 1-$\sigma$ confidence region) is consistent with a essentially all binary fractions. We also show our tangential IVD components in $\alpha$ (green dashed line and region) and $\delta$ (red dotted line and region), indicating that the binary fraction would need to be $\gtrsim$75\% to bring the line of sight component within the range of the tangential components.}
\label{fig:binaryivds}
\end{figure}

We kept the first 10$^5$ models that fit our similarity criteria stated above and generate a distribution of intrinsic velocity dispersions that passed the similarity criteria of our observed RV distribution. We performed this test for binary fractions of 0\%, 25\%, 50\%, 75\%, 100\%. Figure~\ref{fig:binaryivds} shows our simulation results compared to our IVD determined in Section~\ref{ivds}. Figure~\ref{fig:binaryivds} illustrates that our measured IVD is consistent with any level of binarity, however, for the line of sight component to be consistent with the tangential components, the ONC would need a binary fraction $\gtrsim$75\%, which is inconsistent with observations and modeling results. Therefore, the larger measured $\sigma_{v_r}$ is likely due to formation/evolution rather than binary effects, and the apparent elongation in the line of sight component appears to be real rather than a systematic.

\section{Discussion}
\label{sec:discussion}

\subsection{Virial State of the ONC Core}

A number of studies have examined if the ONC is virialzed \citep[e.g.,][]{jones:1988:1755, da-rio:2014:55, da-rio:2017:105, kim:2019:109, kuhn:2019:32}. \citet{da-rio:2014:55} estimated a 1-D mean velocity dispersion of $\sigma_{v, \mathrm{1D}} \approx 1.73$~km s$^{-1}$ if the ONC is in virial equilibrium. To test for a virialized state, we adopt a methodology similar to \citet{kim:2019:109}, computing the 1-D velocity dispersion as a function of radial distance outwards.

\begin{figure}
\centering
\includegraphics[width=\linewidth]{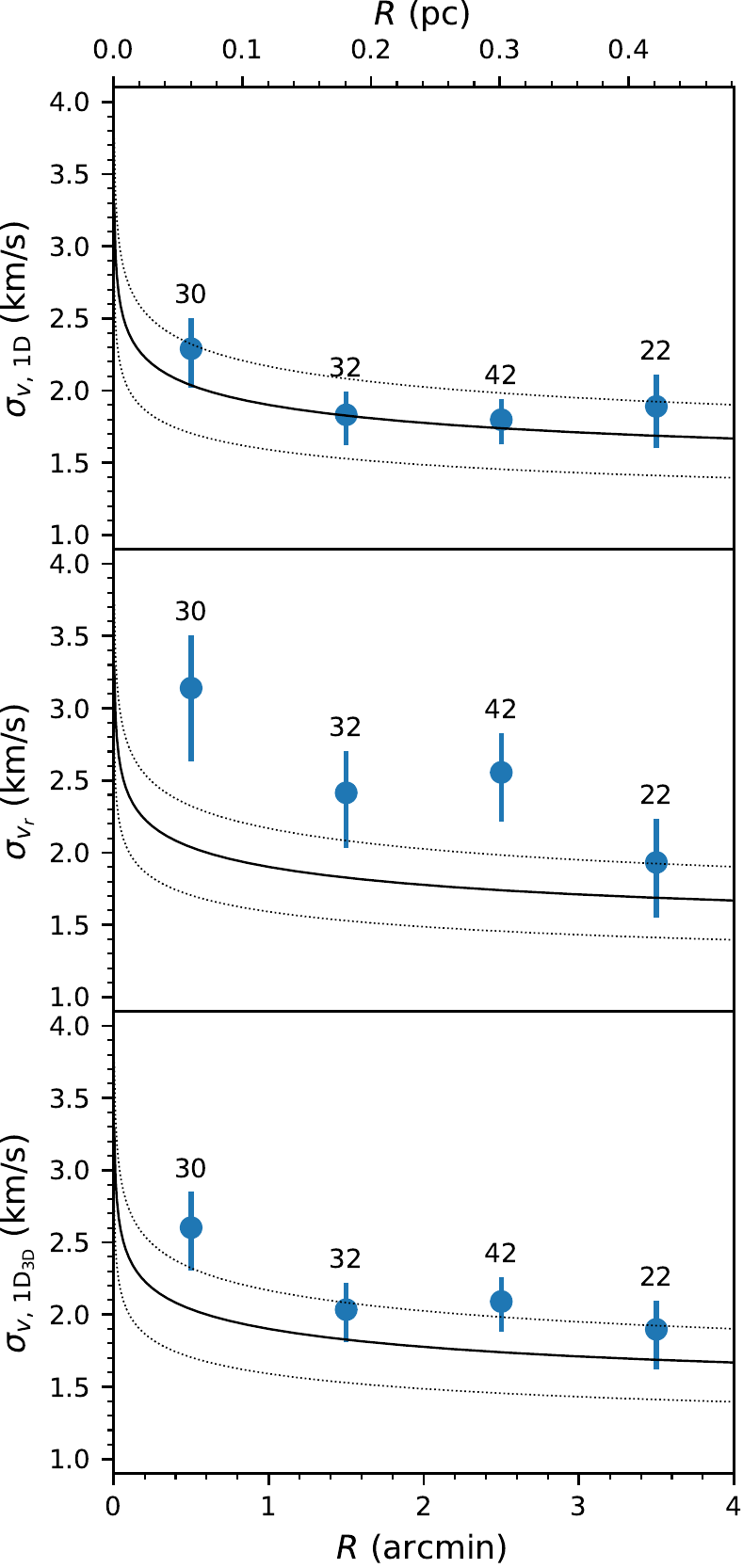}
\caption{ 
Plot of the one-dimensional tangential velocity dispersion $\sigma_{v,\mathrm{1D}}$ (top), the line of sight velocity dispersion $\sigma_{v_r}$ (middle), and the one-dimensional velocity dispersion composed of all three velocity components (assumed isotropic) $\sigma_{v,\mathrm{1D_{3D}}}$ (bottom) as a function of distance from the center of the ONC. Values plotted are listed in Table~\ref{tbl:dispersions}, using an estimated distance to the ONC of $388\pm5$~pc \citep{kounkel:2017:142}. The number of sources in each bin is indicated above that bin. 
%The blue and yellow lines and dashed lines indicate the best-fit and uncertainty to the radial velocity dispersion profiles. 
The black solid line illustrates the one-dimensional velocity dispersion for virial equilibrium predicted from the stellar and gas mass assuming a spherical potential from \citet{da-rio:2014:55}, and the dashed lines mark the uncertainty assuming a 30\% mass error. The Bottom x-axis indicates distance in arcminutes, while the top x-axis indicates distance in pc.
}
\label{fig:veldispersions}
\end{figure}

To balance the small numbers in our sample, we chose radial bins of size 1\arcmin. For each bin, we computed velocity dispersion using the methods outlined in Section~\ref{model}. The values of $\sigma_\alpha$ and $\sigma_\delta$ where then used to compute the 1-D velocity dispersion, i.e., $\sigma_{v, \mathrm{1D}}^2=(\sigma_{v_\alpha}^2 + \sigma_{v_\delta}^2)/2$. Our computed velocity dispersions are listed in Table~\ref{tbl:dispersions} and also plotted in Figure~\ref{fig:veldispersions}. We also show the 1-D velocity dispersion predicted for models of virial equilibrium using the estimated combined gas and stellar mass from \citet[][solid line]{da-rio:2014:55}. We assigned a 30\% mass uncertainty, similar to \citet{kim:2019:109}, which are illustrated with dotted lines. It should be noted that the model from \citet{da-rio:2014:55} assumes a spherical potential; however, there is evidence that the potential is closer to an elongated spheroid \citep[e.g.,][]{hillenbrand:1998:540, carpenter:2000:3139, kuhn:2014:107, kuznetsova:2015:27, megeath:2016:5}.

The 1-D velocity dispersion computed from proper motions is extremely consistent with the results of \citet{kim:2019:109}, which is expected since our subsample originated from their catalog. The 1-D velocity dispersions favor a virialized state for the majority of the ONC. The radial distribution of $\sigma_{v_r}$ is similar to that from \citet[][see Figure 12]{da-rio:2017:105} for the ONC, although, they only show 1 bin from $R\sim0$--0.4~pc. 
However, they also find a dispersion 1-$\sigma$ higher than the virial equilibrium model predicts. We also computed the one-dimensional velocity dispersion using all three velocity components $\sigma^2_{v, \mathrm{1D_{3D}}} = (\sigma^2_{v_\alpha} + \sigma^2_{v_\delta} + \sigma^2_{v_r})/3$, listed in Table~\ref{tbl:dispersions}. This measurement marginalizes over potential differences in a single velocity component. Figure~\ref{fig:veldispersions} (bottom panel) shows that only the bin closest to the ONC core appears elevated from the virialized model, however, this method may wash out features in a single velocity component which deviate from a virialized state. These results add to the growing evidence that the core of the ONC is not fully virialized. Additionally, we do not find evidence of global expansion similar to the results of \citet{kim:2019:109}. Without accurate distances to individual sources, we are not able to explore if there is expansion along the line of sight velocity component.

\begin{deluxetable*}{ccccccccc}
\tablecaption{Velocity Dispersions as a Function of Distance \label{tbl:dispersions}}
\tablecolumns{9}
\tablehead{
\colhead{Radii} & \colhead{$\sigma_{v_\alpha}$} & \colhead{$\sigma_{v_\delta}$} & \colhead{$\sigma_{v, \mathrm{rad}}$} & \colhead{$\sigma_{v, \mathrm{tan}}$} & \colhead{$\sigma_{v, \mathrm{1D}}$} & \colhead{$\sigma_{v_r}$} & \colhead{$\sigma_{v, \mathrm{1D_{3D}}}$} & \colhead{$N$} \\
\colhead{(arcmin)} & \colhead{(km s$^{-1}$)} & \colhead{(km s$^{-1}$)} &\colhead{(km s$^{-1}$)} & \colhead{(km s$^{-1}$)} &
\colhead{(km s$^{-1}$)} & \colhead{(km s$^{-1}$)} & \colhead{(km s$^{-1}$)} & \colhead{} 
}
\startdata
0--1 & $1.98_{-0.27}^{+0.27}$ & $2.55_{-0.31}^{+0.31}$ & $2.22_{-0.29}^{+0.29}$ & $2.38_{-0.31}^{+0.31}$ & $2.28_{-0.29}^{+0.21}$ & $3.17_{-0.39}^{+0.39}$ & $2.61_{-0.31}^{+0.25}$ & 30 \vspace{5pt} \\ 
1--2 & $1.71_{-0.23}^{+0.23}$ & $1.95_{-0.24}^{+0.24}$ & $1.46_{-0.19}^{+0.19}$ & $2.12_{-0.26}^{+0.26}$ & $1.84_{-0.21}^{+0.16}$ & $2.41_{-0.28}^{+0.28}$ & $2.04_{-0.23}^{+0.18}$ & 32 \vspace{5pt} \\ 
2--3 & $1.74_{-0.19}^{+0.19}$ & $1.86_{-0.21}^{+0.21}$ & $1.77_{-0.20}^{+0.20}$ & $1.85_{-0.20}^{+0.20}$ & $1.80_{-0.17}^{+0.14}$ & $2.56_{-0.29}^{+0.29}$ & $2.08_{-0.20}^{+0.17}$ & 42 \vspace{5pt} \\ 
3--4 & $1.45_{-0.22}^{+0.22}$ & $2.22_{-0.33}^{+0.33}$ & $1.88_{-0.29}^{+0.29}$ & $1.94_{-0.27}^{+0.27}$ & $1.87_{-0.29}^{+0.21}$ & $1.92_{-0.30}^{+0.30}$ & $1.89_{-0.28}^{+0.21}$ & 22 \vspace{5pt} \\ 
\hline
\enddata
\end{deluxetable*}

\subsection{Effects from the Integral Shaped Filament}

The Trapezium sits approximately in the middle of the ``integral-shaped filament" \citep[ISF;][]{bally:1987:l45}, a long filament of gas with an approximate ``S" shape, and 0.1--0.3~pc in front of the filament \citep[e.g.,][]{baldwin:1991:580, wen:1995:784, odell:2018:1017, abel:2019:130}. There has been an observed elongation in the line of sight velocity component, which \citet{stutz:2016:a2} attributed to interactions with the ISF using APOGEE data. The mechanism put forth by \citet{stutz:2016:a2} to explain the observed elongation in velocity space is the ``slingshot" mechanism, where stars born along the filament could be ejected due to the filament undergoing transverse acceleration while the protostar continually accretes mass. The reason for the ejection is due to the fact that, ``when the protostar system becomes sufficiently massive to decouple from the filament, it is released" \citep{stutz:2016:a2}. Such a mechanism would provide an additional contribution to a larger velocity dispersion. However, the slingshot mechanism is dependent on the direction of the transverse acceleration of the filament (i.e., along the line of sight or tangential on the plane of the sky). As the filament runs north-south in the region of the Trapezium, this could provide the mechanism for the observed anisotropy in the tangential velocity components, as well as the radial component. This could be an important effect as \citet{stutz:2016:a2} find the gravity of the background filament likely dominates the gravity field from the stars.

From the analysis of \citet{stutz:2016:a2}, the region we have analyzed here contains primarily stars versus protostars, determined based on their excess IR emission.
However, \citet{stutz:2016:a2} only considered the radial component of motion, and not the tangential components. Additionally, their RV sources were obtained from APOGEE, providing very few sources within the central 0.1~pc of the core region where we observe the highest velocity dispersion along the line of sight. 
As such, there is additional work to determine how the ISF might work to influence the measured velocity dispersions. Although we do not provide an in-depth theoretical analysis here to compare to observational data, one is warranted.

\subsection{Velocity Isotropy}

There is a known kinematic anisotropy in the tangential velocities along the north-south direction \citep[e.g.,][]{mcnamara:1976:375, hillenbrand:1998:540, da-rio:2014:55, kim:2019:109}. However, the deviation from tangential to radial (i.e., towards the center of the cluster) anisotropy ($\sigma_\mathrm{tan}/\sigma_\mathrm{rad} - 1$; \citealt{bellini:2018:86}) found by \citet{kim:2019:109} of $0.03 \pm 0.04$ is consistent with isotropic velocities in the tangential-radial velocity space. 
It should be noted that throughout this section, we use ``tangential" ($v_\mathrm{tan}$) to indicate motion on the plane of the sky, ``radial" ($v_\mathrm{rad}$) to indicate motion on the plane of the sky pointing towards the cluster center, and ``line of sight" ($v_r$) to indicate the line of sight (RV) component of motion.

\begin{figure}
\centering
\includegraphics[width=\linewidth]{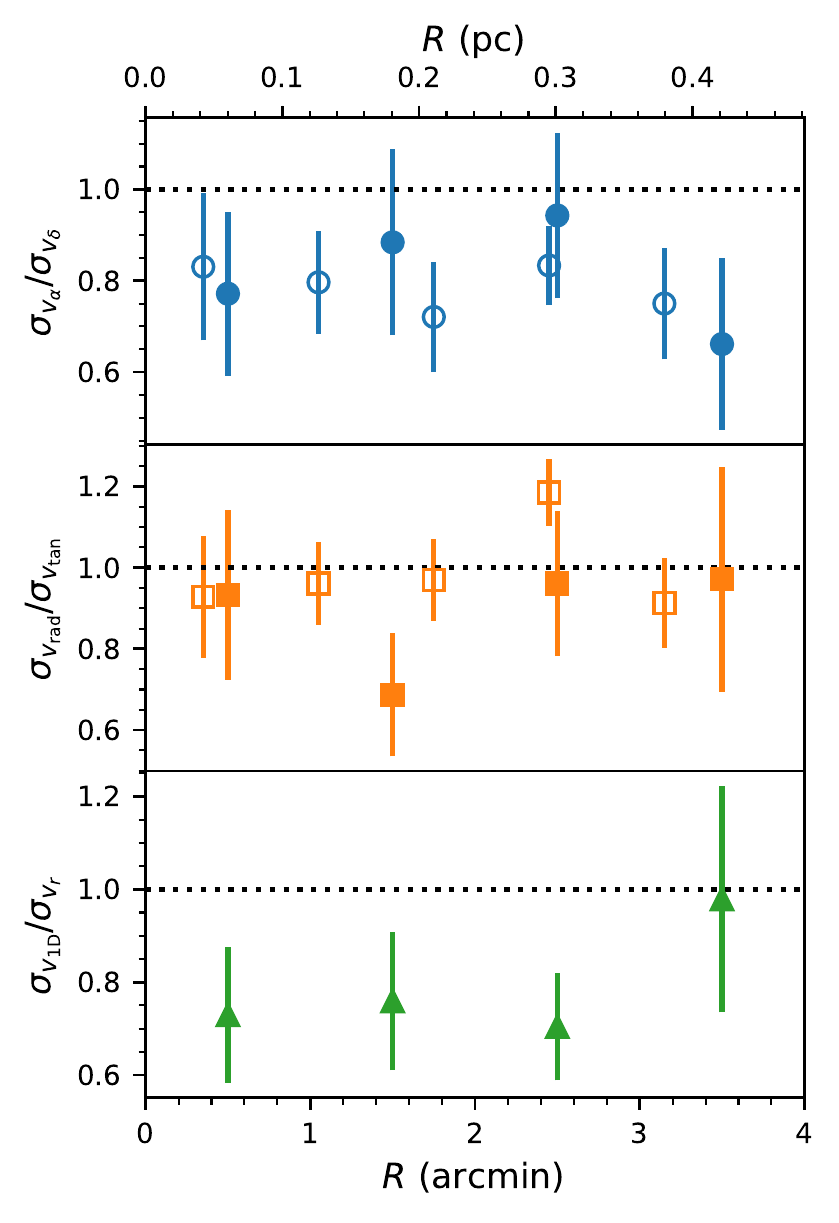}
\caption{\textit{Top}: Ratio of east-west to north-south velocity dispersions ($\sigma_{v_\alpha} / \sigma_{v_\delta}$) as a function of distance from the core of the ONC. Open markers are values from \citet{kim:2019:109}, while closed symbols are from this study. There is a slight north-south elongation that has been found by other studies \citep[e.g.,][]{da-rio:2014:55, kim:2019:109}. 
\textit{Middle}: Ratio of radial to tangential (vectors on the plane of the sky pointing radially towards the center of the ONC and tangential to that vector) velocity dispersions ($\sigma_{v_\mathrm{rad}} / \sigma_{v_\mathrm{tan}}$). These components are primarily isotropic, with the exception of the bin between 1\arcmin--2\arcmin. 
\textit{Bottom}: Ratio of tangential to line of sight velocity dispersions ($\sigma_{v_\mathrm{1D}} / \sigma_{v_r}$). There appears to be a slight elongation in the line of sight direction.}
\label{fig:anisotropy}
\end{figure}

With our three-dimensional kinematic information, we can now compute the ratio of the tangential dispersion to the line of sight dispersion ($\sigma_{v_\mathrm{1D}} / \sigma_{v_r}$). Figure~\ref{fig:anisotropy} shows the velocity ratios for $\sigma_{v_\alpha} / \sigma_{v_\delta}$ (top) and $\sigma_{v_\mathrm{1D}} / \sigma_{v_r}$ (bottom). The $\sigma_{v_\alpha} / \sigma_{v_\delta}$ shows a north-south elongation, which is consistent with previous studies, e.g., $\sigma_{\mu_{\alpha\cos\delta}} / \sigma_{\mu_\delta} = 0.74\pm0.03$; \citep{kim:2019:109}; $b/a \approx 0.7$ \citep{da-rio:2014:55}; see also \citet{jones:1988:1755} and \citet{kuhn:2019:32}. 

We also decomposed tangential velocities to radial ($v_\mathrm{rad}$) and tangential ($v_\mathrm{tan}$) coordinates, both on the plane of the sky, through a change of basis where the radial axis points towards the ONC CoM. \textit{We note that both of these components, $v_\mathrm{rad}$ and $v_\mathrm{tan}$, are strictly on the plane of the sky and do not include any line of sight velocity component within their vectors.} The ratio of $\sigma_{v_\mathrm{rad}} / \sigma_{v_\mathrm{tan}}$ is shown in Figure~\ref{fig:anisotropy} (middle), where the majority of points are consistent with isotropic dispersions. We also compare our results to the parent population from \citet[][open markers]{kim:2019:109}, which shows that our subsample shows variation from the parent population, possibly due to selection effects.
Our 1-D tangential to line of sight velocity dispersions ($\sigma_{v_\mathrm{1D}} / \sigma_{v_r}$) show significant deviation from unity. The combined value for our total sample is $\sigma_{v_\mathrm{1D}} / \sigma_{v_r} = 0.78 \pm 0.19$. These measurements may indicate an elongation in the velocities along the line of sight direction.

\begin{figure}
\centering
\includegraphics[width=0.8\linewidth]{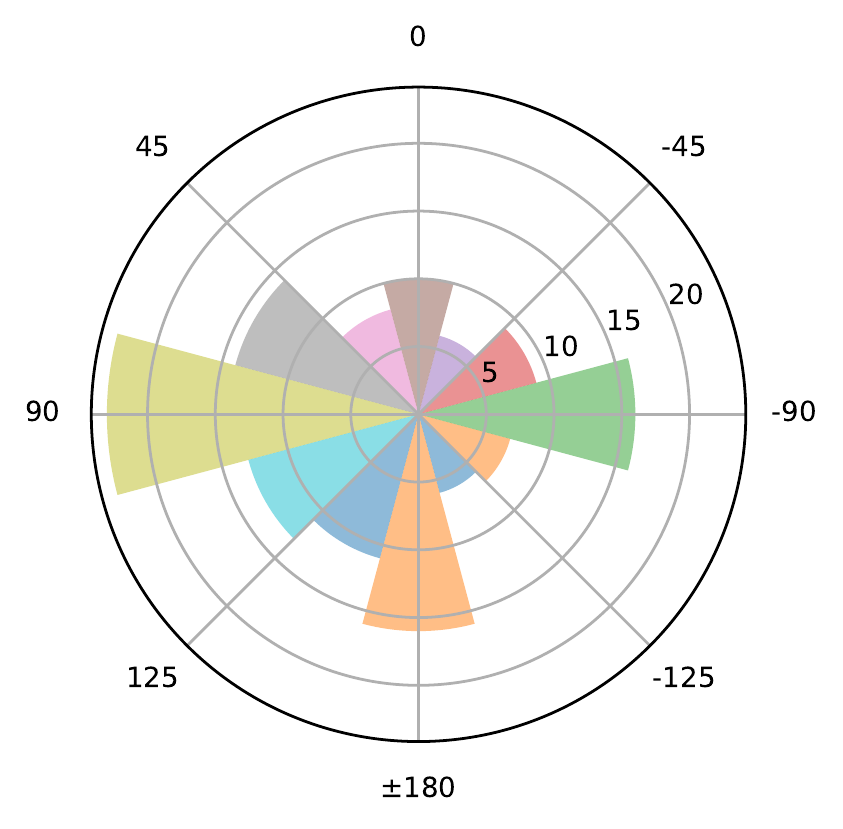}
\includegraphics[width=0.9\linewidth]{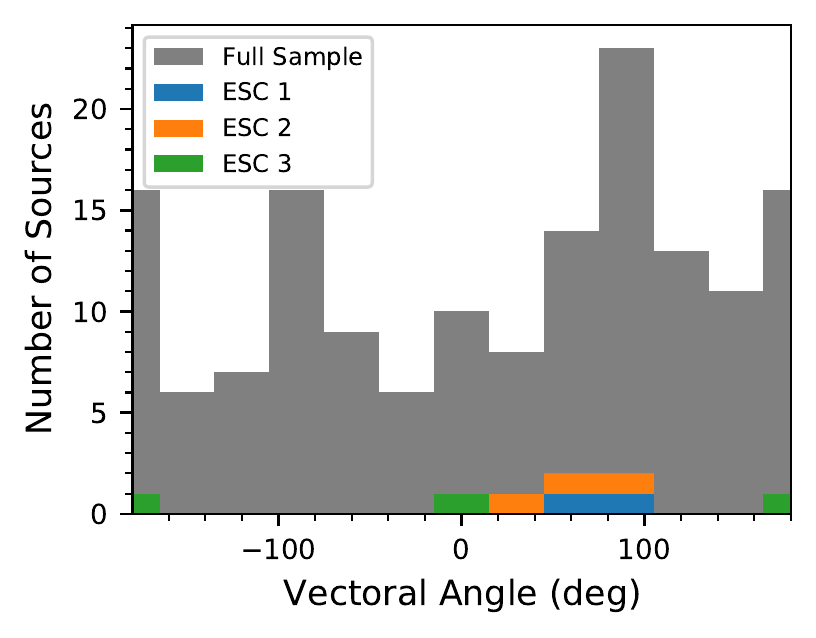}
\includegraphics[width=0.9\linewidth]{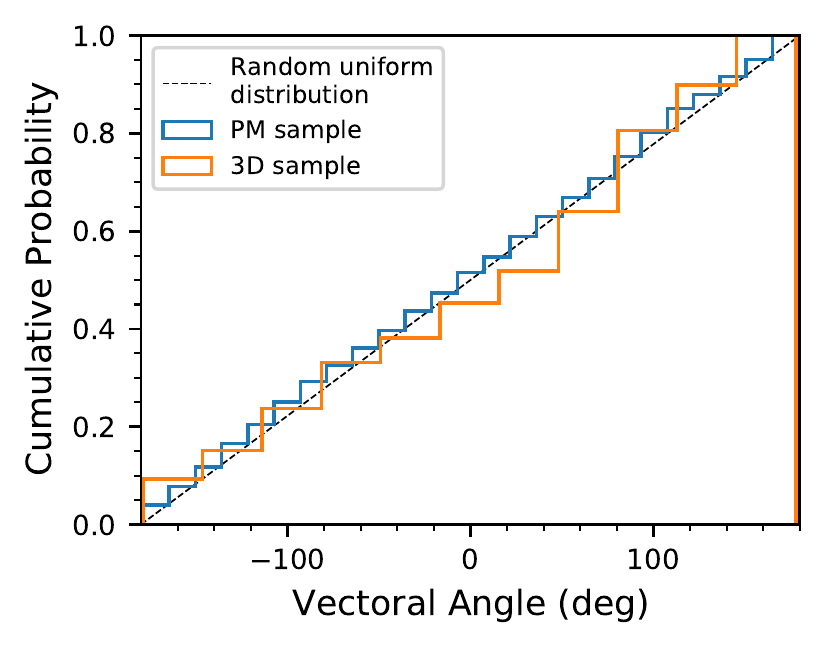}
\caption{
\textit{Top}: Polar histogram of the angle between the proper motion vector and the radial position vector on the plane of the sky, where an angle of 0 corresponds to both the position vector and velocity vector pointing radially away from the cluster center. Each bin is 30$^\circ$ wide. The radial direction indicates the number of sources in each bin. There is an observed preference for sources to have motion tangential to the core of the ONC (${\sim}\pm90^\circ$), with more sources exhibiting motion +$90^\circ$ than $-90^\circ$.
\textit{Middle}: The 1-D histogram corresponding to the bins in the polar histogram. We show the vector angles of the escape groups identified in this study and \citet{kim:2019:109}, noting that few of these sources have angles $\sim$0$^\circ$.
\textit{Bottom}: Cumulative probability distributions for our 3-D sample (orange), the entire PM sample within 4\arcmin\ (blue), and the expected distribution for a random uniform sample (dashed line).
}
\label{fig:vectoralangles}
\end{figure}

\citet{platais:2020:272} used a diagram of the angle of the proper motion vector in polar coordinates with respect to the cluster center to examine the how the position vector from the center of the ONC and tangential velocity vector were related for fast moving sources. The expectation is that runwaway stars should have angles between these two vectors close to 0, that is, both the position vector point and proper motion vector point radially away from the center of the ONC. We can use a similar analysis to see if there is preferential motion in the core of the ONC. Figure ~\ref{fig:vectoralangles} show the distribution of vectoral angles for all \kinematicsample\ sources in our sample with three components of motion. Random orbits should have no preferential angle, however, there is structure seen in Figure~\ref{fig:vectoralangles}. Specifically, there is a preference for sources to have vectoral angles around 90$^\circ$ and $-90^\circ$. This corresponds to sources whose motion is tangential to a vector pointing radially away from the center of the ONC (normal vector), possibly indicating that there is a rotational preference for sources in the central ONC. In Figure~\ref{fig:vectoralangles} (bottom) we compare the cumulative distribution of vectoral angles for our 3-D sample to the vectoral angles for all PM sources within 4\arcmin\ of the ONC CoM from \citet[][671 sources]{kim:2019:109} and a random uniform distribution. The PM sample follows the expected behavior for a random uniform distribution of vectoral angles, which implies a potential bias within our 3-D subsample. 

To test for similarity between our 3-D subsample and the parent PM sample from \citet{kim:2019:109}, we performed the Kolmogorov-Smirnov test (K-S test) and the Anderson-Darling test (A-D test). The K-S test gave a test statistic = 0.115 with a p-value = 0.085, which indicates that we can reject the null hypothesis that the two samples come from the same distribution only at an 8.5\% probability. The A-D test gave a test statistic = 0.802 with a p-value = 0.153, indicating that we cannot reject that these two distributions are significantly different, similar to the result of the K-S test. Neither of these test results are extremely significant, indicating that the resulting preferential motion of our 3-D sample may not be statistically significant. However, it is worth noting the potential biases introduced into our 3-D subsample versus the \citet{kim:2019:109} sample. 

We also performed a comparison test against a random uniform sample using the same tests mentioned earlier. For our 3-D subsample, the K-S test gave a test statistic = 0.126 with a p-value = 0.023, which indicates that we can reject the null hypothesis that the kinematic sample comes from a uniform distribution at a 2.3\% probability. The A-D test gave a test statistic = 1.786 with a p-value = 0.064, which indicates that the null hypothesis can be rejected at the 6.4\% level. Again, neither of these test results are extremely significant, motivating the need for a larger kinematic sample within the core. Comparing the parent sample from \citet{kim:2019:109} to the random distribution gives a K-S test statistic = 0.019 with a p-value = 0.974, and an A-D test statistic = -0.892 with a p-value $> 0.25$ (value capped). Both tests indicate that the parent sample is consistent with a random distribution.

Our sample was selected for sources bright enough to be targeted with NIRSPAO (or APOGEE), likely selecting more ONC sources that are in the foreground portion of the cluster rather than deeply embedded sources, which may indicate motion differences as a function of depth in the ONC. Our target list was also curated from the [HC2000] catalog, selecting sources preferentially thought to have masses $\lesssim 1\,M_\odot$. It is possible that lower-mass sources are showing different kinematics than the bulk motion within the ONC. \citet{kuhn:2019:32} attempted to measure the bulk rotation of the ONC using tangential kinematics, but found no significant rotational preference.

It is important to note that this analysis is done assuming all stars are at the same distance. The \textit{Gaia} satellite \citep{gaia-collaboration:2016:a1} is currently obtaining parallaxes for over 1 billion sources; however, it has been noted in previous studies that \textit{Gaia} measurements in the ONC are unreliable due to the high level of nebulosity \citep[e.g.,][]{kim:2019:109}. Indeed, the number of \textit{Gaia} early Data Release 3 \citep[eDR3;][]{gaia-collaboration:2021:a1} sources within 1\arcsec\ of our sources from this study, and consistent within the 2-$\sigma$ combined uncertainty in $\mu_\alpha$ and $\mu_\delta$ is only three sources. Therefore, a full three-dimensional study of ONC kinematics is not yet possible, and motivates the need for future facilities to obtain parallax measurements for ONC sources.

\section{Conclusions}
\label{sec:conclusions}

Using Keck/NIRSPEC+AO, we have obtained high-precision RV measurements ($\sigma \lesssim 0.5$~km s$^{-1}$) for a large number of sources within the core region of the ONC (41 sources within 1\arcmin). Additionally, we included a reanalysis of \apogeesourcesfourarcmin\ ONC sources observed with SDSS/APOGEE. Using a combined sample of Keck/NIRSPAO, SDSS/APOGEE, and PM measurements from \citep{kim:2019:109}, for a total sample of \kinematicsample\ sources, we presented a 3-D study of the kinematics of the core population of the ONC. Our main takeaways are the following:

\begin{enumerate}%[nolistsep]

\item Our derived tangential intrinsic velocity dispersions of $\sigma_{v_\alpha} = 1.64 \pm 0.12$~km s$^{-1}$ and
$\sigma_{v_\delta} = 2.03 \pm 0.13$~km s$^{-1}$ are consistent with previous results from the literature, and consistent with the virialized model of the ONC from \citet{da-rio:2014:55}.

\item Our derived line of sight velocity dispersion of $\sigma_{v_r} = 2.56_{-0.17}^{+0.16}$~km s$^{-1}$ is slightly higher than literature estimates. This is potentially due to our sources being concentrated more towards the core of the ONC, which may indicate that the core of the ONC is not yet fully virialized.
We explored the possibility that binarity could play a role in creating our larger observed RV dispersions. We simulated different binary fractions, from 0\%--100\%, and their effect on the observed line of sight velocity dispersion. We found that almost any level of binarity could produce our observed velocity dispersion, however, only a very high level of binarity ($\gtrsim$75\%) would make our line of sight velocity dispersion consistent with the observed tangential velocity dispersions. As the binary fraction for the ONC is estimated to be between $\sim$10\%--50\% \citep[e.g.,][]{reipurth:2007:2272, de-furio:2019:95, da-rio:2017:105}, our larger observed RV dispersion is likely not caused by orbital motion in unresolved binaries, and is more likely related to formation/evolution within the ONC.

\item We measure an elongation in the velocity dispersion along the line of sight direction compared to the tangential velocity dispersion, $\sigma_{v_\mathrm{1D}} / \sigma_{v_r} = 0.78 \pm 0.19$. This may indicate that there is structure to the velocities of stars in the ONC, or possibly a result of the ``slingshot" mechanism from the background filament \citep{stutz:2016:a2,stutz:2018:4890}. The ratio of tangential velocity dispersions $\sigma_{v_\alpha} / \sigma_{v_\delta}$ shows a north-south elongation which is consistent with previous studies \citep[e.g.,][]{da-rio:2014:55, kim:2019:109}, and tends to run along the filament. However, the tangential to radial (towards the center of the ONC) velocity dispersions $\sigma_{v_\mathrm{1D}} / \sigma_{v_r}$ appear consistent with unity, as other studies have noted \citep[e.g.,][]{da-rio:2014:55, kim:2019:109}.

\item We observe two additional potential kinematic outlier sources (escaping/evaporating) based on their RVs from APOGEE (2MASS 05351906$-$0523495 and 2MASS 05352321$-$0521357). However, their large line of sight velocities may also be due to binarity, and additional follow up is needed to confirm if their velocities are truly elevated.

\item There is a somewhat low probability that the 3-D sample is drawn from a uniform distribution, which could indicate a rotational preference on the plane of the sky, as indicated by the angles between sources position vectors and tangential motion vectors. However, the parent population of \citet{kim:2019:109} is consistent with a uniform distribution, i.e. no rotational preference. We find that both our 3-D sample and the \citet{kim:2019:109} sample are relatively consistent, therefore, we cannot rule out the null hypothesis that both populations are consistent with a uniform distribution. A larger sample with higher precision measurements is likely required to further investigate the bulk rotation of the ONC core.

\end{enumerate}

Our study indicates that additional AO imaging and/or extremely high-resolution spectroscopy is necessary to increase the sample size and reduce uncertainties to determine if the core of the ONC is virialized and/or shows significant velocity structure. This will allow for a comprehensive study of the 3-D kinematics of the ONC to better understand the difference between tangential velocities and line of sight velocities. 
Future work is required to compare data to detailed simulations \citep[e.g.,][]{kroupa:2000:615, mckee:2002:59, mckee:2003:850, proszkow:2009:1020, krumholz:2011:74, krumholz:2012:71, kuznetsova:2015:27, kuznetsova:2018:2372} to determine the dynamical state of the ONC.

%\clearpage
\begin{acknowledgments}

The authors would like to thank the anonymous referee for the incredibly valuable suggestions and comments that helped improve the impact of this manuscript. The authors would also like to thank observing assistants Joel Aycock, Heather Hershley, Carolyn Jordan, Julie Rivera, Terry Stickel, Gary Puniwai, and Cynthia Wilburn, and supporting astronomers Greg Doppmann, Carlos Alvarez, Josh Walawender, Percy Gomez, Randy Campbell, Al Conrad, Alessandro Rettura, Jim Lyke, Luca Rizzi, and Hien Tran, for their help in obtaining the observations. J.R.L. and D.K. acknowledge support from NSF AST-1764218, HST-AR- 13258, and HST-GO-13826. J.R.L., Q.K., D.K., and C.A.T. are supported by NSF grant AST-1714816. This work is based on observations made with the NASA/ESA \textit{Hubble Space Telescope}, obtained at the Space Telescope Science Institute, which is operated by the Association of Universities for Research in Astronomy, Inc., under NASA contract NAS 5-26555

Support for this work was provided by NASA through the NASA Hubble Fellowship grant HST-HF2-51447.001-A awarded by the Space Telescope Science Institute, which is operated by the Association of Universities for Research in Astronomy, Inc., for NASA, under contract NAS5-26555.

Funding for the Sloan Digital Sky Survey IV has been provided by the Alfred P. Sloan Foundation, the U.S. Department of Energy Office of Science, and the Participating Institutions. SDSS-IV acknowledges support and resources from the Center for High-Performance Computing at the University of Utah. The SDSS web site is www.sdss.org.

SDSS-IV is managed by the Astrophysical Research Consortium for the Participating Institutions of the SDSS Collaboration including the Brazilian Participation Group, the Carnegie Institution for Science, Carnegie Mellon University, the Chilean Participation Group, the French Participation Group, Harvard-Smithsonian Center for Astrophysics, Instituto de Astrof\'isica de Canarias, The Johns Hopkins University, Kavli Institute for the Physics and Mathematics of the Universe (IPMU) / University of Tokyo, the Korean Participation Group, Lawrence Berkeley National Laboratory, Leibniz Institut f\"ur Astrophysik Potsdam (AIP), Max-Planck-Institut f\"ur Astronomie (MPIA Heidelberg), Max-Planck-Institut f\"ur Astrophysik (MPA Garching), Max-Planck-Institut f\"ur Extraterrestrische Physik (MPE), National Astronomical Observatories of China, New Mexico State University, New York University, University of Notre Dame, Observat\'ario Nacional / MCTI, The Ohio State University, Pennsylvania State University, Shanghai Astronomical Observatory, United Kingdom Participation Group, Universidad Nacional Aut\'onoma de M\'exico, University of Arizona, University of Colorado Boulder, University of Oxford, University of Portsmouth, University of Utah, University of Virginia, University of Washington, University of Wisconsin, Vanderbilt University, and Yale University.

Some of the data presented herein were obtained at the W. M. Keck Observatory, which is operated as a scientific partnership among the California Institute of Technology, the University of California and the National Aeronautics and Space Administration. The Observatory was made possible by the generous financial support of the W. M. Keck Foundation.

The authors recognize and acknowledge the very significant cultural role and reverence that the summit of Maunakea has with the indigenous Hawaiian community, and that the W.\ M.\ Keck Observatory stands on Crown and Government Lands that the State of Hawai'i is obligated to protect and preserve for future generations of indigenous Hawaiians. 

This research has made use of the Keck Observatory Archive (KOA), which is operated by the W. M. Keck Observatory and the NASA Exoplanet Science Institute (NExScI), under contract with the National Aeronautics and Space Administration.

Portions of this work were conducted at the University of California, San Diego, which was built on the unceded territory of the Kumeyaay Nation, whose people continue to maintain their political sovereignty and cultural traditions as vital members of the San Diego community. 
\end{acknowledgments}

%\clearpage
\facilities{HST(ACS), 
Keck:II (NIRSPEC, NIRC2)
Sloan (APOGEE)
}
\software{\textit{Astropy} \citep{astropy-collaboration:2013:a33}, 
\textit{emcee} \citep{foreman-mackey:2013:306}, 
\textit{corner} \citep{foreman-mackey:2016:}, 
\textit{NumPy} \citep{oliphant:2006:},
\textit{SciPy} \citep{virtanen:2020:261},
\textit{Matplotlib} \citep{hunter:2007:90},
\textit{apogee} \citep{bovy:2016:49}
}

%\bibliography{MyLibrary.bib}
\bibliography{main_arxiv}

\appendix

\section{APOGEE Results}
\label{apogeetable}

Table~\ref{tbl:apogeeresults} shows the results of our forward-modeling pipeline applied to our APOGEE sources (see Section~\ref{sec:apogee}).

\begin{longrotatetable}
\begin{deluxetable*}{ccccccccccc}
\tablecaption{APOGEE Forward-Modeling Results \label{tbl:apogeeresults}}
\tabletypesize{\scriptsize}
%\tabletypesize{\footnotetsize}
\tablecolumns{11}
%\tablenum{5}
%\tablewidth{0pt}
\tablehead{
\colhead{APOGEE ID} & \colhead{ID\tablenotemark{a}} & \colhead{$\alpha_{J2000}$} & \colhead{$\delta_{J2000}$} & \colhead{$\overline{RV}$} & \colhead{$\mu_{\alpha\cos\delta}$} & \colhead{$\mu_\delta$} &
\colhead{$T_\mathrm{eff}$\tablenotemark{b}} & \colhead{$v\sin{i}$} & \colhead{Veiling\tablenotemark{c}} & \colhead{Note\tablenotemark{d}} \\
\colhead{} & \colhead{} & \colhead{(deg)} &
\colhead{(deg)} & \colhead{(km s$^{-1}$)} & \colhead{(mas yr$^{-1}$)} & 
\colhead{(mas yr$^{-1}$)} & \colhead{(K)} &  \colhead{(km s$^{-1}$)} & \colhead{$\left(\dfrac{F_{H, \mathrm{cont}}}{F_{H, \ast}}\right)$} & \colhead{}
}
\startdata
2M05350101-0524103 & 663 & $83.75420833$ & $-5.40286111$ & $31.46 \pm 0.19$ & $0.16 \pm 0.07$ & $-0.62 \pm 0.14$ & $3825.01 \pm 12.53$ & $38.23 \pm 0.92$ & 0.56 & C \\
2M05350117-0524067 & \nodata & $83.75487500$ & $-5.40186111$ & $25.59 \pm 0.18$ & \nodata & \nodata & $3979 \pm 6$ & $5.40 \pm 1.00$ & 0.51 &  \\
2M05350160-0524101 & 655 & $83.75666667$ & $-5.40280556$ & $27.82 \pm 0.24$ & $1.21 \pm 0.29$ & $0.25 \pm 0.27$ & $2908.02 \pm 9.21$ & $35.18 \pm 0.88$ & 0.36 &  \\
2M05350284-0522082 & 140 & $83.76183333$ & $-5.36894444$ & $25.49 \pm 0.14$ & $-0.96 \pm 0.15$ & $-1.85 \pm 0.31$ & $5100.45 \pm 1.56$ & $27.43 \pm 0.36$ & 0.58 &  \\
2M05350309-0522378 & 190 & $83.76287500$ & $-5.37716667$ & $29.98 \pm 1.67$ & $-0.20 \pm 0.38$ & $1.01 \pm 0.05$ & $3057.64 \pm 30.73$ & $15.65 \pm 0.39$ & 0.39 &  \\
2M05350370-0522457 & 189 & $83.76541667$ & $-5.37936111$ & $31.37 \pm 0.57$ & $-0.13 \pm 0.48$ & $-0.16 \pm 0.19$ & $3138.91 \pm 23.35$ & $12.69 \pm 0.50$ & 0.50 &  \\
2M05350437-0523138 & 11 & $83.76820833$ & $-5.38716667$ & $24.65 \pm 1.29$ & $0.42 \pm 0.16$ & $0.95 \pm 0.05$ & $4107.19 \pm 63.65$ & $17.67 \pm 1.21$ & 1.04 &  \\
2M05350450-0523565 & \nodata & $83.76875000$ & $-5.39902778$ & $33.82 \pm 0.92$ & \nodata & \nodata & $4591 \pm 7$ & $27.77 \pm 1.63$ & 0.85 &  \\
2M05350461-0524424 & 657 & $83.76920833$ & $-5.41177778$ & $26.44 \pm 0.44$ & $0.12 \pm 0.10$ & $-0.86 \pm 0.08$ & $3988.02 \pm 9.04$ & $16.79 \pm 0.40$ & 0.42 &  \\
2M05350481-0522387 & 67 & $83.77004167$ & $-5.37741667$ & $26.95 \pm 1.16$ & $0.63 \pm 0.01$ & $1.26 \pm 0.11$ & $3544.42 \pm 5.64$ & $10.52 \pm 2.27$ & 0.67 &  \\
2M05350487-0520574 & 542 & $83.77029167$ & $-5.34927778$ & $29.34 \pm 0.57$ & $0.25 \pm 0.15$ & $-0.23 \pm 0.25$ & $3733.67 \pm 7.92$ & $11.91 \pm 0.53$ & 0.62 &  \\
2M05350495-0521092 & 3 & $83.77062500$ & $-5.35255556$ & $29.75 \pm 0.41$ & $-0.43 \pm 0.06$ & $2.12 \pm 0.18$ & $4119.19 \pm 13.19$ & $13.71 \pm 1.44$ & 0.59 &  \\
2M05350513-0520244 & 564 & $83.77137500$ & $-5.34011111$ & $26.75 \pm 0.28$ & $-0.38 \pm 0.46$ & $-0.42 \pm 0.18$ & $4126.16 \pm 10.04$ & $9.15 \pm 0.44$ & 0.53 &  \\
2M05350537-0524105 & 77 & $83.77237500$ & $-5.40291667$ & $26.33 \pm 0.87$ & $0.14 \pm 0.36$ & $-0.19 \pm 0.08$ & $4249.94 \pm 50.38$ & $16.20 \pm 0.36$ & 0.95 &  \\
2M05350540-0524150 & 176 & $83.77250000$ & $-5.40416667$ & $24.40 \pm 0.28$ & $0.95 \pm 0.25$ & $1.04 \pm 0.23$ & $3686.05 \pm 23.07$ & $4.84 \pm 1.85$ & 1.51 &  \\
2M05350563-0525195 & \nodata & $83.77345833$ & $-5.42208333$ & $28.83 \pm 0.52$ & \nodata & \nodata & $4582 \pm 10$ & $42.24 \pm 0.81$ & 1.05 &  \\
2M05350568-0525058 & 656 & $83.77366667$ & $-5.41827778$ & $27.17 \pm 0.29$ & $-1.38 \pm 0.27$ & $-0.16 \pm 0.11$ & $4205.50 \pm 8.81$ & $99.00 \pm 0.87$ & 0.16 &  \\
2M05350571-0523540 & 39 & $83.77379167$ & $-5.39833333$ & $23.77 \pm 0.22$ & $-1.04 \pm 0.47$ & $-0.84 \pm 0.05$ & $3029.62 \pm 6.43$ & $21.64 \pm 0.16$ & 0.42 &  \\
2M05350617-0522124 & 264 & $83.77570833$ & $-5.37011111$ & $29.26 \pm 0.24$ & $-0.85 \pm 0.39$ & $-0.75 \pm 0.16$ & $3828.14 \pm 7.31$ & $9.69 \pm 1.16$ & 0.51 & C \\
2M05350627-0522027 & 129 & $83.77612500$ & $-5.36741667$ & $27.96 \pm 0.23$ & $-0.32 \pm 0.24$ & $0.11 \pm 0.34$ & $4450.92 \pm 6.15$ & $27.39 \pm 0.27$ & 0.24 &  \\
2M05350651-0524414 & \nodata & $83.77712500$ & $-5.41150000$ & $25.72 \pm 0.22$ & \nodata & \nodata & $4269 \pm 7$ & $21.56 \pm 0.07$ & 0.24 &  \\
2M05350727-0522266 & 131 & $83.78029167$ & $-5.37405556$ & $25.99 \pm 0.27$ & $0.24 \pm 0.09$ & $-0.66 \pm 0.78$ & $3163.11 \pm 37.48$ & $13.13 \pm 0.47$ & 0.39 &  \\
2M05350739-0525481 & 585 & $83.78079167$ & $-5.43002778$ & $25.80 \pm 0.30$ & $-0.16 \pm 0.06$ & $0.27 \pm 0.05$ & $2814.95 \pm 17.31$ & $9.62 \pm 0.98$ & 0.41 &  \\
2M05350773-0521014 & 629 & $83.78220833$ & $-5.35038889$ & $28.70 \pm 0.30$ & $-0.56 \pm 0.16$ & $-1.30 \pm 0.13$ & $4326.53 \pm 15.17$ & $24.86 \pm 0.97$ & 1.14 &  \\
2M05350822-0524032 & 679 & $83.78425000$ & $-5.40088889$ & $26.43 \pm 0.26$ & $1.14 \pm 0.13$ & $1.47 \pm 0.22$ & $3168.52 \pm 7.79$ & $6.48 \pm 1.30$ & 0.76 &  \\
2M05350829-0524348 & 17 & $83.78454167$ & $-5.40966667$ & $25.05 \pm 0.29$ & $-0.57 \pm 0.46$ & $0.12 \pm 0.01$ & $3064.95 \pm 8.25$ & $7.43 \pm 0.69$ & 0.51 &  \\
2M05350853-0524411 & 171 & $83.78554167$ & $-5.41141667$ & $31.54 \pm 1.53$ & $-1.82 \pm 0.73$ & $-0.07 \pm 0.18$ & $3048.41 \pm 28.89$ & $10.95 \pm 0.62$ & 0.52 & C \\
2M05350853-0525179 & \nodata & $83.78554167$ & $-5.42163889$ & $23.35 \pm 0.19$ & \nodata & \nodata & $4016 \pm 8$ & $5.69 \pm 0.50$ & 0.51 &  \\
2M05350859-0526194 & \nodata & $83.78579167$ & $-5.43872222$ & $26.43 \pm 0.43$ & \nodata & \nodata & $2977 \pm 11$ & $14.62 \pm 0.66$ & 0.29 &  \\
2M05350873-0522566 & 50 & $83.78637500$ & $-5.38238889$ & $26.21 \pm 0.26$ & $1.05 \pm 0.42$ & $-1.26 \pm 0.26$ & $4198.10 \pm 10.63$ & $71.75 \pm 0.89$ & 0.32 &  \\
2M05350976-0521282 & 58 & $83.79066667$ & $-5.35783333$ & $27.32 \pm 0.40$ & $-1.15 \pm 0.21$ & $-2.57 \pm 0.29$ & $4372.17 \pm 6.81$ & $27.20 \pm 0.14$ & 0.16 &  \\
2M05350992-0521433 & 359 & $83.79133333$ & $-5.36202778$ & $28.57 \pm 0.71$ & $-1.17 \pm 0.06$ & $-0.23 \pm 0.48$ & $4131.49 \pm 32.85$ & $14.92 \pm 1.73$ & 1.03 &  \\
2M05351014-0522326 & 46 & $83.79225000$ & $-5.37572222$ & $28.82 \pm 0.27$ & $-1.02 \pm 0.20$ & $0.73 \pm 0.26$ & $4021.55 \pm 10.76$ & $24.02 \pm 1.33$ & 0.64 &  \\
2M05351021-0523215 & 169 & $83.79254167$ & $-5.38930556$ & $25.51 \pm 0.34$ & $1.31 \pm 0.63$ & $-0.86 \pm 0.43$ & $4329.25 \pm 7.55$ & $15.92 \pm 0.45$ & 0.98 &  \\
2M05351029-0519563 & 550 & $83.79287500$ & $-5.33230556$ & $26.54 \pm 0.32$ & $-1.68 \pm 0.11$ & $1.67 \pm 0.10$ & $3449.75 \pm 9.32$ & $24.33 \pm 0.54$ & 0.54 &  \\
2M05351031-0521130 & 519 & $83.79295833$ & $-5.35361111$ & $27.14 \pm 0.15$ & $1.14 \pm 0.07$ & $-0.36 \pm 0.12$ & $4165.05 \pm 7.13$ & $8.73 \pm 1.04$ & 0.69 &  \\
2M05351041-0519523 & 552 & $83.79337500$ & $-5.33119444$ & $27.39 \pm 0.55$ & $0.82 \pm 0.12$ & $-1.53 \pm 0.17$ & $3445.06 \pm 43.97$ & $19.36 \pm 0.24$ & 0.54 &  \\
2M05351047-0526003 & 595 & $83.79362500$ & $-5.43341667$ & $27.74 \pm 2.66$ & $-0.59 \pm 0.35$ & $-0.20 \pm 0.52$ & $2722.45 \pm 220.03$ & $19.91 \pm 1.37$ & 0.04 &  \\
2M05351050-0522455 & 6 & $83.79375000$ & $-5.37930556$ & $24.86 \pm 0.29$ & $0.14 \pm 0.46$ & $-1.53 \pm 0.06$ & $4125.77 \pm 7.78$ & $44.03 \pm 0.13$ & 0.36 &  \\
2M05351050-0526183 & \nodata & $83.79375000$ & $-5.43841667$ & $22.61 \pm 0.26$ & \nodata & \nodata & $4504 \pm 8$ & $23.52 \pm 0.23$ & 0.50 &  \\
2M05351053-0522166 & 262 & $83.79387500$ & $-5.37127778$ & $27.55 \pm 0.28$ & $-0.91 \pm 0.25$ & $0.89 \pm 0.10$ & $3728.33 \pm 8.85$ & $23.61 \pm 0.36$ & 0.66 &  \\
2M05351061-0522559 & 251 & $83.79420833$ & $-5.38219444$ & $26.01 \pm 1.01$ & $-0.37 \pm 0.15$ & $-0.24 \pm 0.10$ & $4244.42 \pm 76.70$ & $22.11 \pm 1.01$ & 0.89 &  \\
2M05351073-0526280 & \nodata & $83.79470833$ & $-5.44111111$ & $28.03 \pm 0.26$ & \nodata & \nodata & $3804 \pm 9$ & $18.25 \pm 0.47$ & 0.64 &  \\
2M05351083-0525569 & 594 & $83.79512500$ & $-5.43247222$ & $27.93 \pm 0.96$ & $0.58 \pm 0.30$ & $2.74 \pm 0.45$ & $2924.92 \pm 41.45$ & $15.96 \pm 1.17$ & 0.56 &  \\
2M05351094-0524486 & 489 & $83.79558333$ & $-5.41350000$ & $24.98 \pm 1.48$ & $2.04 \pm 0.72$ & $1.94 \pm 0.31$ & $3716.95 \pm 51.72$ & $28.31 \pm 0.28$ & 0.90 & C \\
2M05351149-0526023 & \nodata & $83.79787500$ & $-5.43397222$ & $19.04 \pm 0.18$ & \nodata & \nodata & $4370 \pm 6$ & $42.42 \pm 1.60$ & 0.16 &  \\
2M05351163-0522515 & 75 & $83.79845833$ & $-5.38097222$ & $25.36 \pm 0.24$ & $-0.04 \pm 0.79$ & $-0.20 \pm 0.24$ & $3054.10 \pm 16.36$ & $11.06 \pm 1.01$ & 0.99 &  \\
2M05351165-0524213 & 191 & $83.79854167$ & $-5.40591667$ & $25.25 \pm 0.14$ & $0.34 \pm 0.13$ & $-0.52 \pm 0.14$ & $3029.60 \pm 7.19$ & $12.41 \pm 0.51$ & 0.86 &  \\
2M05351178-0521555 & 150 & $83.79908333$ & $-5.36541667$ & $26.84 \pm 1.28$ & $-0.03 \pm 0.66$ & $-0.30 \pm 0.94$ & $2929.69 \pm 8.11$ & $12.07 \pm 1.17$ & 0.51 &  \\
2M05351188-0521032 & 38 & $83.79950000$ & $-5.35088889$ & $27.48 \pm 0.24$ & $-0.25 \pm 0.15$ & $0.33 \pm 0.52$ & $4292.95 \pm 5.88$ & $10.65 \pm 0.48$ & 0.74 & C \\
2M05351197-0522541 & 248 & $83.79987500$ & $-5.38169444$ & $93.82 \pm 39.46$ & $0.82 \pm 0.57$ & $-0.00 \pm 0.53$ & $3781.89 \pm 82.67$ & $255.27 \pm 23.13$ & 7.21 &  \\
2M05351227-0520452 & 526 & $83.80112500$ & $-5.34588889$ & $28.05 \pm 1.13$ & $1.26 \pm 0.34$ & $0.46 \pm 0.36$ & $3351.49 \pm 12.02$ & $7.51 \pm 0.52$ & 0.78 &  \\
2M05351227-0523479 & \nodata & $83.80112500$ & $-5.39663889$ & $27.37 \pm 0.23$ & \nodata & \nodata & $4216 \pm 6$ & $12.34 \pm 0.31$ & 0.37 &  \\
2M05351259-0523440 & \nodata & $83.80245833$ & $-5.39555556$ & $27.42 \pm 1.27$ & \nodata & \nodata & $4445 \pm 8$ & $41.44 \pm 0.64$ & 0.45 &  \\
2M05351269-0519353 & \nodata & $83.80287500$ & $-5.32647222$ & $30.15 \pm 0.19$ & \nodata & \nodata & $3713 \pm 13$ & $7.39 \pm 0.87$ & 0.58 &  \\
2M05351277-0520349 & 56 & $83.80320833$ & $-5.34302778$ & $21.73 \pm 0.65$ & $0.71 \pm 0.15$ & $0.71 \pm 0.20$ & $3054.25 \pm 12.07$ & $13.47 \pm 0.89$ & 1.08 &  \\
2M05351304-0520302 & 554 & $83.80433333$ & $-5.34172222$ & $30.84 \pm 0.43$ & $-1.30 \pm 0.19$ & $1.12 \pm 0.31$ & $4249.74 \pm 6.80$ & $19.38 \pm 0.23$ & 0.31 &  \\
2M05351305-0521532 & 261 & $83.80437500$ & $-5.36477778$ & $29.92 \pm 0.24$ & $-1.65 \pm 0.05$ & $-1.42 \pm 0.04$ & $3825.16 \pm 14.48$ & $9.99 \pm 1.21$ & 1.40 &  \\
2M05351319-0524554 & \nodata & $83.80495833$ & $-5.41538889$ & $25.21 \pm 0.33$ & \nodata & \nodata & $4296 \pm 9$ & $26.69 \pm 0.38$ & 0.95 &  \\
2M05351330-0520189 & 109 & $83.80541667$ & $-5.33858333$ & $29.60 \pm 0.21$ & $-0.02 \pm 0.26$ & $1.56 \pm 2.06$ & $4048.63 \pm 5.49$ & $42.23 \pm 0.23$ & 0.44 &  \\
2M05351336-0522261 & 53 & $83.80566667$ & $-5.37391667$ & $25.87 \pm 1.45$ & $2.45 \pm 0.36$ & $1.29 \pm 0.33$ & $3631.97 \pm 60.91$ & $13.81 \pm 1.69$ & 1.30 &  \\
2M05351343-0521073 & 377 & $83.80595833$ & $-5.35202778$ & $30.99 \pm 0.22$ & $0.88 \pm 0.05$ & $-0.94 \pm 0.07$ & $3814.00 \pm 12.61$ & $3.85 \pm 1.61$ & 1.27 &  \\
2M05351352-0522195 & 146 & $83.80633333$ & $-5.37208333$ & $31.09 \pm 0.54$ & $-0.70 \pm 0.14$ & $-1.02 \pm 0.32$ & $4058.32 \pm 17.56$ & $16.49 \pm 0.39$ & 0.85 &  \\
2M05351362-0519548 & \nodata & $83.80675000$ & $-5.33188889$ & $30.67 \pm 1.39$ & \nodata & \nodata & $4405 \pm 30$ & $23.53 \pm 0.87$ & 0.41 &  \\
2M05351379-0519254 & \nodata & $83.80745833$ & $-5.32372222$ & $23.99 \pm 0.24$ & \nodata & \nodata & $2838 \pm 7$ & $15.29 \pm 0.92$ & 0.37 &  \\
2M05351397-0521233 & 117 & $83.80820833$ & $-5.35647222$ & $24.02 \pm 0.21$ & $1.21 \pm 0.49$ & $-0.58 \pm 0.23$ & $4053.25 \pm 13.71$ & $9.09 \pm 0.41$ & 0.95 &  \\
2M05351405-0519520 & \nodata & $83.80854167$ & $-5.33111111$ & $23.78 \pm 0.15$ & \nodata & \nodata & $4267 \pm 7$ & $12.32 \pm 0.28$ & 0.68 &  \\
2M05351405-0523383 & 507 & $83.80854167$ & $-5.39397222$ & $24.91 \pm 0.33$ & $0.07 \pm 0.12$ & $0.44 \pm 0.01$ & $3921.76 \pm 8.73$ & $60.27 \pm 0.44$ & 0.35 &  \\
2M05351421-0520042 & \nodata & $83.80920833$ & $-5.33450000$ & $26.68 \pm 0.30$ & \nodata & \nodata & $4091 \pm 12$ & $12.38 \pm 0.40$ & 0.85 &  \\
2M05351427-0524246 & \nodata & $83.80945833$ & $-5.40683333$ & $27.94 \pm 0.80$ & \nodata & \nodata & $4285 \pm 20$ & $50.89 \pm 0.46$ & 0.41 &  \\
2M05351439-0523335 & 521 & $83.80995833$ & $-5.39263889$ & $26.01 \pm 0.79$ & $-0.11 \pm 0.23$ & $-1.65 \pm 0.03$ & $4130.95 \pm 25.34$ & $27.82 \pm 0.55$ & 0.66 &  \\
2M05351465-0523018 & 236 & $83.81104167$ & $-5.38383333$ & $30.78 \pm 0.31$ & $1.17 \pm 0.69$ & $-0.94 \pm 0.35$ & $4038.46 \pm 12.69$ & $14.62 \pm 0.46$ & 0.78 &  \\
2M05351471-0521063 & 319 & $83.81129167$ & $-5.35175000$ & $29.72 \pm 0.85$ & $-0.58 \pm 0.02$ & $-0.82 \pm 0.05$ & $4354.13 \pm 28.40$ & $17.89 \pm 0.69$ & 0.43 &  \\
2M05351498-0521598 & 567 & $83.81241667$ & $-5.36661111$ & $29.14 \pm 0.38$ & $2.25 \pm 0.09$ & $0.96 \pm 0.11$ & $4417.59 \pm 8.65$ & $21.56 \pm 0.06$ & 0.25 & V \\
2M05351554-0525140 & \nodata & $83.81475000$ & $-5.42055556$ & $36.71 \pm 1.67$ & \nodata & \nodata & $4277 \pm 9$ & $95.71 \pm 0.60$ & 0.21 & \\
2M05351561-0524030 & 37 & $83.81504167$ & $-5.40083333$ & $29.62 \pm 1.23$ & $0.54 \pm 0.11$ & $-1.80 \pm 0.02$ & $4389.08 \pm 10.23$ & $60.37 \pm 0.94$ & 0.19 &  \\
2M05351563-0522565 & \nodata & $83.81512500$ & $-5.38236111$ & $30.29 \pm 1.22$ & \nodata & \nodata & $4468 \pm 6$ & $28.65 \pm 0.32$ & 0.54 &  \\
2M05351567-0525331 & 616 & $83.81529167$ & $-5.42586111$ & $25.57 \pm 0.48$ & $0.09 \pm 0.20$ & $-0.74 \pm 0.11$ & $3951.44 \pm 9.34$ & $22.09 \pm 0.74$ & 1.38 &  \\
2M05351571-0526283 & \nodata & $83.81545833$ & $-5.44119444$ & $25.79 \pm 1.34$ & \nodata & \nodata & $3875 \pm 37$ & $8.89 \pm 1.08$ & 0.67 &  \\
2M05351587-0522328 & 241 & $83.81612500$ & $-5.37577778$ & $30.63 \pm 1.45$ & $0.24 \pm 0.15$ & $1.21 \pm 0.17$ & $3022.39 \pm 39.10$ & $8.83 \pm 1.63$ & 1.35 &  \\
2M05351606-0520363 & 517 & $83.81691667$ & $-5.34341667$ & $23.56 \pm 0.22$ & $0.55 \pm 0.25$ & $1.44 \pm 0.24$ & $4163.85 \pm 8.74$ & $16.04 \pm 1.49$ & 0.83 &  \\
2M05351609-0524112 & 173 & $83.81704167$ & $-5.40311111$ & $23.01 \pm 0.24$ & $-1.73 \pm 0.13$ & $-0.79 \pm 0.12$ & $3463.66 \pm 13.22$ & $33.20 \pm 0.69$ & 0.90 &  \\
2M05351661-0519357 & \nodata & $83.81920833$ & $-5.32658333$ & $28.54 \pm 0.36$ & \nodata & \nodata & $4098 \pm 11$ & $19.85 \pm 0.33$ & 0.94 &  \\
2M05351671-0522309 & 175 & $83.81962500$ & $-5.37525000$ & $29.02 \pm 0.28$ & $0.65 \pm 0.11$ & $1.33 \pm 0.18$ & $3659.32 \pm 26.94$ & $11.82 \pm 1.24$ & 1.99 &  \\
2M05351694-0525469 & 451 & $83.82058333$ & $-5.42969444$ & $29.40 \pm 0.37$ & $0.31 \pm 0.04$ & $-0.91 \pm 0.06$ & $3660.92 \pm 13.75$ & $54.77 \pm 1.18$ & 0.95 &  \\
2M05351697-0521452 & \nodata & $83.82070833$ & $-5.36255556$ & $47.39 \pm 3.56$ & \nodata & \nodata & $6998 \pm 2$ & $299.48 \pm 0.88$ & 0.10 &  \\
2M05351697-0523009 & 560 & $83.82070833$ & $-5.38358333$ & $33.28 \pm 0.30$ & $0.78 \pm 0.13$ & $1.15 \pm 0.04$ & $4228.79 \pm 6.23$ & $29.65 \pm 0.73$ & 0.59 &  \\
2M05351712-0524585 & 331 & $83.82133333$ & $-5.41625000$ & $26.60 \pm 0.25$ & $-0.80 \pm 0.29$ & $-0.85 \pm 0.07$ & $3023.52 \pm 7.83$ & $15.66 \pm 0.29$ & 0.39 &  \\
2M05351722-0520277 & 529 & $83.82175000$ & $-5.34102778$ & $24.01 \pm 1.76$ & $-0.22 \pm 0.09$ & $-0.27 \pm 0.12$ & $3812.67 \pm 11.91$ & $13.15 \pm 1.46$ & 0.82 &  \\
2M05351734-0522357 & \nodata & $83.82225000$ & $-5.37658333$ & $28.46 \pm 2.12$ & \nodata & \nodata & $4504 \pm 71$ & $32.52 \pm 2.03$ & 1.00 &  \\
2M05351736-0520149 & 505 & $83.82233333$ & $-5.33747222$ & $27.15 \pm 0.44$ & $-0.71 \pm 0.16$ & $-0.31 \pm 0.12$ & $4078.60 \pm 11.35$ & $7.13 \pm 0.85$ & 0.64 &  \\
2M05351736-0525446 & 619 & $83.82233333$ & $-5.42905556$ & $27.12 \pm 0.40$ & $0.93 \pm 0.10$ & $0.24 \pm 0.10$ & $4098.05 \pm 10.50$ & $8.37 \pm 1.64$ & 0.47 & C \\
2M05351754-0525427 & 618 & $83.82308333$ & $-5.42852778$ & $24.34 \pm 0.25$ & $-1.00 \pm 0.15$ & $-0.39 \pm 0.17$ & $4185.24 \pm 10.90$ & $23.47 \pm 0.32$ & 0.57 &  \\
2M05351778-0523155 & 183 & $83.82408333$ & $-5.38763889$ & $25.69 \pm 0.55$ & $-1.61 \pm 0.17$ & $0.60 \pm 0.38$ & $4098.62 \pm 13.74$ & $18.31 \pm 0.61$ & 0.53 & C \\
2M05351794-0525061 & 440 & $83.82475000$ & $-5.41836111$ & $31.16 \pm 19.24$ & $-0.19 \pm 0.28$ & $-1.11 \pm 0.43$ & $2739.27 \pm 172.56$ & $17.45 \pm 117.44$ & 0.87 &  \\
2M05351794-0525338 & 453 & $83.82475000$ & $-5.42605556$ & $26.73 \pm 0.50$ & $0.50 \pm 0.04$ & $1.11 \pm 2.80$ & $3638.90 \pm 8.83$ & $27.37 \pm 0.38$ & 0.73 &  \\
2M05351797-0526506 & \nodata & $83.82487500$ & $-5.44738889$ & $25.16 \pm 0.32$ & \nodata & \nodata & $2961 \pm 14$ & $7.41 \pm 0.63$ & 0.51 &  \\
2M05351820-0524302 & 112 & $83.82583333$ & $-5.40838889$ & $25.21 \pm 1.62$ & $0.79 \pm 0.19$ & $2.27 \pm 0.12$ & $2858.10 \pm 13.45$ & $39.54 \pm 0.91$ & 0.75 &  \\
2M05351836-0524267 & 88 & $83.82650000$ & $-5.40741667$ & $25.50 \pm 0.26$ & $-1.00 \pm 0.31$ & $0.31 \pm 0.10$ & $3120.85 \pm 10.14$ & $2.44 \pm 1.38$ & 0.90 &  \\
2M05351846-0524068 & 506 & $83.82691667$ & $-5.40188889$ & $27.13 \pm 0.21$ & $-0.17 \pm 0.42$ & $0.43 \pm 0.17$ & $3849.71 \pm 9.16$ & $19.76 \pm 0.22$ & 0.84 &  \\
2M05351851-0520427 & 119 & $83.82712500$ & $-5.34519444$ & $31.06 \pm 1.62$ & $-1.70 \pm 0.12$ & $0.60 \pm 0.22$ & $3685.33 \pm 15.17$ & $50.66 \pm 0.60$ & 0.73 &  \\
2M05351858-0526248 & \nodata & $83.82741667$ & $-5.44022222$ & $24.92 \pm 7.93$ & \nodata & \nodata & $3832 \pm 102$ & $3.11 \pm 28.24$ & 1.36 &  \\
2M05351866-0523139 & \nodata & $83.82775000$ & $-5.38719444$ & $31.91 \pm 0.28$ & \nodata & \nodata & $5103 \pm 5$ & $39.12 \pm 1.17$ & 1.21 &  \\
2M05351884-0522229 & 20 & $83.82850000$ & $-5.37302778$ & $23.73 \pm 0.91$ & $0.14 \pm 0.20$ & $-0.04 \pm 0.15$ & $3721.33 \pm 235.56$ & $15.83 \pm 7.06$ & 1.58 &  \\
2M05351894-0520522 & 549 & $83.82891667$ & $-5.34783333$ & $24.41 \pm 0.15$ & $1.72 \pm 0.29$ & $-0.99 \pm 0.15$ & $4035.02 \pm 13.79$ & $11.85 \pm 1.22$ & 2.06 &  \\
2M05351906-0523495 & 148 & $83.82941667$ & $-5.39708333$ & $45.52 \pm 1.25$ & $1.86 \pm 0.36$ & $-1.37 \pm 1.24$ & $4429.27 \pm 17.62$ & $53.25 \pm 3.39$ & 0.83 & E3 \\
2M05351913-0520387 & \nodata & $83.82970833$ & $-5.34408333$ & $37.94 \pm 8.57$ & \nodata & \nodata & $6996 \pm 4$ & $299.17 \pm 0.99$ & 0.47 &  \\
2M05351930-0520078 & \nodata & $83.83041667$ & $-5.33550000$ & $32.90 \pm 1.78$ & \nodata & \nodata & $6273 \pm 91$ & $117.19 \pm 10.98$ & 0.75 &  \\
2M05351965-0524266 & 573 & $83.83187500$ & $-5.40738889$ & $23.19 \pm 0.20$ & $0.33 \pm 0.34$ & $-0.29 \pm 0.09$ & $3946.35 \pm 9.24$ & $18.18 \pm 0.34$ & 1.10 &  \\
2M05352004-0521059 & \nodata & $83.83350000$ & $-5.35163889$ & $37.13 \pm 0.21$ & \nodata & \nodata & $4437 \pm 7$ & $48.90 \pm 0.42$ & 0.27 &  \\
2M05352004-0525375 & \nodata & $83.83350000$ & $-5.42708333$ & $26.74 \pm 0.66$ & \nodata & \nodata & $4135 \pm 33$ & $11.75 \pm 0.79$ & 0.56 &  \\
2M05352016-0526390 & \nodata & $83.83400000$ & $-5.44416667$ & $29.18 \pm 1.03$ & \nodata & \nodata & $4422 \pm 8$ & $53.50 \pm 1.09$ & 0.15 &  \\
2M05352017-0523085 & 441 & $83.83404167$ & $-5.38569444$ & $29.24 \pm 1.64$ & $-2.14 \pm 0.69$ & $-0.87 \pm 0.29$ & $3867.80 \pm 12.47$ & $40.21 \pm 1.46$ & 0.95 &  \\
2M05352054-0524208 & 104 & $83.83558333$ & $-5.40577778$ & $27.47 \pm 0.21$ & $-0.58 \pm 0.20$ & $-0.92 \pm 0.05$ & $3572.29 \pm 50.85$ & $7.87 \pm 1.48$ & 1.64 &  \\
2M05352056-0520431 & 192 & $83.83566667$ & $-5.34530556$ & $29.39 \pm 1.53$ & $-0.29 \pm 0.05$ & $0.28 \pm 0.82$ & $3498.69 \pm 252.68$ & $17.07 \pm 1.19$ & 0.55 &  \\
2M05352067-0523531 & 72 & $83.83612500$ & $-5.39808333$ & $25.70 \pm 0.20$ & $0.09 \pm 0.12$ & $0.31 \pm 0.05$ & $2978.36 \pm 7.07$ & $10.68 \pm 1.18$ & 0.30 &  \\
2M05352076-0521550 & \nodata & $83.83650000$ & $-5.36527778$ & $25.24 \pm 1.53$ & \nodata & \nodata & $4288 \pm 14$ & $10.22 \pm 0.71$ & 0.58 &  \\
2M05352082-0521216 & 34 & $83.83675000$ & $-5.35600000$ & $32.46 \pm 0.26$ & $0.70 \pm 2.17$ & $0.83 \pm 0.91$ & $4020.81 \pm 16.31$ & $35.50 \pm 1.37$ & 1.42 &  \\
2M05352103-0522250 & 86 & $83.83762500$ & $-5.37361111$ & $27.16 \pm 0.18$ & $-1.02 \pm 0.09$ & $1.41 \pm 0.72$ & $3725.53 \pm 8.10$ & $16.62 \pm 0.72$ & 0.83 &  \\
2M05352104-0523490 & \nodata & $83.83766667$ & $-5.39694444$ & $31.25 \pm 0.23$ & \nodata & \nodata & $4503 \pm 10$ & $54.80 \pm 0.31$ & 0.12 & V \\
2M05352115-0525569 & 612 & $83.83812500$ & $-5.43247222$ & $22.19 \pm 0.23$ & $1.04 \pm 0.26$ & $-0.11 \pm 0.16$ & $4107.03 \pm 7.56$ & $29.68 \pm 0.54$ & 0.45 &  \\
2M05352139-0526439 & \nodata & $83.83912500$ & $-5.44552778$ & $26.03 \pm 0.21$ & \nodata & \nodata & $4386 \pm 7$ & $19.74 \pm 0.17$ & 0.13 &  \\
2M05352162-0526576 & \nodata & $83.84008333$ & $-5.44933333$ & $30.06 \pm 1.01$ & \nodata & \nodata & $3061 \pm 9$ & $72.25 \pm 0.95$ & 0.33 &  \\
2M05352166-0525264 & \nodata & $83.84025000$ & $-5.42400000$ & $25.74 \pm 0.18$ & \nodata & \nodata & $4209 \pm 12$ & $12.44 \pm 1.36$ & 0.59 &  \\
2M05352172-0526443 & 617 & $83.84050000$ & $-5.44563889$ & $26.49 \pm 0.63$ & $1.05 \pm 0.06$ & $1.38 \pm 0.03$ & $3749.12 \pm 8.63$ & $16.95 \pm 0.86$ & 0.61 &  \\
2M05352181-0523539 & 139 & $83.84087500$ & $-5.39830556$ & $25.79 \pm 1.28$ & $0.98 \pm 0.09$ & $0.84 \pm 0.21$ & $4535.92 \pm 8.94$ & $48.97 \pm 0.50$ & 0.14 &  \\
2M05352184-0522082 & 89 & $83.84100000$ & $-5.36894444$ & $26.71 \pm 0.27$ & $0.15 \pm 0.29$ & $1.13 \pm 0.13$ & $4067.68 \pm 11.01$ & $23.87 \pm 0.50$ & 0.44 &  \\
2M05352208-0524327 & 145 & $83.84200000$ & $-5.40908333$ & $23.37 \pm 0.20$ & $1.36 \pm 0.47$ & $-1.32 \pm 0.33$ & $3257.96 \pm 8.18$ & $12.89 \pm 0.68$ & 0.71 &  \\
2M05352219-0526373 & \nodata & $83.84245833$ & $-5.44369444$ & $29.56 \pm 0.31$ & \nodata & \nodata & $4246 \pm 7$ & $34.67 \pm 0.10$ & 0.13 &  \\
2M05352226-0520292 & \nodata & $83.84275000$ & $-5.34144444$ & $23.37 \pm 0.34$ & \nodata & \nodata & $4526 \pm 6$ & $13.41 \pm 0.83$ & 0.08 &  \\
2M05352244-0522010 & 558 & $83.84350000$ & $-5.36694444$ & $22.86 \pm 0.29$ & $0.03 \pm 0.09$ & $-1.20 \pm 0.11$ & $4467.55 \pm 7.58$ & $19.88 \pm 1.08$ & 0.36 &  \\
2M05352246-0525451 & 652 & $83.84358333$ & $-5.42919444$ & $30.53 \pm 0.41$ & $-2.23 \pm 0.31$ & $-1.26 \pm 0.45$ & $3544.14 \pm 12.54$ & $14.72 \pm 0.88$ & 0.91 &  \\
2M05352274-0519478 & \nodata & $83.84475000$ & $-5.32994444$ & $30.20 \pm 0.22$ & \nodata & \nodata & $4292 \pm 12$ & $31.28 \pm 0.39$ & 0.46 &  \\
2M05352296-0522415 & 551 & $83.84566667$ & $-5.37819444$ & $28.32 \pm 1.53$ & $-0.59 \pm 0.13$ & $2.26 \pm 0.07$ & $4477.92 \pm 8.26$ & $66.41 \pm 0.82$ & 0.07 &  \\
2M05352317-0522283 & 623 & $83.84654167$ & $-5.37452778$ & $25.57 \pm 0.31$ & $-0.28 \pm 0.19$ & $-0.09 \pm 0.21$ & $3933.29 \pm 12.09$ & $3.56 \pm 1.35$ & 1.10 &  \\
2M05352321-0521357 & 535 & $83.84670833$ & $-5.35991667$ & $47.65 \pm 0.56$ & $-1.26 \pm 0.21$ & $-1.92 \pm 0.20$ & $3132.12 \pm 14.43$ & $53.83 \pm 1.35$ & 1.11 & E3 \\
2M05352332-0521254 & 556 & $83.84716667$ & $-5.35705556$ & $29.54 \pm 0.34$ & $1.51 \pm 0.23$ & $-0.34 \pm 0.22$ & $4235.67 \pm 10.56$ & $11.89 \pm 1.41$ & 1.17 &  \\
2M05352349-0520016 & \nodata & $83.84787500$ & $-5.33377778$ & $26.49 \pm 1.78$ & \nodata & \nodata & $4058 \pm 27$ & $53.43 \pm 1.19$ & 0.55 &  \\
2M05352433-0522322 & 98 & $83.85137500$ & $-5.37561111$ & $25.14 \pm 1.61$ & $-1.58 \pm 0.25$ & $0.12 \pm 0.08$ & $4290.35 \pm 48.94$ & $16.00 \pm 0.59$ & 1.39 &  \\
2M05352433-0526003 & 342 & $83.85137500$ & $-5.43341667$ & $26.95 \pm 1.57$ & $-0.65 \pm 0.36$ & $-0.46 \pm 0.28$ & $2822.52 \pm 9.19$ & $31.32 \pm 3.75$ & 0.22 &  \\
2M05352443-0524398 & 508 & $83.85179167$ & $-5.41105556$ & $23.50 \pm 0.28$ & $0.85 \pm 0.07$ & $-1.81 \pm 0.17$ & $4178.56 \pm 9.65$ & $39.41 \pm 0.55$ & 0.84 &  \\
2M05352445-0524010 & 269 & $83.85187500$ & $-5.40027778$ & $31.09 \pm 0.39$ & $-0.97 \pm 0.24$ & $2.31 \pm 0.04$ & $2761.69 \pm 21.16$ & $20.71 \pm 1.05$ & 0.91 &  \\
2M05352445-0526314 & \nodata & $83.85187500$ & $-5.44205556$ & $24.08 \pm 1.20$ & \nodata & \nodata & $3260 \pm 197$ & $19.86 \pm 0.92$ & 0.69 &  \\
2M05352465-0522425 & 194 & $83.85270833$ & $-5.37847222$ & $32.95 \pm 0.96$ & $-0.26 \pm 0.12$ & $-0.13 \pm 0.03$ & $3737.85 \pm 46.02$ & $2.23 \pm 2.43$ & 0.61 &  \\
2M05352469-0524357 & 177 & $83.85287500$ & $-5.40991667$ & $24.77 \pm 0.34$ & $0.18 \pm 0.65$ & $1.11 \pm 0.32$ & $4270.39 \pm 7.66$ & $35.59 \pm 0.49$ & 0.79 &  \\
2M05352488-0525101 & 450 & $83.85366667$ & $-5.41947222$ & $25.83 \pm 1.74$ & $-0.34 \pm 0.25$ & $-0.39 \pm 0.20$ & $2889.50 \pm 65.12$ & $13.39 \pm 0.66$ & 0.24 &  \\
2M05352505-0522585 & \nodata & $83.85437500$ & $-5.38291667$ & $29.88 \pm 0.21$ & \nodata & \nodata & $4431 \pm 7$ & $27.19 \pm 0.08$ & 0.38 &  \\
2M05352508-0523467 & \nodata & $83.85450000$ & $-5.39630556$ & $31.86 \pm 0.29$ & \nodata & \nodata & $4325 \pm 7$ & $109.87 \pm 0.77$ & 0.05 &  \\
2M05352512-0522252 & 504 & $83.85466667$ & $-5.37366667$ & $25.85 \pm 0.97$ & $-0.02 \pm 0.22$ & $-0.32 \pm 0.28$ & $3028.95 \pm 7.75$ & $8.51 \pm 0.82$ & 0.42 &  \\
2M05352517-0523536 & 565 & $83.85487500$ & $-5.39822222$ & $27.14 \pm 1.76$ & $-0.10 \pm 0.21$ & $0.56 \pm 0.15$ & $3980.26 \pm 58.62$ & $15.50 \pm 2.19$ & 1.60 &  \\
2M05352534-0525295 & 325 & $83.85558333$ & $-5.42486111$ & $30.22 \pm 0.24$ & $-1.16 \pm 0.08$ & $-1.66 \pm 0.38$ & $2861.28 \pm 8.23$ & $16.46 \pm 0.49$ & 0.49 &  \\
2M05352537-0524114 & 13 & $83.85570833$ & $-5.40316667$ & $27.17 \pm 1.00$ & $-0.14 \pm 0.69$ & $-0.42 \pm 0.08$ & $4223.46 \pm 17.45$ & $18.93 \pm 2.36$ & 0.94 &  \\
2M05352543-0521515 & \nodata & $83.85595833$ & $-5.36430556$ & $28.40 \pm 0.23$ & \nodata & \nodata & $4188 \pm 9$ & $29.41 \pm 0.72$ & 0.47 &  \\
2M05352547-0521349 & 91 & $83.85612500$ & $-5.35969444$ & $31.67 \pm 0.25$ & $0.30 \pm 0.15$ & $-1.08 \pm 0.17$ & $4379.39 \pm 7.24$ & $16.07 \pm 0.31$ & 0.17 & C \\
2M05352571-0523094 & 111 & $83.85712500$ & $-5.38594444$ & $26.37 \pm 0.31$ & $-0.15 \pm 0.17$ & $0.53 \pm 0.09$ & $4349.06 \pm 5.53$ & $17.91 \pm 0.20$ & 0.43 &  \\
2M05352600-0525477 & 436 & $83.85833333$ & $-5.42991667$ & $26.44 \pm 0.19$ & $-0.02 \pm 0.08$ & $-0.33 \pm 0.06$ & $4151.75 \pm 5.88$ & $16.04 \pm 0.25$ & 0.49 &  \\
2M05352605-0521210 & \nodata & $83.85854167$ & $-5.35583333$ & $31.20 \pm 0.19$ & \nodata & \nodata & $4392 \pm 7$ & $66.77 \pm 1.61$ & 0.17 &  \\
2M05352615-0522570 & 106 & $83.85895833$ & $-5.38250000$ & $26.50 \pm 0.26$ & $0.72 \pm 0.13$ & $1.17 \pm 0.03$ & $3812.07 \pm 8.28$ & $12.48 \pm 0.58$ & 0.68 &  \\
2M05352618-0525203 & 350 & $83.85908333$ & $-5.42230556$ & $28.33 \pm 0.28$ & $0.68 \pm 0.08$ & $0.91 \pm 0.12$ & $3519.43 \pm 47.07$ & $25.72 \pm 0.63$ & 0.85 &  \\
2M05352634-0525401 & \nodata & $83.85975000$ & $-5.42780556$ & $29.74 \pm 1.79$ & \nodata & \nodata & $4616 \pm 10$ & $104.47 \pm 1.27$ & 0.20 &  \\
2M05352696-0524005 & 182 & $83.86233333$ & $-5.40013889$ & $26.47 \pm 0.42$ & $1.23 \pm 0.38$ & $-0.27 \pm 0.02$ & $3345.34 \pm 16.18$ & $10.65 \pm 0.85$ & 1.94 & C \\
2M05352730-0523366 & \nodata & $83.86375000$ & $-5.39350000$ & $29.70 \pm 0.34$ & \nodata & \nodata & $4284 \pm 8$ & $30.96 \pm 0.15$ & 0.46 &  \\
2M05352813-0523064 & 120 & $83.86720833$ & $-5.38511111$ & $26.61 \pm 0.89$ & $0.19 \pm 0.35$ & $1.16 \pm 0.12$ & $3670.84 \pm 9.54$ & $58.01 \pm 1.07$ & 0.54 &  \\
2M05352838-0525033 & \nodata & $83.86825000$ & $-5.41758333$ & $25.70 \pm 0.23$ & \nodata & \nodata & $3910 \pm 11$ & $11.07 \pm 0.38$ & 0.46 &  \\
2M05352958-0524568 & \nodata & $83.87325000$ & $-5.41577778$ & $29.81 \pm 6.89$ & \nodata & \nodata & $3906 \pm 19$ & $14.25 \pm 1.47$ & 0.84 &  \\
2M05352985-0523555 & \nodata & $83.87437500$ & $-5.39875000$ & $26.93 \pm 0.18$ & \nodata & \nodata & $3978 \pm 18$ & $19.56 \pm 0.53$ & 0.59 &  \\
2M05352986-0523484 & \nodata & $83.87441667$ & $-5.39677778$ & $27.24 \pm 0.20$ & \nodata & \nodata & $3873 \pm 12$ & $14.21 \pm 0.38$ & 0.94 &  \\
2M05353048-0524230 & \nodata & $83.87700000$ & $-5.40638889$ & $33.13 \pm 0.36$ & \nodata & \nodata & $4319 \pm 8$ & $83.73 \pm 0.62$ & 0.13 &  \\
2M05353071-0524312 & 654 & $83.87795833$ & $-5.40866667$ & $28.30 \pm 1.35$ & $-2.04 \pm 1.86$ & $1.68 \pm 1.63$ & $2879.93 \pm 20.03$ & $19.93 \pm 0.42$ & 0.25 &  \\
2M05353074-0521466 & 548 & $83.87808333$ & $-5.36294444$ & $24.72 \pm 0.30$ & $0.15 \pm 0.43$ & $1.10 \pm 0.53$ & $4033.18 \pm 43.25$ & $8.04 \pm 1.48$ & 0.58 &  \\
2M05353099-0522013 & 538 & $83.87912500$ & $-5.36702778$ & $34.11 \pm 2.61$ & $-1.19 \pm 0.08$ & $-1.41 \pm 0.36$ & $4273.76 \pm 42.75$ & $80.53 \pm 9.56$ & 0.29 &  \\
2M05353124-0523400 & \nodata & $83.88016667$ & $-5.39444444$ & $24.73 \pm 0.17$ & \nodata & \nodata & $4288 \pm 5$ & $12.28 \pm 0.15$ & 0.29 &   
\enddata
\tablenotetext{a}{ID number from \citet{kim:2019:109}.}
\tablenotetext{b}{Continuum veiling causes extreme degeneracies with $T_\mathrm{eff}$. Caution should be taken when using derived temperatures with high veiling.}
\tablenotetext{c}{$H$-band veiling parameter.}
\tablenotetext{d}{In this column, 
C = close companion to source ($\lesssim 1\arcsec$);
E3 = escape group 3;
V = RV variable source.}
\end{deluxetable*}
\end{longrotatetable}

\end{document}